\documentclass{JFM-FLM_Au}

\usepackage{bm}
\usepackage{cancel}
\usepackage{upgreek}
\usepackage{subcaption}
\usepackage{afterpage}
\usepackage{nameref}
\usepackage{indentfirst}
\usepackage{enumerate}
\usepackage{tikz}
\usepackage{mathtools, mathrsfs}
\mathtoolsset{showonlyrefs}

\usepackage{etoolbox}

\makeatletter

\patchcmd{\NAT@citex}
  {\@citea\NAT@hyper@{%
     \NAT@nmfmt{\NAT@nm}%
     \hyper@natlinkbreak{\NAT@aysep\NAT@spacechar}{\@citeb\@extra@b@citeb}%
     \NAT@date}}
  {\@citea\NAT@nmfmt{\NAT@nm}%
   \NAT@aysep\NAT@spacechar\NAT@hyper@{\NAT@date}}{}{}

\patchcmd{\NAT@citex}
  {\@citea\NAT@hyper@{%
     \NAT@nmfmt{\NAT@nm}%
     \hyper@natlinkbreak{\NAT@spacechar\NAT@@open\if*#1*\else#1\NAT@spacechar\fi}%
       {\@citeb\@extra@b@citeb}%
     \NAT@date}}
  {\@citea\NAT@nmfmt{\NAT@nm}%
   \NAT@spacechar\NAT@@open\if*#1*\else#1\NAT@spacechar\fi\NAT@hyper@{\NAT@date}}
  {}{}

\makeatother

\hypersetup{
    colorlinks = true,
    urlcolor   = blue,
    citecolor  = blue,
    linkcolor = blue
}

 \newtheorem{theorem}{Theorem}
 \newtheorem{lemma}{Lemma}

\newtheorem{definition}{Definition}

\def\d{\mathrm{d}}
\def\e{\mathrm{e}}
\def\i{\mathrm{i}}
\def\g{\mathrm{g}}

\newcommand\eps           {\varepsilon}

\renewcommand\p            {\partial}

\def\rr{\mathscr{R}}

\DeclareMathOperator{\E}{e}
\DeclareMathOperator{\sign}{sign}

\newcommand{\ep}{\mathbf{e}_{\varphi}}
\newcommand{\et}{\mathbf{e}_{\theta}}
\newcommand{\er}{\mathbf{e}_r}
\newcommand{\s}{\mathbb{S}}

\newcommand{\bu}{\boldsymbol{u}}

\def\ds{\,{\rm d}s}
\def\dt{\,{\rm d}t}
\def\dz{\,{\rm d}z}
\def\dtau{\,{\rm d}\tau}

\def\a{{\rm a}}
\def\R{\mathbb{R}}
\def\N{\mathbb{N}}
\def\C{\mathbb{C}}

\lefttitle{C. Puntini, L. Roberti and E. Stefanescu}
\righttitle{Journal of Fluid Mechanics}
\title{On large-scale wind-drift ocean currents: An asymptotic approach in spherical coordinates}

\author{Christian Puntini\aff{1}, Luigi Roberti\aff{2} \and Eduard Stefanescu\aff{3}}

\affiliation{\aff{1}Faculty of Mathematics, University of Vienna, Oskar--Morgenstern--Platz 1, 1090 Vienna, Austria
\aff{2}Institut f\"ur Angewandte Mathematik, Leibniz Universit\"at Hannover,  Welfengarten 1, 30167 Hannover, Germany
\aff{3} Institut f\"ur Analysis und Zahlentheorie, TU Graz, Steyrergasse 30, 8010 Graz, Austria}

\corresau{Christian Puntini, \email{christian.puntini@univie.ac.at}}
\begin{document}
\maketitle

\begin{abstract} 
Starting from the Navier--Stokes equations in rotating spherical coordinates with depth-varying density and eddy viscosity, we derive an asymptotic model describing non-equatorial wind-generated ocean drift currents. Our approach allows for large-scale flows that cannot be captured by classical tangent-plane approximations. The strategy is to perform a careful scaling and to perform a double asymptotic expansion with respect to two small parameters arising from the scaling: the Rossby number and the ratio between the Ekman depth and the Earth's radius. We obtain a system of linear ordinary differential equations with nonlinear boundary conditions governing the leading-order dynamics, highlighting that the dynamics is governed by the linear terms, whereas the nonlinear ones, related to the injection and dissipation of kinetic energy, appear only at higher order. We use the leading-order equations to compare our model with the simplest theory of ocean circulation due to Sverdrup and note that, even at this level of simplification, our equations have the potential to provide deeper insight. Subsequently, focusing on Ekman flows, we prove existence and uniqueness of the leading-order solution, which retains the classical Ekman spiral structure for arbitrary eddy viscosity profiles. Finally, we compute the surface deflection angle of the wind-driven current for three explicit eddy viscosity profiles, obtaining results consistent with observations. In addition, we derive the governing equations for the first-order correction with respect to the Rossby number and provide \emph{a priori} bounds for its solution.
\end{abstract}

\noindent{\keyfont {\bf MSC Codes:}  35Q30; 35Q35; 76D05; 76U60; 86A05.} \\

\begin{keywords}

\end{keywords}

\section{Introduction}
\noindent
One of the classic problems in physical oceanography concerns the effects of wind forcing on the ocean. Its roots trace back to the pioneering work of \citet{Ekm05}, who identified the crucial role of Earth's rotation in explaining the observation made by Nansen during the \emph{Fram} Arctic expedition of 1893--1896 (see \citet{Nan97} for his own account of the voyage): steady wind-driven ocean currents are deflected to the right of the prevailing wind direction in the Northern Hemisphere. This deflection arises from the momentum balance between the Coriolis acceleration acting on the upper ocean layers and the frictional forces generated by the turbulent stress exerted by the wind.

The study of wind-driven flows has also played a fundamental role in the development of modern theories of large-scale ocean circulation. In particular, the seminal work of \citet{sverdrup47} established the relation between wind stress curl and meridional mass transport in the interior of the ocean. Building on earlier ideas of \citet{Ros36} concerning lateral eddy viscosity, \citet{Munk50} derived a solution for the wind-driven circulation in a closed ocean basin and provided a theoretical explanation for the structure of large-scale ocean gyres. Together with the theory of western boundary intensification introduced by \citet{Sto48}, these developments laid the foundations of dynamical oceanography and remain a cornerstone of our current understanding of the global ocean circulation. For a comprehensive account of these classical theories and their subsequent developments, we refer the reader to \citet{Ped96}.

Despite the central importance of these ideas, theoretical investigations of wind-driven flows have become progressively less prominent in recent years. Increasing attention has been devoted to data-driven analyses and high-resolution numerical simulations, while analytical approaches based on classical fluid mechanics have received comparatively less emphasis (see, e.g.,  \citet{chereskin95}, \citet{NiilerPaduan}, \citet{Poulain}, \citet{YM}, \citet{YMetal}, \citet{KimETAL}, \citet{RohrsC2015}, \citet{SentchevETAL} and \citet{BertaETAL} for recent data-driven studies of Ekman flows). Needless to say, observational and numerical data are indispensable for testing hypotheses and validating theoretical predictions. However, we contend that a careful and systematic (asymptotic) analysis of the governing equations remains crucial for identifying the fundamental balances underlying the observed dynamics and for gaining a deeper understanding of the mechanisms driving such complex phenomena.

Furthermore, most theoretical studies of Ekman flows are formulated within the $f$-plane or $\beta$-plane approximations; see, for instance, \citet{Gri95}, \citet{lewis2004}, \citet{DriPalCon20}, \citet{Con21}, \citet{Rob22,luigi-time}, \citet{Mar22}, \citet{Ste24} and \citet{mioGoverning, c_nlarwa}. These approximations are obtained through the \emph{ad hoc} neglect of selected terms in the governing equations and, although they have proved useful in many contexts, their range of validity is intrinsically limited, since the underlying assumptions are appropriate only for flows of sufficiently small horizontal extent. Therefore, it is desirable to have a model in full spherical coordinates that may be able to capture large-scale flows as well. Unfortunately, the full Navier--Stokes equations in spherical coordinates are intractable analytically; thus, it is necessary to make some approximations while trying to minimise the assumptions required. In this paper, we develop such a framework and apply it to the classical problem of wind-driven ocean currents.

The approach we will pursue relies on asymptotic expansions and has had a well-documented success in fluid mechanics; see, e.g., \citet{BalmforthLiu}, \citet{nonNewt} and \citet{JFM884A43} for applications to non-Newtonian fluids or \citet{Peregrine}, \citet{J1980}, \citet{CJderivation} and \citet{KP} in the context of water waves. In geophysical fluid dynamics, this method was used by \citet{ConJoh17} for the description of ocean gyres and has subsequently been successfully applied to a variety of problems; see \citet{CJ2018,CJ2018b, CJ2019, CJ2021, CJ2023, CJ2024} and \citet{ZhangZhang}. For a detailed review of similar asymptotic methods in oceanography, we refer to \citet{JohnsonOceanography,J2022}.

In the present work, we proceed as follows. We begin with the Navier--Stokes equations, expressed in a rotating system of spherical coordinates, along with the equation of mass conservation and appropriate boundary conditions. Two assumptions are made \emph{a priori}: we model turbulent stresses using an eddy viscosity parametrisation, with the eddy viscosity assumed to depend only on depth, and we assume the density to vary only with depth too. The relevance of such depth-dependent viscosity profiles has been emphasised by \citet{CroninKessler2009} and \citet{CroninTozuka2016}. Then, we carefully rescale all the quantities appearing in the equations and identify two small parameters, which are independent of each other: the \emph{thin shell parameter} ($\varepsilon$) and the \emph{Rossby number} ($\mathscr{R}$). These two parameters emerge naturally from the Navier-Stokes equations in spherical coordinates via the non-dimensionalisation. The former represents the ratio of a depth scale---in our case, the so-called \textit{Ekman depth}---and the radius of the Earth, and is guaranteed to be small regardless of the nature of the flow; the latter is a well-known parameter in physical oceanography that is typically small for large-scale flows strongly affected by the Coriolis force (cf. the discussion by \citet{NumOcean}). As noted by \citet{CJ2019}, performing an asymptotic expansion first in the thin-shell parameter and then in the Rossby number effectively separates the dynamical effects of the geometry (via the thin-shell parameter) and the properties of the flow due to the dominant time and length scales (via the Rossby number). A further advantage of the asymptotic approach adopted here is that it provides a systematic procedure for computing the governing equations at successive orders of approximation. In the present work, we derive the leading-order equations and obtain several explicit solutions, together with bounds for first-order correction with respect to $\mathscr{R}$. More importantly, the asymptotic hierarchy can, in principle, be continued to arbitrarily high order. Since the equations arising at each order are explicitly computable, numerical approximations of any desired accuracy may be obtained by retaining sufficiently many terms in the expansion.

We believe that the combination of mathematical rigour and physical consistency underlying our derivation and analysis provides a valuable contribution to the modelling of wind-driven ocean currents. In addition, the framework developed here may serve as a foundation for future investigations of large-scale ocean circulation. Indeed, the use of spherical coordinates offers the possibility of overcoming some of the limitations inherent in the classical $f$-plane and $\beta$-plane theories, and may therefore lead to a more accurate description of basin-scale and global ocean dynamics.

\subsection*{Plan of the paper}

The paper is organised as follows. In §~\ref{section governing equations}, after some remarks on the frame of reference and the coordinate system, we first present the governing equations and discuss in detail the boundary conditions, then proceed to rescaling all quantities and non-dimensionalising the equations and the boundary conditions, thereby identifying the thin-shell parameter and the Rossby number. Then, §~\ref{section asymptotics} is devoted to the derivation of the leading order governing equation through asymptotics: first, we make an asymptotic expansion of all variables with respect to the thin-shell parameter; then, within the leading order equations thus derived, we perform a second asymptotic expansion---now with respect to the Rossby number. In §~\ref{section Sverdrup}, we show how our formulation in spherical coordinates can be used to generalise the classic theory of the ocean circulation in the $\beta$-plane due to \citet{sverdrup47} to large-scale flows. In §~\ref{section Ekman}, we derive the leading order problem for the Ekman flow away from the equator by splitting the solution into a geostrophic and an Ekman part. In particular, in §~\ref{section analytical results} we prove the existence and uniqueness of the solution to the problem for the leading order Ekman flow and show that this solution describes a classical Ekman spiral and analyse the surface deflection angle (that is, the angle between the direction of the wind and the induce wind-drift current on the surface); then, in §~\ref{section explicit} we discuss some concrete examples of explicit eddy viscosity profiles. Here we can explicitly solve the equation, draw numerical plots that help to visualise the flow, and calculate the surface deflection angle using the theory developed in §~\ref{section analytical results}. The proofs of the results of §~\ref{section analytical results} and the calculations of §~\ref{section explicit} are deferred to Appendix \ref{appendix proofs} and Appendix \ref{Appendix explicit}, respectively. Appendix \ref{appendix first order} provides the derivation of the governing equations of the first-order correction to the leading order equations with respect to the Rossby number $\mathscr{R}$, along with some ideas on how to provide \emph{a priori} bounds on this first-order correction only in terms of the leading-order solution. Finally, Appendix \ref{spirals appendix} includes some 3D plots of the Ekman spirals of  §~\ref{section explicit}.

\section{Governing equations} \label{section governing equations}

In this section, we will first present the equations of motion and the boundary conditions, then we will introduce an appropriate scaling and non-dimensionalise the problem. In the following, we use primes to denote physical (dimensional) quantities; we will remove them when we rescale the variables.

\begin{figure}
    \centering
    \includegraphics[width=0.55\linewidth]{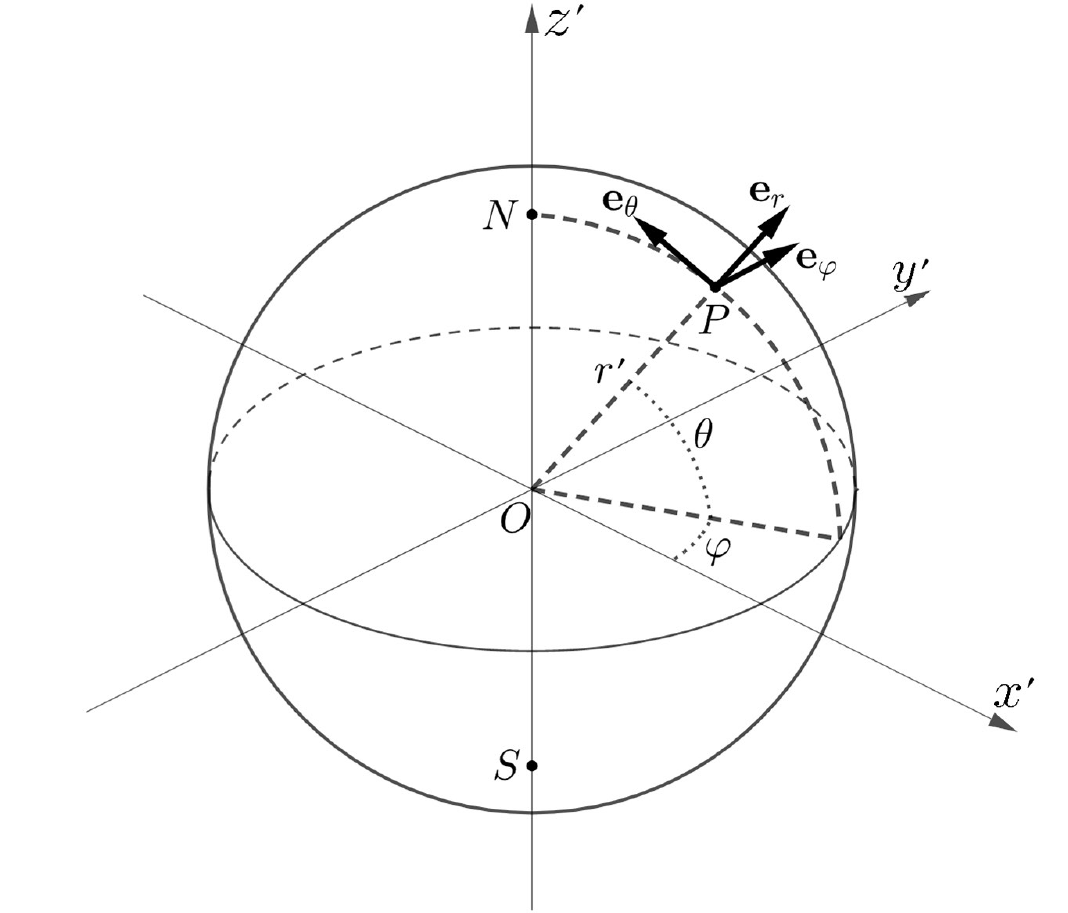}
    \caption{The classical spherical coordinate system.}
    \label{fig-spherical}
\end{figure}

First, let us discuss the frame of reference in which the problem is formulated. A suitable coordinate system is the system of right-handed spherical coordinates $(\varphi, \theta, r')$, where $\theta \in \left[-\frac{\pi}{2},\frac{\pi}{2}\right]$ is the angle of latitude, $\varphi \in [0,2\pi)$ is the angle of longitude, and $r'\in[0,\infty)$ is the radial distance from the origin; see figure~\ref{fig-spherical} for a sketch of these spherical coordinates. The coordinate change from these spherical coordinates to the standard Cartesian coordinates $(x',y',z')$ is given by
\begin{equation}
\left\{\begin{aligned}
    x' &= r' \cos\theta \cos \varphi,\\[0.2em]
    y' &= r' \cos\theta \sin \varphi,\\[0.2em]
    z' &= r' \sin\theta ,
\end{aligned}\right.\qquad \text{with inverse}\qquad 
\left\{ \begin{aligned}
    &\varphi = \tan^{-1}\left(\frac{y'}{x'}\right),\\[0.2em]
    &\theta = \sin^{-1}\left(\frac{z'}{\sqrt{x'^2+y'^2+z'^2}}\right).\\[0.2em]
    & r'=\sqrt{x'^2+y'^2+z'^2}.
\end{aligned}\right.\end{equation}
To each point for which $r>0$ and $\theta \notin \{\pm\frac{\pi}{2}\}$ we can associate the usual local orthonormal basis for $\R^3$ of unit vectors $(\ep, \et, \er)$: $\ep $ points eastward, $\et$ northward, and $\er $ radially. The velocity field is denoted by $\bu'=u'\ep + v'\et + w'\er$. Note that we must take care to exclude the poles, where the tangent vectors $\ep$ and $\et$ are not defined, as a consequence of the Hairy Ball Theorem \citep{Ren13}; flows that extend to the poles may be treated, for instance, by rotating the coordinate system and moving the singularities to the equator, as done by \citet{Con24} and \citet{mioGoverning}, but we will not pursue this in the present paper. The vectors $(\ep, \et, \er)$ are related to the Cartesian unit basis vectors $(\mathbf{e}_1, \mathbf{e}_2, \mathbf{e}_3)$ through
\begin{equation} \label{base sferiche}
\left\{ \begin{aligned}
    &\mathbf{e_{\varphi}}=-\sin\varphi\,\mathbf{e}_1+\cos\varphi\,\mathbf{e}_2,\\[0.2em]
    &\mathbf{e_{\theta}}=-\cos\varphi\sin\theta\,\mathbf{e}_1-\sin\varphi\sin\theta\,\mathbf{e}_2+\cos\theta\,\mathbf{e}_3,\\[0.2em]
    &\mathbf{e}_{r}=\cos\varphi\cos\theta\,\mathbf{e}_1+\sin\varphi\cos\theta\,\mathbf{e}_2+\sin\theta\,\mathbf{e}_3.
\end{aligned}\right.
\end{equation}

Moreover, it is crucial to point out that, denoting by $\bm{\Omega}'$ the Earth's constant angular velocity, with magnitude $\Omega' = |\bm{\Omega}'| \approx 7.29\cdot10^{-5} \,\mathrm{rad\,s^{-1}}$, we choose the reference system so that it rotates about the $z'$ axis, which is aligned with $\bm{\Omega}'$, at angular speed $\Omega'$; as a result, the Earth appears stationary. Although this choice is advantageous in many ways, it has the drawback of not being inertial, which---as discussed, for example, by \citet{Pedlosky}, \citet{Phillips} and \citet{Vallis}---will give rise to additional terms (corresponding to apparent forces) in the momentum equations, namely, the Coriolis force $2\bm{\Omega}'\times\bu'$ and the centrifugal force $\bm{\Omega}' \times (\bm{\Omega}' \times \mathbf{r}')$, where $\mathbf{r}'$ is the position vector.

Since, for most geophysical flows, the Earth's oblateness is not relevant (see the discussions by \citet{White2002} and \citet{CJ2021}), we would like to approximate the Earth's surface as a sphere of radius $R' \approx 6371\, \mathrm{km}$. However, some care is needed in order to minimise the dynamical error due to this approximation. Let us first define the \emph{geopotential} $\Phi'$ as the sum of the gravitational potential and the term $-\frac{1}{2}\Omega'^2\ell^2$, where $\ell$ is the distance of $\mathbf{r}'$ from the axis of rotation. Over time, Earth has developed an equatorial bulge to counteract the centrifugal force, making the \emph{geopotential force} (or \emph{effective gravity}) $\boldsymbol{g}' = -\nabla\Phi'$ act everywhere nearly perpendicularly to the Earth's surface; in other words, the Earth's surface is essentially a level set of the geopotential $\Phi'$. Therefore, we choose to treat the level sets of the geopotential $\Phi$ (which, in reality, would be rotation ellipsoids of tiny eccentricity) as if they were perfect spheres. Thus, in this approximation, the horizontal component of effective gravity vanishes, and we replace a potentially significant dynamic error with a minor geometric error; the geopotential $\Phi'$ then depends solely on the vertical coordinate, and, for many applications, we can approximate $\Phi'(r')=g'r'$. Given that the Earth's oblateness is minimal, the spherical approximation provides a highly accurate representation of the spheroid.

\subsection{The equations of motion} \label{section 2.1}
We can now present the equations of motion. In addition to the velocity field $\bu'=u'\ep + v'\et + w'\er$ introduced previously, we denote by $\rho'$ the density and by $p'$ the pressure. We assume that the density varies only with depth, while still treating the fluid as incompressible, since a pressure change of about $500\,\mathrm{kPa}$ leads to only a $0.025\%$ variation in the density of water (see \cite{maslowe} and \citet{Vallis}). Therefore, the mass conservation equation reduces to the incompressibility condition, which in spherical coordinates reads
\begin{equation}\label{continuity2}
  \frac{1}{r' \cos \theta} \frac{\partial u'}{\partial \varphi} + \frac{1}{r' \cos \theta} \frac{\partial}{\partial \theta} \left( v' \cos \theta  \right) + \frac{1}{r'^2} \frac{\partial}{\partial r'} \bigl( r'^2 w' \bigr)=0.
\end{equation}
In addition to incompressibility, in the equations of motions we assume the fluid flow to be laminar, with the turbulent effects parametrised by the dynamic (effective) horizontal and vertical viscosities $\mu'_{\rm h}$ and $\mu'_{\rm v}$, which, like the density, are assumed to vary only with respect to the depth. We note that the concept of eddy viscosity represents a simple model for the closure of the turbulence problem, based on the hypothesis that Reynolds stresses are proportional to the mean velocity gradient (and thus depend on flow conditions) but do not represent a physical characteristic of the fluid \citep{Pedlosky, Pope}; we will make additional remarks on the eddy viscosity in §~\ref{section scaling} and §~\ref{section discussion}. With these assumptions, the Navier--Stokes equations in spherical rotating coordinates are (see the derivation by \citet{mioGoverning}):
\begin{equation}\label{NS 2}
\begin{aligned}
&\left(\frac{\partial}{\partial t'}+\frac{u'}{r'\cos\theta}\frac{\partial}{\partial\varphi}+\frac{v'}{r' }\frac{\partial}{\partial\theta}+w'\frac{\partial}{\partial r'}\right)\begin{pmatrix}
u' \\[0.2em]
v' \\[0.2em]
w'
\end{pmatrix} \\
&\quad+\frac{1}{r'}\begin{pmatrix}-u'v'\tan \theta + u'w' \\[0.2em]
u'^2\tan\theta+v'w'\\[0.2em]
-u'^2-v'^2
\end{pmatrix}+2\Omega'\begin{pmatrix}
    -v'\sin\theta+w'\cos\theta\\[0.2em]
    u'\sin\theta\\[0.2em]
    -u'\cos\theta
\end{pmatrix}\\
&\quad= -\frac{1}{\rho'}\begin{pmatrix}
   \frac{1}{r'\cos\theta}\frac{\partial p'}{\partial\varphi}\\[0.2em]
    \frac{1}{r'}\frac{\partial p'}{\partial\theta}\\[0.2em]
    \frac{\partial p'}{\partial r}
\end{pmatrix} -\frac{g'R^{\prime 2}}{r^{\prime 2}} \begin{pmatrix}
    0\\[0.2em] 0\\[0.2em] 1
\end{pmatrix} + \frac{\mu'_{\rm v}}{\rho'} \left(\frac{\partial^2}{\partial r'^2}+ \frac{2}{r'}\frac{\partial}{\partial r'}\right)\begin{pmatrix}
    u'\\[0.2em] v'\\[0.2em] w'
\end{pmatrix}\\
&\quad\quad + \frac{\mu'_{\rm h}}{\rho' r'^2}\left(\frac{1}{ \cos^2 \theta} \frac{\partial^2 }{\partial \varphi^2}+\frac{\partial^2}{\partial \theta^2}   -  \tan\theta\frac{\partial}{\partial \theta}\right)
\begin{pmatrix}
    u'\\[0.2em] v'\\[0.2em] w'
\end{pmatrix}\\
&\quad\quad -\frac{1}{\rho'r'^2\cos^2\theta}\begin{pmatrix}
    \mu'_{\rm h} u\\[0.2em]
    \mu'_{\rm h} v\\[0.2em]
    2\mu'_{\rm v}(w'\cos^2\theta-v'\sin\theta\cos\theta)
\end{pmatrix}+\frac{2\mu'_{\rm h}}{\rho'r'^2}\frac{\partial}{\partial \theta}\begin{pmatrix}
0\\[0.2em]
 w'\\[0.2em]
   -v'
\end{pmatrix}\\
&\quad\quad+\frac{2\mu'_{\rm h}}{\rho'r'^2\cos\theta}\frac{\partial}{\partial \varphi}\begin{pmatrix}
w'-v'\tan\theta\\[0.2em]
u'\tan\theta\\[0.2em]
-u'
\end{pmatrix}+\frac{1}{\rho'} \frac{\d \mu'_{\rm v}}{\d r'} \begin{pmatrix}
  r'  \frac{\partial}{\partial r'}\bigl(\frac{u'}{r'}\bigr)\\[0.5em]
      r' \frac{\partial}{\partial r'}\bigl(\frac{v'}{r'}\bigr)\\[0.5em]
        2 \frac{\partial w'}{\partial r'}  
\end{pmatrix}+\frac{1}{\rho'} \frac{\d\mu'_{\rm h}}{\d r'} \begin{pmatrix}
    \frac{1}{r'\cos \theta}\frac{\partial  {w'}}{\partial \varphi}\\[0.5em]
     \frac{1}{r'} \frac{\partial {w'}}{\partial \theta}\\[0.5em]
{0}
\end{pmatrix},
\end{aligned}
\end{equation}
where $R' \approx 6371\, \mathrm{km}$ is the Earth's average radius and $g' \approx 9.81 \, \mathrm{m\, s^{-2}}$ is the average gravitational acceleration at ground level. It is worth remarking that the Navier-Stokes equation \eqref{NS 2}, derived by \cite{mioGoverning} to correct some errors generally present in the literature, could be further generalised. Indeed, the tensor related to the eddy viscosity considered by \cite{mioGoverning} is isotropic in the horizontal directions and is assumed to vary only with depth. Accounting for a more general eddy viscosity tensor (in the vein of recent works in the $f$-plane approximation, such as those by \citet{Wirth2010} or \citet{BykovGordin}), while technically possible and undoubtedly interesting, requires extensive calculations that would exceed the scope of this work. Therefore, for the time being, we content ourselves with the simpler stress tensor from \citet{mioGoverning} and leave the more general case for future investigations.

\subsection{Boundary conditions}

Let us now discuss the boundary conditions that characterise the dynamics of wind-driven ocean currents. At the free surface of the ocean, which we assume to be described by the equation
\[r'=R'+h'(\varphi,\theta, t')\]
for an unknown function $h'$, we impose the dynamic boundary condition
\begin{equation}\label{dynamicBX}
    p'=p_{\rm s}\qquad  \text{on $\{r'=R'+h'\}$},
\end{equation}
where $p_{\rm s}$ is the given (and, in general, non-constant) surface (atmospheric) pressure, and the kinematic boundary condition
\begin{equation}\label{KBC}
    w'=\frac{\p h'}{\p t'}+\frac{u'}{r'\cos\theta}\frac{\p h'}{\p \varphi}+\frac{v'}{r'}\frac{\p h'}{\p \theta}\qquad  \text{on $\{r'=R'+h'\}$},
\end{equation}
which ensures that no fluid particles escape the fluid via the free surface. On the rigid bottom, which for simplicity we assume to be flat, we require the velocity field to vanish:
\begin{equation} \label{bottom BC}
    \bu' = 0 \qquad \text{on $\{r'=R'+H'\}$},
\end{equation}
for a constant $H' < h'$.
  
\begin{figure}[t!]
    \centering
    \includegraphics[width=0.8\linewidth]{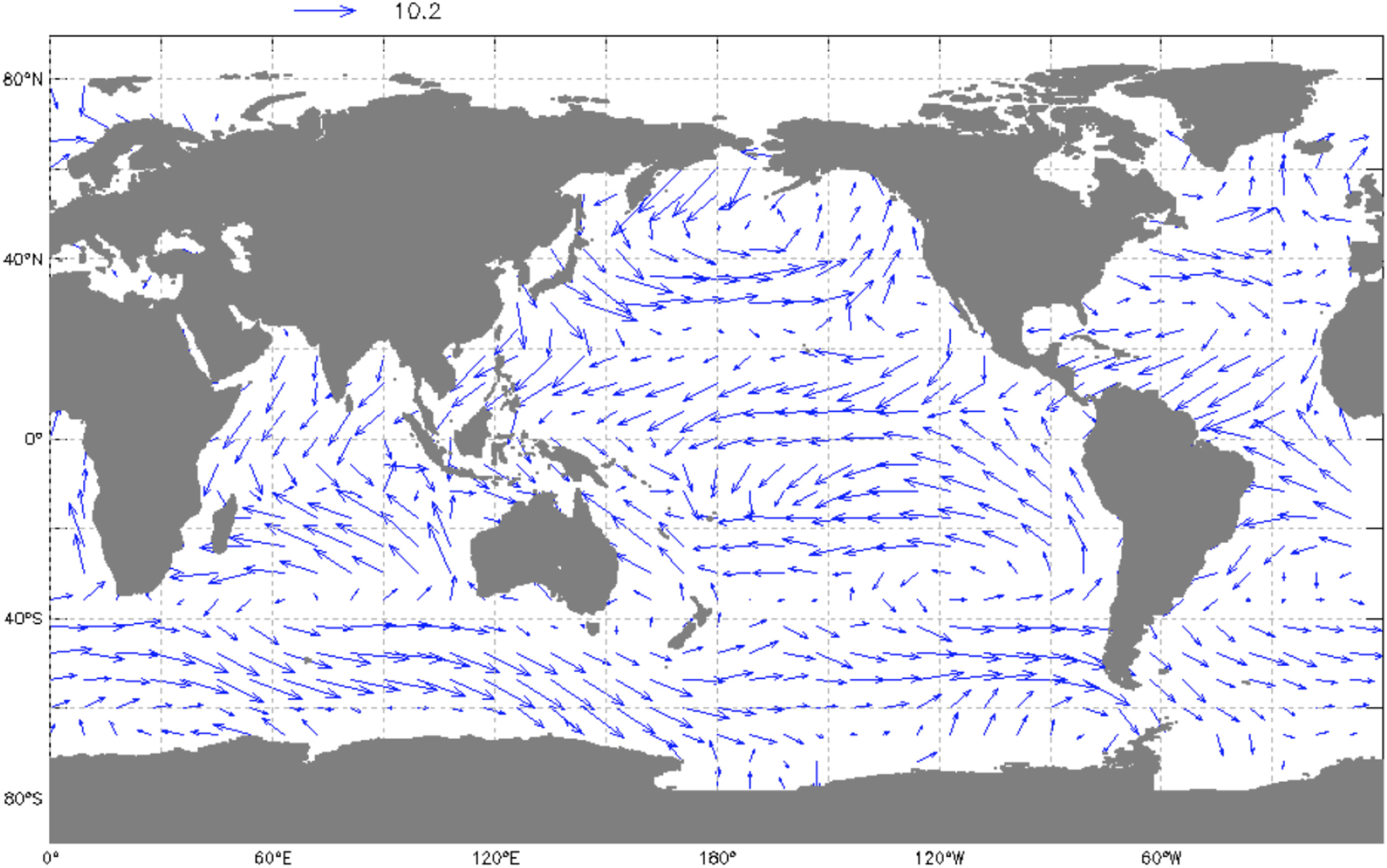}
    \caption{Monthly-averaged ocean wind speed and direction vectors, with vector lengths proportional to the reference scale (in $\mathrm{m\, s^{-1}}$), based on observations from NASA's QuikSCAT satellite. Image credit: NOAA.}
    \label{wind}
\end{figure}

The stress generated by the wind blowing over the ocean's surface $\{r = R'+h'\}$ is expressed by the bulk formula \citep{S-LBook}
\begin{equation}\label{stress2}
    \boldsymbol{\tau}'_{\rm a}=\rho'_{\rm a} C_{\rm aw} |\bu'_{\rm a}-\bu'|(\bu'_{\rm a}-\bu') \qquad \text{on $\{r'=R'+h'\}$},
\end{equation}
where $\rho'_\a \approx 1.2\,$kg\,m$^{-3}$ is the density of the air, $C_{\rm aw}\approx10^{-3}$ is the non-dimensional drag coefficient of air over water and $\bu'_{\rm a}= u'_{\rm a}\ep + v'_{\rm a}\et$ is the horizontal wind velocity, conventionally measured at $10\,\mathrm{m}$ above the sea surface. See figure~\ref{wind} and \citet{annualWind} for data on the mean wind velocity over the ocean.

The bulk formula \eqref{stress2} provides us with an additional boundary condition for our problem. In fact, on the surface of the ocean, the shear stress is given by
\begin{equation}\label{stress1}
\boldsymbol{\tau}'_{\rm w}=\mu'_{\mathrm{v}}\left(\frac{\p u'}{\p r'}\ep + \frac{\p v'}{\p r'}\et\right) \qquad  \text{on $\{r'=R'+h'\}$},
\end{equation}
where $\mu'_{\rm v}$ is the vertical dynamic viscosity, and consequently, in view of \eqref{stress2}, the continuity of the stress across the water surface yields
\begin{equation}\label{stress3}
    \mu'_{\mathrm{v}}\begin{pmatrix}
    \dfrac{\p u'}{\p r'}\\[1em]
    \dfrac{\p v'}{\p r'}
    \end{pmatrix} =\rho'_{\rm a} C_{\rm aw} \sqrt{(u_{\rm a}'-u')^2+(v_{\rm a}'-v')^2}\begin{pmatrix}
    u'_\a-u'\\[0.2em]
    v'_\a-v'
\end{pmatrix} \qquad  \text{on $\{r'=R'+h'\}$}.
\end{equation}

\subsection{Scaling and non-dimensionalisation} \label{section scaling}
Clearly, the Navier--Stokes equations \eqref{NS 2} are intractable by analytical means in their full generality. In order to simplify them, we will be guided by the scaling of the variables appearing in the equations; through this, we will be able to identify the terms in the equations that contribute the most and those that, on the contrary, may be neglected with little error. To this end, we must identify the relevant scales of the flow.

Before we proceed, some clarifications on the concept of eddy viscosity are necessary, because the physical interpretation of a local eddy viscosity at the ocean surface is not straightforward. Near the air-sea interface, wave breaking and other surface processes can induce non-local momentum transport, so that turbulent stresses are not necessarily related to the local mean shear through a simple diffusive closure \citep{Craig}. Furthermore, the specification of an eddy viscosity at the ocean surface is inherently ambiguous. As an example, KPP turbulence closures predict that the mixing length, and thus the eddy viscosity, vanishes as one approaches the surface \citep{Large94}, whereas field observations suggest that wave-induced processes may substantially enhance near-surface mixing beyond the predictions of such models \citep{Fisher}. At the same time, direct observations in the immediate vicinity of the air-sea interface are extremely challenging. Consequently, no single universally accepted value can be assigned to the surface eddy viscosity, and estimates may vary depending on the modelling framework and the representation of near-surface processes. In this work, the adopted value should therefore be interpreted as a convenient reference scale for non-dimensionalisation within the present modelling framework, rather than as a directly measurable physical quantity.

With this \emph{caveat}, let us denote by $\bar{\mu}'_{\rm h}$ and $\bar{\mu}'_{\rm v}$ the reference values for the horizontal and vertical dynamic eddy viscosity on the surface, respectively. A typical value for $\bar{\mu}'_{\rm v}$ in the near-surface ocean is $\bar{\mu}'_{\rm v} \approx 10\,\mathrm{kg\,m^{-1}\,s^{-1}}$ \citep{KBook,WMcPL2014}, although observations indicate substantial variability \citep{Huang1979}. A reference value for the horizontal (dynamic) eddy viscosity is even more difficult to determine; nevertheless, horizontal mixing greatly exceeds vertical mixing, with $\bar{\mu}'_{\rm h}/\bar{\mu}'_{\rm v}\sim 10^3$ \citep{Pedlosky, Talley, CJ2018, C2022}, implying $\bar{\mu}'_{\rm h}\gg \bar{\mu}'_{\rm v}$. However, because we non-dimensionalise the equations, their precise values are not essential. Furthermore, we denote by 
\begin{equation}\label{D scale}
    D'=\sqrt{\frac{\Bar{\mu'_{\rm v}} }{\bar{\rho}'\Omega'}}\approx 12\, \mathrm{m}
\end{equation}
the \emph{Ekman depth} and by $\bar{\rho}' = 1.025\,{\rm kg\,m^{-3}}$ the average seawater density at the surface. The choice for the speed scale $U'$ arises naturally from the equation \eqref{stress3} for the wind stress on the surface. In fact, using \eqref{scaling transformation} and setting $(u_\a',v_\a') = U_\a'(u_\a,v_\a)$ for an average wind speed $U_\a$ (for instance, $U_\a' = 10\,{\rm m\,s^{-1}}$), it follows immediately that the appropriate scale for $U'$ is
\begin{equation} \label{U'}
    U' = \frac{\rho'_{\rm a} C_{\rm aw} U_\a^{\prime 2}D'}{\bar{\mu}'_{\rm v}}.
\end{equation}
Substituting the values mentioned above and taking $U'_\a = 10\,\mathrm{m\,s^{-1}}$, we find
\begin{equation}
    U' \approx 0.14\,\mathrm{m\,s^{-1}}.
\end{equation}
\noindent
Based on these scales, we introduce two non-dimensional parameters
\begin{equation}
    \varepsilon=\frac{D'}{R'}\approx 2\cdot 10^{-6} \qquad \text{and} \qquad \rr = \frac{U'}{\Omega' R'}\approx 2\cdot 10^{-4};
\end{equation}
the former is the \emph{thin-shell parameter} (the ratio of the Ekman length and the Earth's average radius), and the latter is the \emph{Rossby number} (a well-known parameter in oceanography, defined as the ratio between the scales of the advective terms and the Coriolis terms in the Navier--Stokes equations). Intuitively, a small Rossby number indicates that the advective terms are dominated by the Coriolis terms; this is usually the case for large-scale oceanic flows (cf. the discussion by \citet{NumOcean} and \citet{Vallis}). 

In view of these considerations, we can non-dimensionalise the equations \eqref{continuity2} and \eqref{NS 2} by setting
\begin{equation} \label{scaling transformation}
\begin{gathered}
    r'=R'+D'z, \quad (u',v',w')=U'(u,v,\varepsilon w), \quad \rho'(r') = \bar{\rho}'\rho(z), \\ 
    p'= \bar{\rho}'U'\Omega'R'p, \quad t'=\frac{R'}{U'}\,t,\quad  \mu'_{\rm h}(r')=\Bar{\mu}'_{\rm h} n(z), \quad   \mu'_{\rm v}(r')=\Bar{\mu}'_{\rm v}m(z).  
\end{gathered}
\end{equation}
The factor $\varepsilon$ in front of $w$ ensures the consistency of the scaling. Indeed, given the particle trajectories $t'\mapsto \upPhi(t') = (\varphi(t'),\theta(t'),r'(t'))$ in spherical coordinates, we have $w'(\upPhi(t'),t') = \frac{\d r'(t')}{\d t'}$, hence, taking \eqref{scaling transformation} into account and setting $w = \frac{\d z}{\d t}$, it follows $w' = \eps U'w$.

With \eqref{scaling transformation}, the first two components of \eqref{NS 2} are
\begin{equation}\label{NS uv red}
\begin{aligned}
&\rr\left(\frac{\partial}{\partial t}+\frac{u}{(1+\eps z)\cos\theta}\frac{\partial}{\partial\varphi}+\frac{v}{1+\eps z}\frac{\partial}{\partial\theta}+w\frac{\partial}{\partial z}\right)\begin{pmatrix}
u \\[0.2em]
v 
\end{pmatrix} \\
&+\frac{\rr}{1+\varepsilon z}\begin{pmatrix}-uv\tan \theta + \eps uw \\[0.2em]
u^2\tan\theta+\eps vw
\end{pmatrix}+2 \begin{pmatrix}
    -v\sin\theta+\eps w\cos\theta\\[0.2em]
    u\sin\theta
\end{pmatrix} \\
&\quad = -\frac{1}{1+\eps z}\begin{pmatrix}\frac{1}{\rho\cos\theta}\frac{\partial p}{\partial\varphi} \\[0.5em] \frac{1}{\rho}\frac{\partial p}{\partial\theta}\end{pmatrix}+ 
 \frac{m}{\rho} \left(\frac{\partial^2}{\partial z^2}+ \frac{2\eps}{(1+\varepsilon z)}\frac{\partial}{\partial z}\right)\begin{pmatrix}
    u \\[0.2em] v 
\end{pmatrix} \\
&\quad\quad +\frac{{\eps^2\upnu} n}{\rho(1+\eps z)^2} \left(\frac{1}{ \cos^2 \theta} \frac{\partial^2 }{\partial \varphi^2}+\frac{\partial^2}{\partial \theta^2}   -  \tan\theta\frac{\partial}{\partial \theta}\right)
\begin{pmatrix}
    u\\[0.2em] v
\end{pmatrix}\\
&\quad\quad -\frac{{\eps^2\upnu} n}{\rho(1+\eps z)^2\cos^2\theta}\begin{pmatrix}
  u\\[0.2em]
     v
\end{pmatrix}+\frac{2 {\eps^2\upnu} n}{\rho(1+\eps z)^2} \left(\frac{\partial}{\partial \theta}\begin{pmatrix}
0\\[0.2em]
 \eps w
\end{pmatrix}+\frac{1}{\cos\theta}\frac{\partial}{\partial \varphi}\begin{pmatrix}
\eps w-v\tan\theta\\[0.2em]
u\tan\theta
\end{pmatrix}\right) \\
&\quad\quad + \frac{1}{\rho}\frac{\d  m}{\d z} \begin{pmatrix}
\frac{\partial u}{\partial z}-\frac{u\eps}{(1+\eps z)^2}\\[0.5em]
    \frac{\partial v}{\partial z}-\frac{v\eps}{(1+\eps z)^2}
\end{pmatrix}+ \frac{\eps^{2}\upnu}{\rho (1+\eps z)}\frac{\d  n}{\d z} \begin{pmatrix}
    \frac{1}{\cos \theta}\frac{\partial  {w}}{\partial \varphi}\\[0.5em]
     \frac{\partial {w}}{\partial \theta}
\end{pmatrix},
\end{aligned}
\end{equation}
where we denoted the ratio between the scales of the dynamic eddy viscosities by
$\upnu=\frac{\bar{\mu'_{\rm h}} }{\bar{\mu'_{\rm v}} }$, while, writing
\begin{equation}
    g = \frac{g'D'}{U'\Omega'R'} =\mathcal{O}(1),
\end{equation}
the third component of \eqref{NS 2} is
\begin{equation}\label{NS w red}
\begin{aligned}
&\eps^2\rr\left(\frac{\partial w}{\partial t}+\frac{u}{(1+\eps z)\cos\theta}\frac{\partial w}{\partial\varphi}+\frac{v}{(1+\eps z)}\frac{\partial w}{\partial\theta}+ w\frac{\partial w}{\partial z}\right) -\frac{\eps\rr}{1+\varepsilon z} (u^2+v^2)-2  u \eps\cos\theta\\
&\quad= - \frac{1}{\rho}\frac{\partial p}{\partial z} - \frac{g}{(1+\eps z)^2} + \eps^2 \frac{m}{\rho} \left(\frac{\partial^2 w}{\partial z^2}+ \frac{2\eps}{(1+\varepsilon z)}\frac{\partial w}{\partial z}\right) \\
&\quad\quad+\frac{{\eps^{4}\upnu}n}{\rho(1+\eps z)^2}\left(\frac{1}{ \cos^2 \theta} \frac{\partial^2 w }{\partial \varphi^2}+\frac{\partial^2 w}{\partial \theta^2} - \tan\theta\frac{\partial w}{\partial \theta}\right)\\
&\quad\quad-\frac{2\eps^3m}{\rho(1+\eps z)^2\cos^2\theta}(\eps  w \cos^2\theta-v\sin\theta\cos\theta) \\
&\quad\quad-\frac{2 {\eps^3\upnu} n}{\rho(1+\eps z)^2} \left(\frac{\partial v}{\partial \theta} +\frac{1}{\cos\theta}\frac{\partial u}{\partial \varphi} \right) + 2\frac{\eps^2}{\rho} \frac{\d  m}{\d z} \frac{\partial w}{\partial z}.
\end{aligned}
\end{equation}
Similarly, from the continuity equation \eqref{continuity2} we end up with the non-dimensional equation
\begin{equation}\label{continuity adim 1}
  (1+\eps z)\left[  \frac{\partial u}{\partial \varphi} + \frac{\partial}{\partial \theta} \left( v \cos \theta  \right)\right] + \cos\theta \frac{\partial}{\partial z} \bigl( (1+\eps z)^2 w \bigr)=0.
\end{equation}

Analogously to the equations of motion, we have to non-dimensionalise the boundary conditions as well. In order to do so, we start from the kinematic boundary condition. Setting $h'=D'h$ and $H' = D'H$, \eqref{KBC} reads
\begin{equation}\label{KBC Adim}
w = \frac{\p h}{\p t}+\frac{u}{(1+\varepsilon z)\cos\theta}\frac{\p h}{\p \varphi}+\frac{v}{(1+\varepsilon z)}\frac{\p h}{\p \theta}\qquad  \text{on $\{z=h\}$},
\end{equation}
while, setting $\bu = u\ep + v\et + w\er$, \eqref{bottom BC} becomes
\begin{equation} \label{bottom BC nondim}
    \bu = 0 \qquad \text{on $\{z=H\}$}.
\end{equation}
For the dynamic boundary condition, in view of \eqref{dynamicBX} and \eqref{scaling transformation}, on the free surface we have
\begin{equation} \label{DBC nondim}
    p = p_{\rm s} \qquad \text{on $\{z=h\}$},
\end{equation}
where
\begin{equation}
    p_{\rm s} = \frac{p'_{\rm s}}{\rho'U'\Omega'R'}.
\end{equation}
Finally, setting
\begin{equation}
    \sigma = \frac{U'}{U_\a'} \approx 0.014,
\end{equation}
from \eqref{U'} we see that the non-dimensional form of \eqref{stress3} is
\begin{equation}\label{stress adimensionalizzato}
    m\begin{pmatrix}
        \dfrac{\p u}{\p z}\\[1em]
        \dfrac{\p v}{\p z}
    \end{pmatrix} = \sqrt{(u_{\rm a}-\sigma u)^2+(v_{\rm a}-\sigma v')^2}\begin{pmatrix}
        u_\a-\sigma u\\[0.2em]
        v_\a-\sigma v
    \end{pmatrix} \qquad  \text{on $\{z=h\}$}.
\end{equation}

\section{Derivation of the leading-order equations}\label{section asymptotics}

Now that the equations have been non-dimensionalised, we would like to exploit the smallness of the parameters $\eps$ and $\rr$ to simplify the problem using asymptotic expansions, which should provide us at least with viable approximations of the full solution; for more details on the general theory, we refer to \citet{VanDyke1975}, \citet{JBook} and \citet{Holmes}. First, let us forget about $\rr$ for the moment and make the ansatz
\begin{equation} \label{as ansatz}
    q(\varphi,\theta,z,t;\eps) = q_0(\varphi,\theta,z,t) + o(\eps) \qquad \text{as $\eps\searrow 0$},
\end{equation}
where $q$ (and, correspondingly, $q_0$) represents each of the variables $u,\,v,\,w,\,P,\,h$. This ansatz is motivated by the structure of \eqref{NS uv red} and \eqref{NS w red}, where $\eps$ appears only in integer powers, as well as through $(1+\eps z)^{-1} = 1 - \eps z + o(\eps)$. Plugging \eqref{as ansatz} into \eqref{NS uv red} and equating the coefficients of the various powers of $\eps$, we obtain, at leading order with respect to $\eps$,
\begin{equation} \label{uv asymptotic 1}
\begin{aligned}
&\rr\left\{\left(\frac{\partial}{\partial t}+\frac{u_0}{\cos\theta}\frac{\partial}{\partial\varphi}+v_0\frac{\partial}{\partial\theta}+w_0\frac{\partial}{\partial z}\right)\begin{pmatrix}
u_0 \\[0.2em]
v_0
\end{pmatrix} + \begin{pmatrix}-u_0v_0\tan \theta \\[0.2em]
u_0^2\tan\theta
\end{pmatrix}\right\} \\ 
&+ 2 \begin{pmatrix}
    -v_0\sin\theta\\[0.2em]
    u_0\sin\theta
\end{pmatrix} + \begin{pmatrix}\frac{1}{\rho\cos\theta}\frac{\partial p_0}{\partial\varphi} \\[0.5em] \frac{1}{\rho}\frac{\partial p_0}{\partial\theta}\end{pmatrix} - \frac{1}{\rho}\frac{\partial}{\partial z}\left(m\frac{\partial}{\partial z}\begin{pmatrix}
    u_0 \\[0.2em] v_0 
\end{pmatrix}\right) = 0,
\end{aligned}
\end{equation}
while from \eqref{NS w red} we get
\begin{equation} \label{hydrostatic 1}
    \frac{\partial p_0}{\partial z} = -g\rho,
\end{equation}
and the continuity equation \eqref{continuity adim 1} yields
\begin{equation} \label{continuity asy}
\frac{\partial u_0}{\partial \varphi} + \frac{\partial}{\partial \theta} \left(v_0 \cos \theta  \right) + \cos\theta \frac{\partial w_0}{\partial z} = 0.
\end{equation}
From the kinematic boundary conditions \eqref{KBC Adim} we obtain
\begin{equation}\label{KBC asymptotic}
w_0 = \frac{\p h_0}{\p t}+\frac{u_0}{\cos\theta}\frac{\p h_0}{\p \varphi}+v_0\frac{\p h_0}{\p \theta} \qquad  \text{on $\{z=h_0\}$}.
\end{equation}
For simplicity, henceforth we assume that $h_0$ is a simple flat surface, that is,
\[h_0\equiv 0,\]
which implies that $w_0=0$ there. Writing $\bu_0 = u_0\ep + v_0\et + w_0\et$, \eqref{bottom BC nondim} becomes
\begin{equation} \label{BC nondim 2}
    \bu_0 = 0 \qquad \text{on $\{z=H\}$}.
\end{equation}

With this assumption on the free surface, the dynamic boundary condition \eqref{DBC nondim} is simply
\begin{equation}
    p_0 = p_{\rm s} \qquad \text{on }\{z=0\},
\end{equation}
and the upper boundary condition \eqref{stress adimensionalizzato} for the wind stress gives
\begin{equation} \label{stress asy}
    \begin{pmatrix}
    \dfrac{\p u_0}{\p z}\\[1em]
    \dfrac{\p v_0}{\p z}
    \end{pmatrix} = \sqrt{(u_{\rm a}-\sigma u_0)^2+(v_{\rm a}-\sigma v_0)^2}\begin{pmatrix}
    u_\a-\sigma u_0\\[0.2em]
    v_\a-\sigma v_0
    \end{pmatrix}\qquad  \text{on $\{z=0\}$},
\end{equation}
noting that, with our scaling, $m=1$ on $\{z=0\}$.

The next step is to perform a second asymptotic expansion, this time with respect to $\rr$, within the leading-order equations with respect to $\eps$ that we have derived. To this end, let us make the ansatz
\begin{equation} \label{as2}
    q_0\bigl(\varphi,\theta,z,t;\rr\bigr)\sim \tilde{q}(\varphi,\theta,y,t) + \rr\hat{q}(\varphi,\theta,y,t) + o(\rr),
\end{equation}
where, as in \eqref{as ansatz}, $q$ denotes each of the variables $u,\,v,\,w,$ and $P$. Plugging \eqref{as2} into \eqref{uv asymptotic 1} and collecting terms of the same order in $\rr$, we see that
\begin{equation} \label{horiziontal expanded}
\begin{aligned}
    &\left\{2 \begin{pmatrix}
    -\tilde{v}\sin\theta\\[0.2em]
    \tilde{u}\sin\theta
\end{pmatrix} + \begin{pmatrix}
    \frac{1}{\rho\cos\theta}\frac{\p\tilde{p}}{\p\varphi}\\[0.5em]
    \frac{1}{\rho}\frac{\p\tilde{p}}{\p\theta}
\end{pmatrix} - \frac{1}{\rho}\frac{\partial}{\partial z}\left(m\frac{\partial}{\partial z}\begin{pmatrix}
    \tilde{u}\\[0.2em]
    \tilde{v}
\end{pmatrix}\right)\right\} \\
&+ \rr\left\{\left(\frac{\partial}{\partial t} + \frac{\tilde{u}}{\cos\theta}\frac{\partial}{\partial\varphi} + \tilde{v}\frac{\partial}{\partial\theta} + \tilde{w}\frac{\p}{\p z}\right)\begin{pmatrix}
    \tilde{u}\\[0.2em]
    \tilde{v}
\end{pmatrix} + \begin{pmatrix}
    -\tilde{u}\tilde{v}\tan\theta\\[0.2em]
    \tilde{u}^2\tan\theta
\end{pmatrix} \right.\\
& +\biggl. 2 \begin{pmatrix}
    -\hat{v}\sin\theta\\[0.2em]
    \hat{u}\sin\theta
\end{pmatrix} + \begin{pmatrix}
    \frac{1}{\rho\cos\theta}\frac{\p\hat{p}}{\p\varphi}\\[0.5em]
    \frac{1}{\rho}\frac{\p\hat{p}}{\p\theta}
\end{pmatrix} - \frac{1}{\rho}\frac{\partial}{\partial z}\left(m\frac{\partial}{\partial z}\begin{pmatrix}
    \hat{u}\\[0.2em]
    \hat{v}
\end{pmatrix}\right)\biggr\} + o(\rr) = 0,
\end{aligned}
\end{equation}
while doing the same in \eqref{hydrostatic 1} yields
\begin{equation} \label{vertical expanded}
\frac{\partial\tilde{p}}{\partial z} + g\rho + \rr\frac{\partial\hat{p}}{\partial z} + o(\rr) = 0.
\end{equation}
Performing the same procedure in the continuity equation \eqref{continuity asy} leads us to
\begin{equation}\label{continuity scaling}
  \frac{\partial \tilde{u}}{\partial \varphi} +  \frac{\partial}{\partial \theta} \left( \tilde{v} \cos \theta  \right) + \cos\theta\frac{\p\tilde{w}}{\p z} + \rr\left\{\frac{\partial \hat{u}}{\partial \varphi} +  \frac{\partial}{\partial \theta} \left(\hat{v} \cos \theta  \right) + \cos\theta\frac{\p\tilde{w}}{\p z}\right\} + o(\rr) = 0,
\end{equation}
whereas, for the dynamic boundary condition, we have
\begin{equation} \label{DBC expanded 2}
\tilde{p} + \rr\hat{p} + o(\rr) = 0 \qquad \text{on $\{z=0\}$}.
\end{equation}
For the wind stress, inserting \eqref{as2} into \eqref{stress asy}, and using the Taylor expansion $\sqrt{1+x} = 1 + \frac{1}{2}x + o(x)$ as $x\to 0$, we see that
\begin{equation} \label{stress expanded}
\begin{aligned}
&\sqrt{(u_{\rm a}-\sigma u_0)^2+(v_{\rm a}-\sigma v_0)^2} \\
&\hspace{2cm}= \sqrt{(u_{\rm a}-\sigma\tilde{u})^2+(v_{\rm a}-\sigma\tilde{v})^2} \\
&\hspace{2cm} - \rr\sigma\frac{\hat{u}(u_{\rm a} - \sigma\tilde{u}) + \hat{v}(v_{\rm a} - \sigma\tilde{v})}{\sqrt{(u_{\rm a}-\sigma\tilde{u})^2+(v_{\rm a}-\sigma\tilde{v})^2}} + o(\rr) \qquad \text{on $\{z=0\}$}.
\end{aligned}
\end{equation}

Setting coefficients to zero in \eqref{horiziontal expanded}, \eqref{vertical expanded}, and \eqref{continuity scaling}, we obtain the leading-order set of equations
\begin{equation}\label{governing 00}
\left.\begin{aligned}
& \frac{1}{\rho}\frac{\partial}{\partial z}\left(m\frac{\partial\tilde{u}}{\partial z}\right) + 2\tilde{v}\sin\theta = \frac{1}{\rho\cos\theta}\frac{\p\tilde{p}}{\p\varphi} \\
&\frac{1}{\rho}\frac{\partial}{\partial z}\left(m\frac{\partial\tilde{v}}{\partial z}\right) - 2\tilde{u}\sin\theta = \frac{1}{\rho}\frac{\p\tilde{p}}{\p\theta} \\
& \frac{\partial\tilde{p}}{\partial z} = -g\rho \\
&\frac{\partial \tilde{u}}{\partial \varphi} +  \frac{\partial}{\partial \theta} \left( \tilde{v} \cos \theta  \right) + \cos\theta\frac{\p\tilde{w}}{\p z} = 0
\end{aligned}\;\right\}  \quad \text{in $\{H < z < 0\};$}
\end{equation}
recall that here $m$ denotes the non-dimensional eddy viscosity. The condition \eqref{BC nondim 2} reads simply
\begin{equation} \label{bottom BC leading order}
    \tilde{u}\ep + \tilde{v}\et + \tilde{w}\er = 0 \qquad \text{on $\{z=H\}$};
\end{equation}
the dynamic boundary condition \eqref{DBC expanded 2} with \eqref{DBC nondim} yields
\begin{equation} \label{dynamic final}
\tilde{p} = p_{\rm s} \qquad \text{on $\{z=0\}$}
\end{equation}
and, from \eqref{stress expanded}, we obtain at leading order:
\begin{equation}\label{stressADIM 1st order}
    \begin{pmatrix}
    \dfrac{\p \tilde{u}}{\p z}\\[1em]
    \dfrac{\p \tilde{v}}{\p z}
    \end{pmatrix} = \sqrt{(u_{\rm a}-\sigma\tilde{u})^2+(v_{\rm a}-\sigma\tilde{v})^2}\begin{pmatrix}
    u_\a-\sigma\tilde{u}\\[0.2em]
    v_\a-\sigma\tilde{v}
    \end{pmatrix} \eqqcolon \begin{pmatrix}
        \tau_\varphi \\[0.2em] \tau_\theta
    \end{pmatrix} \qquad  \text{on $\{z=0\}$}.
\end{equation}
Finally, \eqref{BC nondim 2} yields
\begin{equation}
    \tilde{u}\ep + \tilde{v}\et + \tilde{w}\er = 0 \qquad \text{on $\{z=H\}$}.
\end{equation}
The first two equations in \eqref{governing 00} are essentially a second-order system of ordinary differential equations with respect to the variable $z$, in which the other variables $\varphi$ and $\theta$ appear only parametrically. Physically, they describe a three-way balance between the Coriolis force, the horizontal pressure gradient force, and the viscous stress caused by the wind.

\section{Sverdrup's theory of the ocean circulation: Comparison with \texorpdfstring{$\beta$}{TEXT}-plane theory} \label{section Sverdrup}
Before presenting results on the Ekman currents, we briefly digress to consider another problem for which the present asymptotic expansion of the equations of motion in spherical coordinates may prove particularly useful: the understanding of large-scale ocean circulation. An extensive application of the present framework to this problem is ongoing and will be the subject of future works. Nevertheless, to provide some insight into its potential, we consider the simplest model of ocean circulation, namely that of \citet{sverdrup47}. Even at the level of this idealised model, the present equations---which retain the spherical geometry of the Earth at leading order by allowing both $\theta$ and $\varphi$ to vary---lead to results that differ in certain respects from the classical picture proposed by \citet{sverdrup47}.

Following \cite{sverdrup47} (see also \cite{stewart2009introduction}), we define $\Pi$, $M_1$ and $M_2$ by
\begin{equation}
   \frac{\p\Pi}{\p\varphi} = \int^0_H\frac{\p\tilde{p}}{\p\varphi}\dz, \quad  \frac{\p\Pi}{\p\theta} = \int^0_H\frac{\p\tilde{p}}{\p\theta}\dz,\quad M_1=\int_H^0 \rho \tilde{u}\dz, \quad \text{and} \quad M_2=\int_H^0 \rho \tilde{v}\dz,
\end{equation}
respectively, where, as before, $H$ is the bottom of the ocean (corresponding, in dimensional variables, to some thousands of metres). Following Sverdrup, we may argue that at such depth $H$ the stress is negligible, as the total current goes to zero, which, coupled with \eqref{stressADIM 1st order} gives
\begin{equation}\label{Stew11.3}
\begin{aligned}
     \int_H^0 \frac{\partial}{\partial z}\left(m\frac{\partial\tilde{u}}{\partial z}\right)\d z= \tau_\varphi \qquad \text{and} \qquad
     \int_H^0 \frac{\partial}{\partial z}\left(m\frac{\partial\tilde{v}}{\partial z}\right)\d z= \tau_\theta,
\end{aligned}
\end{equation}
where $\tau_\varphi\ep + \tau_\theta\et$ is the surface wind stress (cf. \eqref{stressADIM 1st order}). Integrating \eqref{governing 00}, so that it is possible to work with a baroclinic ocean without having to specify the vertical distribution of the density, and using \eqref{Stew11.3}, it follows that
\begin{align}
        \frac{1}{\cos\theta} \frac{\p\Pi}{\p\varphi}&=\tau_\varphi+2\sin\theta\, M_2,\label{11.1.5}\\[0.2em]
        \frac{\p\Pi}{\p\theta}&=\tau_\theta-2\sin\theta\, M_1.\label{11.1.6}
\end{align}
Analogously, the vertically integrated continuity equation reads
\begin{equation}\label{cont M}
    \frac{1}{\cos\theta}\frac{\partial M_1}{\partial \varphi} +  \frac{\partial M_2}{\partial \theta} -M_2\tan\theta=0.
\end{equation}
Applying $\frac{\partial }{\partial \theta} $ to \eqref{11.1.5} and $\frac{1}{\cos\theta}\frac{\partial }{\partial \varphi} $  to \eqref{11.1.6},  subtracting the resulting equations, and then 
using \eqref{cont M} and \eqref{11.1.5}, gives
\begin{equation}\label{11.1.7 v2}
2\cos\theta\, M_2 
  = \frac{1}{\cos\theta}\frac{\p \tau_\theta}{\p \varphi}-\frac{\p \tau_\varphi}{\p\theta} +\tan\theta\, \tau_\varphi.
\end{equation}
As discussed in \cite{stewart2009introduction}, over large regions of the open ocean, particularly in the tropics, the wind stress is predominantly zonal, so that (in our notation) $\frac{\partial \tau_\theta}{\partial \varphi}\approx 0.$
Under this assumption, \eqref{11.1.7 v2} reduces to
\begin{equation}
2\cos\theta M_2
\approx
-\frac{\partial \tau_\varphi}{\partial\theta}
+\tan\theta\,\tau_\varphi,
\end{equation}
and combining this relation with \eqref{cont M} yields
\begin{equation}
\frac{1}{\cos\theta}\frac{\partial M_1}{\partial\varphi}
\approx
\frac{1}{2\cos\theta}
\left(
\frac{\partial^2\tau_\varphi}{\partial\theta^2}
-\tan\theta\,\frac{\partial\tau_\varphi}{\partial\theta}
-\sec^2\theta\,\tau_\varphi
\right).
\end{equation}

For comparison, the corresponding expression obtained within the tangent-plane approximation is \citep{sverdrup47,stewart2009introduction}
\begin{equation}\label{11.10}
\frac{\partial M_x}{\partial x}
\approx
-\frac{1}{2\Omega'\cos\theta}
\left(
\frac{\partial\tau_x}{\partial y}\tan\theta-
R'\frac{\partial^2\tau_x}{\partial y^2}
\right),
\end{equation}
where we have retained the notation of \citet{sverdrup47} and \citet{stewart2009introduction}; the parameters $\Omega'$ and $R'$ appear because \eqref{11.10} is written in dimensional form. The most noticeable difference between the two expressions is the presence of the term $\sec^2\theta\,\tau_\varphi$ in the spherical formulation. This contribution has no counterpart in the Cartesian approximation and arises solely from the exact spherical geometry retained in the present derivation.

We also note that the right-hand side of \eqref{11.1.7 v2} can be recognised as the spherical curl of the wind-stress field on the unit sphere $\mathbb{S}^2$, and the continuity equation \eqref{cont M} implies the existence of a mass-transport stream function $\Psi$ satisfying
\begin{equation}
M_1=-\frac{\p \Psi}{\p \theta} \qquad \text{and} \qquad M_2=\frac{1}{\cos\theta}\frac{\p \Psi}{\p\varphi},
\end{equation}
leading to 
\begin{equation}\label{WOC}
    2\frac{\p \Psi}{\p\varphi}=\text{curl}_{\s^2}\boldsymbol{\tau}, \qquad \text{where} \qquad \boldsymbol{\tau}=\tau_\varphi\ep + \tau_\theta\et.
\end{equation}
The equations derived above are closely related to those obtained under the classical $\beta$-plane approximation \citep{sverdrup47,stewart2009introduction}. However, a crucial difference is that the present formulation fully retains the spherical geometry of the Earth. For instance, the curl operator appearing in \eqref{WOC} is not the standard Cartesian curl, but rather the curl operator on the sphere $\mathbb{S}^2$.

\section{Ekman flows} \label{section Ekman}
From now on, we will restrict our attention to flows away from the equatorial region. The reason is that, for $\theta \neq 0$, it is convenient to perform the usual splitting of the horizontal velocity into a \emph{geostrophic} and an \emph{Ekman} component:
\begin{equation}
\begin{pmatrix}
    \tilde{u}\\[0.2em]
    \tilde{v}
\end{pmatrix} = \begin{pmatrix}
    \tilde{u}_\g\\[0.2em]
    \tilde{v}_\g
\end{pmatrix} + \begin{pmatrix}
    \tilde{u}_\e\\[0.2em]
    \tilde{v}_\e
\end{pmatrix},
\end{equation}
where the geostrophic flow $(\tilde{u}_\g,\tilde{v}_\g)$ is defined as the solution to the system
\begin{equation}\label{geostrophic}
\left.\begin{aligned}
    2\tilde{v}_\g\sin\theta &= \frac{1}{\rho\cos\theta}\frac{\p\tilde{p}}{\p\varphi} \\
    2\tilde{u}_\g\sin\theta &= -\frac{1}{\rho}\frac{\p\tilde{p}}{\p\theta}
\end{aligned}\;\right\} \quad \text{in $\{H < z < 0\}$}.
\end{equation}
Moreover, the equatorial region exhibits dynamical features that are absent at higher latitudes. Under the action of the trade winds, Ekman transports on either side of the equator are directed away from it, producing a strong divergence of surface waters and intense equatorial upwelling. The associated pressure and thermocline anomalies are not confined to the forcing region but are redistributed across ocean basins by equatorially trapped Kelvin and Rossby waves (see \cite{Gil82} and \citet{boyd}). As a consequence, the oceanic response cannot be described solely through a local balance between pressure gradients, Coriolis forces, and wind stress. These wave-mediated adjustment processes play a central role in large-scale tropical variability, including the El Niño-Southern Oscillation (ENSO). For these reasons, the equatorial band requires a distinct theoretical framework and will not be considered here.

Note that, integrating \eqref{governing 00} and using \eqref{dynamic final}, it follows
\begin{equation} \label{pressure formula}
    \tilde{p}(\varphi,\theta,z) = p_{\rm s}(\varphi,\theta) + g\int_z^0\rho(s)\ds,
\end{equation}
so the geostrophic component of the velocity is uniquely determined by the surface pressure:
\[2\tilde{v}_\g \sin\theta = \frac{1}{\rho\cos\theta}\frac{\p p_{\rm s}}{\p\varphi} \qquad \text{and} \qquad 2\tilde{u}_\g\sin\theta = -\frac{1}{\rho}\frac{\p p_{\rm s}}{\p\theta} \qquad \text{in $\{H < z < 0\}$}.\]
In particular, we see that, in the presence of a flat surface, the geostrophic flow vanishes when the atmospheric pressure $p_{\rm s}$ is constant. Furthermore, note that
\begin{equation} \label{geostrophic divergence}
\frac{\p\tilde{u}_\g}{\p\varphi} + \frac{\p}{\p\theta}(\tilde{v}_\g\cos\theta) = 0.
\end{equation}
Let us introduce the complex notation
\begin{equation} \label{complex notation}
    W=\tilde{u}_\e+\i\tilde{v}_\e, \qquad W_\g = \tilde{u}_\g+\i\tilde{v}_\g, \qquad \text{and} \qquad W_{\rm w} = u_\a - \tilde{u}_\g + \i(v_\a - \tilde{v}_\g).
\end{equation}
The Ekman velocity $W$ describes a boundary layer correction of the geostrophic velocity $W_\g$ due to the wind stress. We then obtain from \eqref{governing 00}, \eqref{bottom BC leading order}, \eqref{stressADIM 1st order} and \eqref{geostrophic}:
\begin{equation}\label{ode inhom}
    \left\{\begin{aligned}
    & \frac{\p}{\p z}\left(m\frac{\p W}{\p z}\right) = 2\i\sin\theta\,\rho W - \frac{\p}{\p z}\left(m\frac{\p W_\g}{\p z}\right) & &\quad \text{in $\{H < z < 0\}$}, \\[0.2em]
    & \frac{\p W}{\p z} = -\frac{\p W_\g}{\p z} + |W_{\rm w}-\sigma W|(W_{\rm w}-\sigma W) & &\quad \text{on $\{z=0\}$}, \\[0.2em]
    & W + W_\g = 0 & &\quad \text{on $\{z=H\}$}
    \end{aligned}\right.
\end{equation}
for the boundary value problem that describes the Ekman component of the flow. As we pointed out previously, if the atmospheric pressure is constant, then the geostrophic flow vanishes and \eqref{ode inhom} reduces to
\begin{equation}\label{ode z1}
    \left\{\begin{aligned}
    & \frac{\p}{\p z}\left(m\frac{\p W}{\p z}\right) = 2\i\sin\theta\,\rho W & &\quad \text{in $\{H < z < 0\}$}, \\[0.2em]
    & \frac{\p W}{\p z} = |W_{\rm w}-\sigma W|(W_{\rm w}-\sigma W) & &\quad \text{on $\{z=0\}$}, \\[0.2em]
    & W = 0 & &\quad \text{on $\{z=H\}$}.
    \end{aligned}\right.
\end{equation}
Henceforth, we will focus only on the problem \eqref{ode z1}, leaving \eqref{ode inhom} to future work.

Throughout this section, to keep the notation as agile as possible, we will denote only the dependence on $z$ explicitly unless necessary, since the other variables enter the equation only parametrically.

\subsection{Some analytical results} \label{section analytical results}

Interestingly, the problem \eqref{ode z1} is nonlinear, because even though the equation itself is linear, the upper boundary condition is nonlinear. Therefore, it is not clear \emph{a priori} whether this problem is solvable and, if so, whether the solution is unique. This is the subject of the next theorem, which not only establishes the existence of a unique solution to \eqref{ode z1} for given $W_{\rm w}$, $m$ and $\rho$, but also provides some information on the qualitative behaviour of the solution.

\begin{theorem} \label{theorem Ekman spiral}
Let $W_{\rm w} \in \C\setminus\{0\}$ be arbitrary, $\theta \in (-\frac{\pi}{2},\frac{\pi}{2})\setminus\{0\}$, and suppose that $\rho\in C([H,0])$ and $m\in C^1((H,0))\cap C([H,0])$ are both positive functions. Then there exists a unique function $W\in C^2((H,0)) \cap C([H,0])$ that solves \eqref{ode z1}. Moreover, writing
\begin{equation} \label{W exponential}
W(z) = M(z)\E^{\i\vartheta(z)},
\end{equation}
where $M(z) = |W(z)|$, we have that, for all $z\in (H,0)$,
\begin{equation} \label{Ekman spiral}
M'(z) \geq 0 \qquad \text{and} \qquad \sign(\theta)\vartheta'(z) >0,
\end{equation}
that is, the solution behaves as a classical Ekman spiral.
\end{theorem}

The proof can be found in Appendix \ref{appendix proofs}. Theorem \ref{theorem Ekman spiral} remains valid if the eddy viscosity $m$ is only piecewise continuously differentiable (but continuous everywhere), as one can see upon inspection of the proof.

It is also possible to infer a formula for the \emph{surface deflection angle}, which is defined as the angle between the surface Ekman current $W(0)$ and the difference $W_{\rm w} = W_\a - W_\g$ between the wind and the geostrophic current. This is the object of the next result:

\begin{theorem} \label{theorem deflection}
With
\begin{equation} \label{lambda def}
    \lambda = \frac{\frac{\p W}{\p z}(0)}{W(0)},
\end{equation}
let $\chi$ be the unique positive root of the polynomial
\begin{equation} \label{polynomial}
    p(x) = \mu^2x^4 + 2\mu{\rm Re}(\lambda)x^3 + |\lambda|^2x^2 - |\lambda W_{\rm w}|^2.
\end{equation}
Denote
\begin{equation}
    W(0) = |W(0)|\E^{\i\Theta} \qquad \text{and} \qquad W_{\rm w} = |W_{\rm w}|\E^{\i\Gamma}.
\end{equation}
Then the surface deflection angle $\omega \coloneqq \Gamma - \Theta$ is determined by the formula
\begin{equation} \label{deflection formula}
    \tan(\omega) = \frac{{\rm Im}(\lambda)}{\sigma\chi + {\rm Re}(\lambda)},
\end{equation}
hence, since $\sign(\theta){\rm Im}(\lambda) > 0$ and ${\rm Re}(\lambda) > 0$, the deflection is to the right in the northern hemisphere and to the left in the southern hemisphere. In particular, we have the bounds
\begin{equation} \label{deflection bounds}
    \frac{|{\rm Im}(\lambda)|}{\alpha + {\rm Re}(\lambda)} \leq |\tan(\omega)| \leq \frac{|{\rm Im}(\lambda)|}{\beta+{\rm Re}(\lambda)},
\end{equation}
where
\begin{equation} \label{bounds formulae}
    \alpha = \frac{-{\rm Re}(\lambda) + \sqrt{{\rm Re}(\lambda)^2 + 4\mu|\lambda W_{\rm w}|}}{2} \qquad \text{and} \qquad \beta = \frac{-|\lambda| + \sqrt{|\lambda|^2 + 4\mu|\lambda W_{\rm w}|}}{2}.
\end{equation}
\end{theorem}

\begin{figure}
    \centering
    \includegraphics[width=0.5\linewidth]{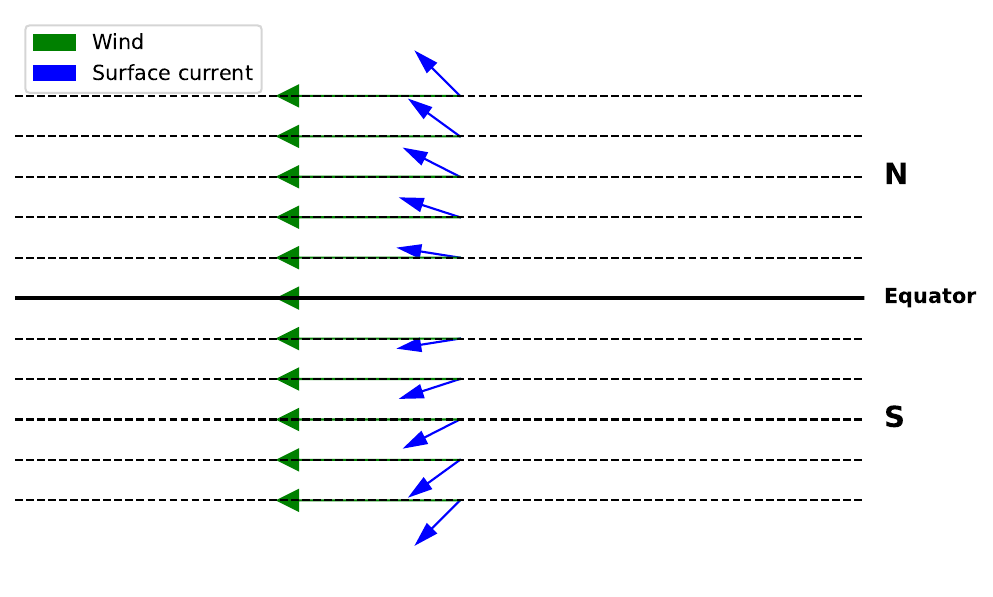}
    \caption{Conceptual diagram showing the progressive decrease in the wind–current deflection angle toward the equator as the Coriolis force approaches zero. Not to scale.}
    \label{equatorial arrows}
\end{figure}

Moreover, although our solution is not defined for $\theta=0$, where the definition of the geostrophic flow breaks down, we recover the classical observation that the deflection angle vanishes as one approaches the equator \citep{W24}; see figure~\ref{equatorial arrows} for a sketch.

\begin{theorem} \label{deflection equator}
With the notation of Theorem \ref{theorem deflection}, for the surface deflection angle $\omega(\theta)$ we have
\begin{equation}
    \lim_{\theta\to 0} \omega(\theta) = 0.
\end{equation}
\end{theorem}
Finally, we define the Ekman transport (denoted by ${\rm Ek}$) as the integrated mass transport induced by the Ekman flow:
\begin{equation}
    \mathrm{Ek} = \int^0_{H}W(z)\, \d z.
\end{equation}
Using \eqref{ode z1}, we immediately see that
\begin{equation}
\begin{aligned}
    {\rm Ek} &= -\i\,\frac{m(0)}{2\sin\theta}|W_{\rm w}-\sigma W(0)|(W_{\rm w}-\sigma W(0)) + \underbrace{\i\frac{m(H)}{2\sin\theta}\frac{\p W}{\p z}(H)}_{\approx0}\\
    &\approx -\,\frac{\i}{2\sin\theta}|W_{\rm w}-\sigma W(0)|(W_{\rm w}-\sigma W(0)),
\end{aligned}
\end{equation}
since $m(H)$ is very small and we expect $\frac{\p W}{\p z}$ to be negligible at great depths. In other words, the Ekman transport is always essentially at a right angle to the vector $W_{\rm w}-\sigma W(0)$ (or, equivalently, to the wind stress on the surface, in view of \eqref{stress2}).

\subsection{Explicit solutions to the leading-order problem} \label{section explicit}
This section is devoted to gaining additional insight into the qualitative behaviour of the leading-order Ekman flow by investigating some concrete examples of explicit eddy viscosity profiles, along the lines of \citet{CJ2019} and \citet{zikanovETAL}. These choices allow us to write down the solution explicitly, calculate the surface deflection angle, compare our results with some available measurements (see table~\ref{tab:deflections}), and visualise the flow by means of numerical plots. For simplicity, throughout this section we will assume the density to be constant: $\rho(z) = 1$ for all $z$.

For clarity and ease of reference later, let us recall the leading-order problem \eqref{ode z1} here, in the special case $\rho(z) = 1$. This problem consists of the complex-valued linear ordinary differential equation
\begin{equation} \label{ode z}
    \frac{\p}{\p z}\left(m\frac{\p W}{\p z}\right) = 2\i W\sin\theta \qquad \text{in $\{H < z < 0\}$},
\end{equation}
with boundary conditions
\begin{equation} \label{BC stress}
   \frac{\p W}{\p z} = |W_{\rm w}-\sigma W|(W_{\rm w}-\sigma W) \qquad \text{on $\{z=0\}$}
\end{equation}
and
\begin{equation} \label{BC lower}
    W = 0 \qquad \text{on $\{z=H\}$},
\end{equation}
where $W_{\rm w}\in\C$ corresponds to the non-dimensional wind velocity (see \eqref{complex notation}) and can thus be assumed to be given.

We will consider an eddy viscosity of the form
\begin{equation} \label{eddy piecewise}
    m(z) = \begin{cases}
        1+\dfrac{M-1}{{z_{\star}}}\,z, &\quad z \in [{z_{\star}},0], \\[0.2em]
        m_1(z), &\quad z\in [z_0,{z_{\star}}], \\[0.2em]
        \mathfrak{m}, &\quad z \in [H,z_0],
    \end{cases}
\end{equation}
where $M,\mathfrak{m}>0$ and the function $m_1$ is positive, continuous, and satisfies $m_1({z_{\star}}) = M$ and $m_1(z_0) = \mathfrak{m}$. In other words, the eddy viscosity increases linearly to the depth ${z_{\star}}$, where it takes the value $M$, and is given by $m_1$ in the interval $[z_0,z_\star]$, below which it coincides with the molecular viscosity $\mathfrak{m}$. This increasing--decreasing profile is consistent with the so-called KPP parametrisation (see, e.g., \cite{Large94} and \citet{McWilliamsHuckle}) and, despite the absence of wave dynamics in our model, can be regarded as a simplified counterpart of the formulation proposed by \cite{K61} (see also the discussion by \cite{Weber81}). There, the eddy viscosity is taken to be proportional to the shear of the wave orbital velocity and to the square of the mixing length.

The value $M$ can be interpreted as the maximum efficiency of turbulent momentum transport within the water column, with a larger $M$ corresponding to more vigorous mixing and a more efficient vertical redistribution of the momentum imparted by the wind, while the depth $z_{\star}$ characterises the depth at which the wind-generated turbulence is at its strongest. We will return to this important fact in §~\ref{section discussion}.

Recalling that we are setting $m(0)=1$, it can be assumed that $M\approx10$ (see \citet{ McWilliamsHuckle} and \citet{McW2012}). In some works (see, e.g., \citet{Weber81} and \citet{zikanovETAL}), the value of $M$ appears to be greater than $10$, even if its value is not easily quantifiable. In the latter case, we will set $M=50$, in order to analyse how a higher value of $M$ affects the Ekman flow. Moreover, we will set $\mathfrak{m} = 0.01$ for the molecular viscosity. Finally, according to \citet{zikanovETAL}, \citet{McWilliamsHuckle} and \citet{McW2012}, one can estimate $z_{\star}\approx z_0/4$, $z_{\star}\approx z_0/5$ and $z_{\star}\approx z_0/6$, respectively, with $z_0\approx 5$ (corresponding to $60$\,m) according to \citet{McWilliamsHuckle} and $z_0\approx 18$ (corresponding to $200$\,m) according to \citet{McW2012}. For particular choices of $m_1$, the equations can be solved explicitly; see Appendix~\ref{Appendix explicit} for a brief discussion of the general solving method in this case.

In the sequel, we will consider some of such situations and compare the results with available measurements. First, we consider the easiest possible case, that is, that of a constant $m_1$. To avoid dealing with discontinuous eddy viscosity, we take $z_0=H$ in this case; while this example is physically of dubious relevance, it is nevertheless interesting to make a comparison with a model similar to that of \citet{Ekm05}. Next, we investigate the cases of linearly decaying eddy viscosity, which can be identified as an approximation of the viscosity profiles in \citet{zikanovETAL}, \citet{McWilliamsHuckle} and \citet{McW2012}. Finally, we turn our attention to the exponentially decaying eddy viscosity proposed by \citet{WenegratMcP2016} to match the measurements by \citet{PetersGT1988} and \citet{Dillon1989}, where the solution can be written in terms of modified Bessel functions. For details on the computations involved, see Appendix \ref{Appendix explicit}.

\subsubsection{Constant \texorpdfstring{$m_1$}{TEXT}} \label{section constant}

The first case we consider is that of a constant $m_1$, where we take only two layers (that is, $z_0=H$):
\begin{equation}
    m_1(z) = M, \qquad z\in [H,{z_{\star}}].
\end{equation}
Although the assumption of constant eddy viscosity may seem unrealistic, it has been observed that it may provide a good fit to certain wind-driven drift currents \citep{DrakePassage}. In this case, the solution is explicit (see Appendix \ref{Appendix constant}) and is depicted in figure~\ref{fig constant spiral} (see also figure~\ref{fig constant spiral app} in Appendix~\ref{spirals appendix}). Figure~\ref{fig constant deflection} presents the computed deflection angles. All values remain below $45^\circ$, owing to the linear portion of the eddy-viscosity profile adopted in the model.

\begin{figure}
    \centering
    \begin{subfigure}{0.48\textwidth}
        \centering
        \includegraphics[width=0.49\linewidth]{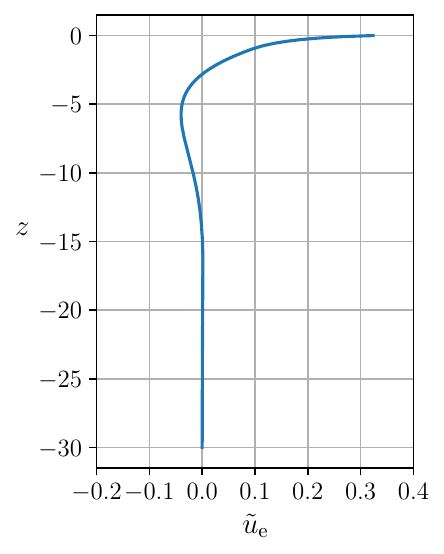}
        \includegraphics[width=0.49\linewidth]{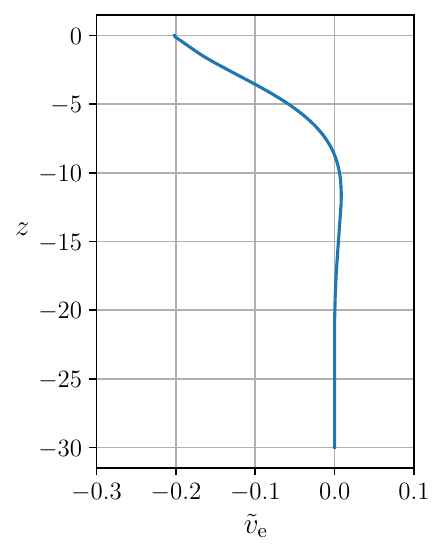}
        \caption{ $M = 10$, ${z_{\star}} = -1$.}
    \end{subfigure}
    \hfill
    \begin{subfigure}{0.48\textwidth}
        \centering
        \includegraphics[width=0.49\linewidth]{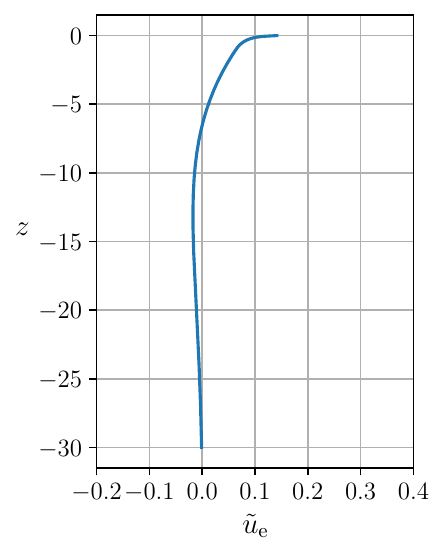}
        \includegraphics[width=0.49\linewidth]{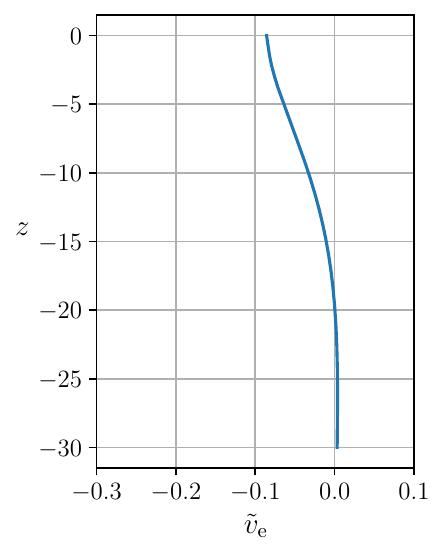}
        \caption{ $M = 50$, ${z_{\star}} = -1$.}
    \end{subfigure}
  \caption{The velocity field $W = \tilde{u}_\e + \i\tilde{v}_\e$ in the case of constant $m_1$ for $|W_{\rm w}|=1$, $\sigma = 0.014$ (corresponding to $U_\a' = 10\,{\rm m\,s^{-1}}$), $\theta=45^\circ$, and $H=-80$. In all cases, the plots are displayed only up to $z=-30.$}
    \label{fig constant spiral}
\end{figure}

    \begin{figure}
    \begin{subfigure}{0.49\textwidth}
        \centering
        \includegraphics[width=0.75\linewidth]{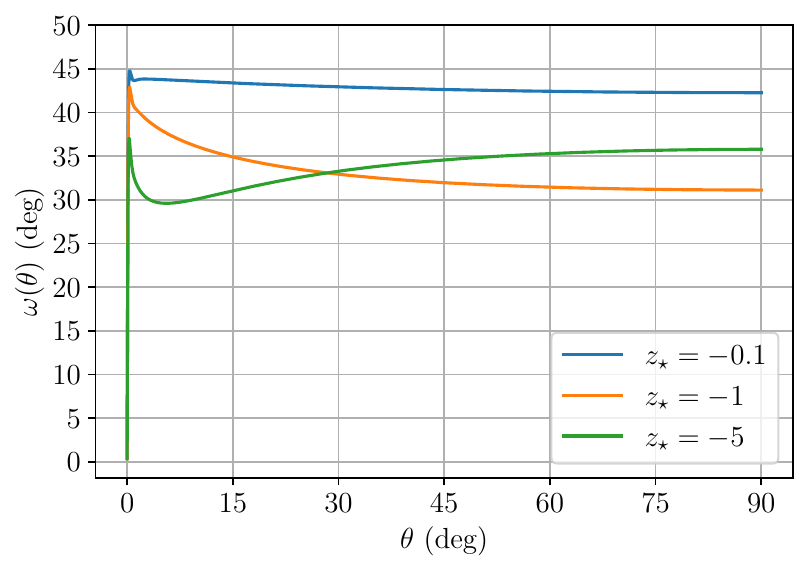}
        \caption{ $M = 10$.}
    \end{subfigure}
    \begin{subfigure}{0.49\textwidth}
        \centering
        \includegraphics[width=0.75\linewidth]{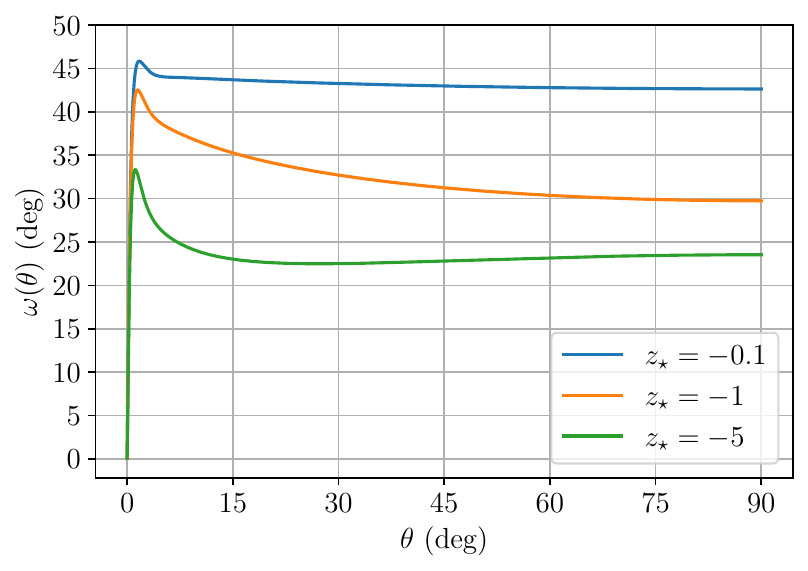}
        \caption{ $M = 50$.}
    \end{subfigure}
    \caption{The surface deflection angle in the case of constant $m_1$ for $|W_{\rm w}|=1$, $\sigma = 0.014$, and $H=-80$. }
    \label{fig constant deflection}
\end{figure}

\subsubsection{Linearly decaying \texorpdfstring{$m_1$}{TEXT}} \label{section linear}

\begin{figure}
    \centering
    \begin{subfigure}{0.48\textwidth}
        \centering
        \includegraphics[width=0.49\linewidth]{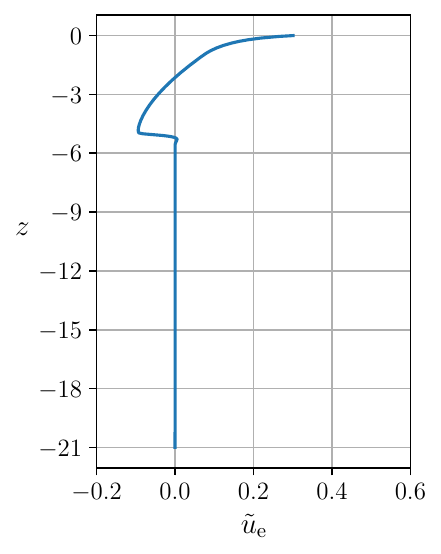}
        \includegraphics[width=0.49\linewidth]{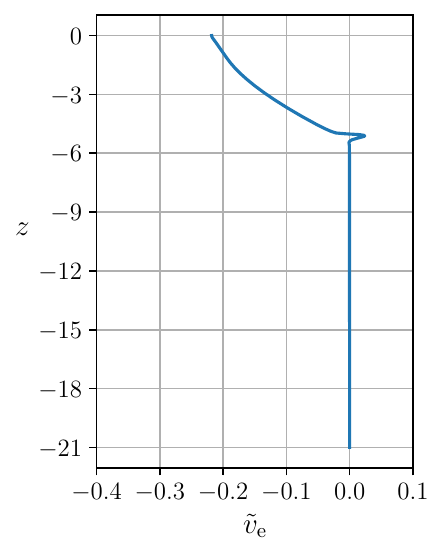}
        \caption{$M=10$, $z_0=-5$, $z_{\star}=-1$.}
    \end{subfigure}
    \hfill
    \begin{subfigure}{0.48\textwidth}
        \centering
        \includegraphics[width=0.49\linewidth]{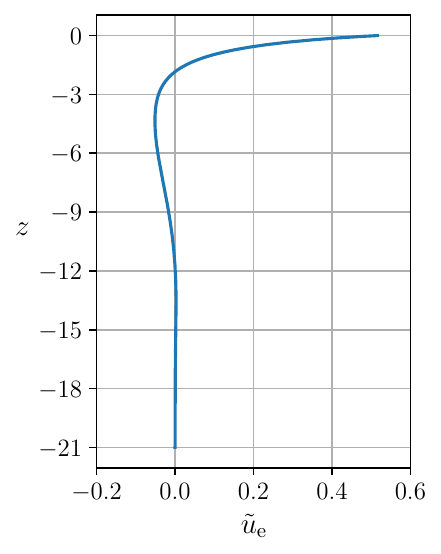}
        \includegraphics[width=0.49\linewidth]{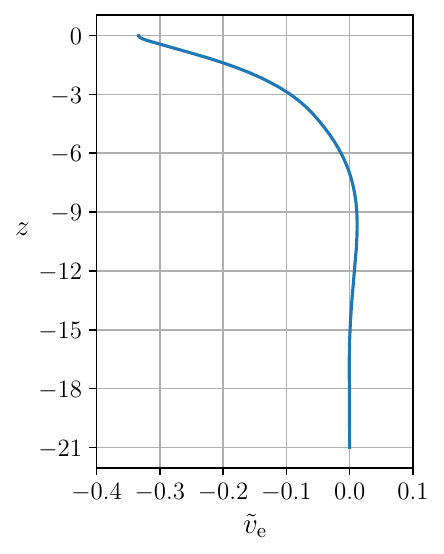}
        \caption{ $M=10$, $z_0=-20$, $z_{\star}=-4$.}
    \end{subfigure}
       \begin{subfigure}{0.48\textwidth}
        \centering
        \includegraphics[width=0.49\linewidth]{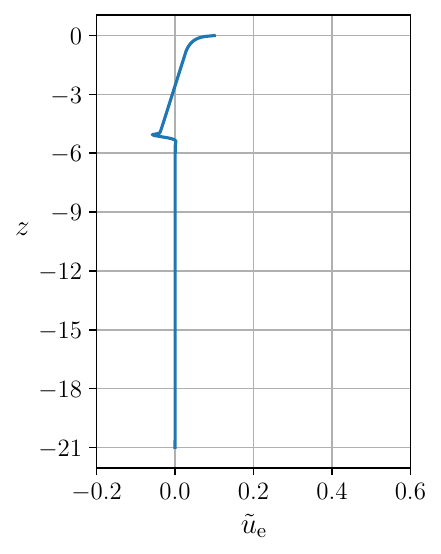}
        \includegraphics[width=0.49\linewidth]{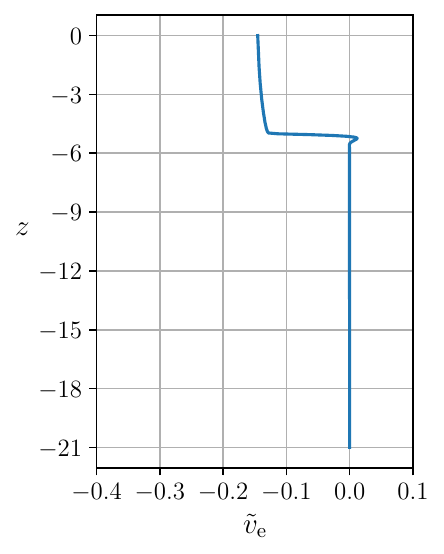}
        \caption{$M=50$, $z_0=-5$, $z_{\star}=-1$.}
    \end{subfigure}
    \hfill
    \begin{subfigure}{0.48\textwidth}
        \centering
        \includegraphics[width=0.49\linewidth]{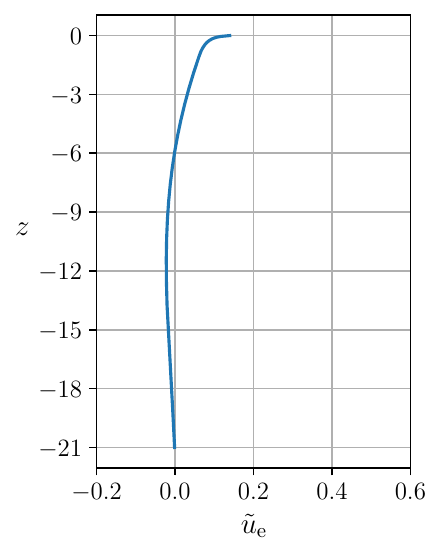}
        \includegraphics[width=0.49\linewidth]{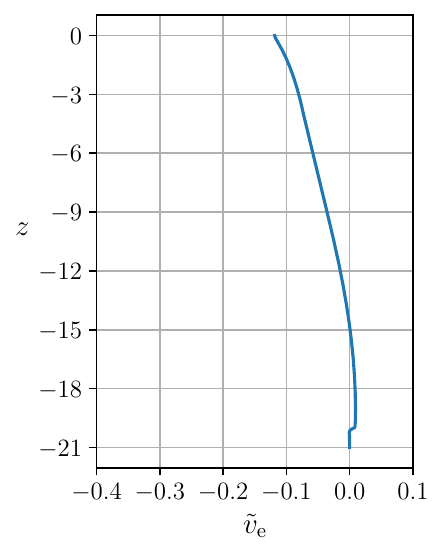}
        \caption{ $M=50$, $z_0=-20$, $z_{\star}=-4$.}
    \end{subfigure}
    \caption{The velocity field $W = \tilde{u}_\e + \i\tilde{v}_\e$ in the case of linearly decaying $m_1$ for $|W_{\rm w}|=1$, $\sigma = 0.014$, $H=-80$ and $\mathfrak{m}=0.01$. In all cases, the plots are displayed only up to $z=-21.$}
    \label{fig linear spiral}
\end{figure}

\begin{figure}
    \centering
    \begin{subfigure}{0.49\textwidth}
        \centering
        \includegraphics[width=.75\linewidth]{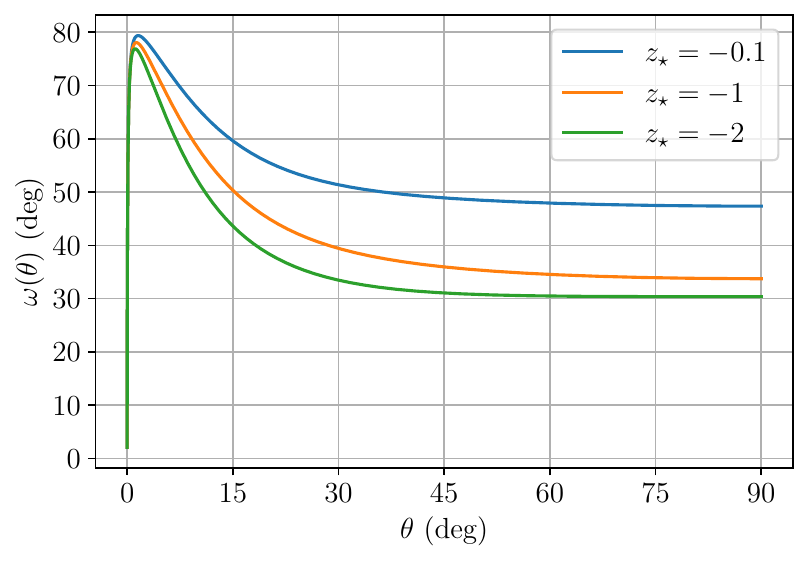}
        \caption{$M=10$, $z_0=-5$.}
    \end{subfigure}
    \begin{subfigure}{0.49\textwidth}
        \centering
        \includegraphics[width=0.75\linewidth]{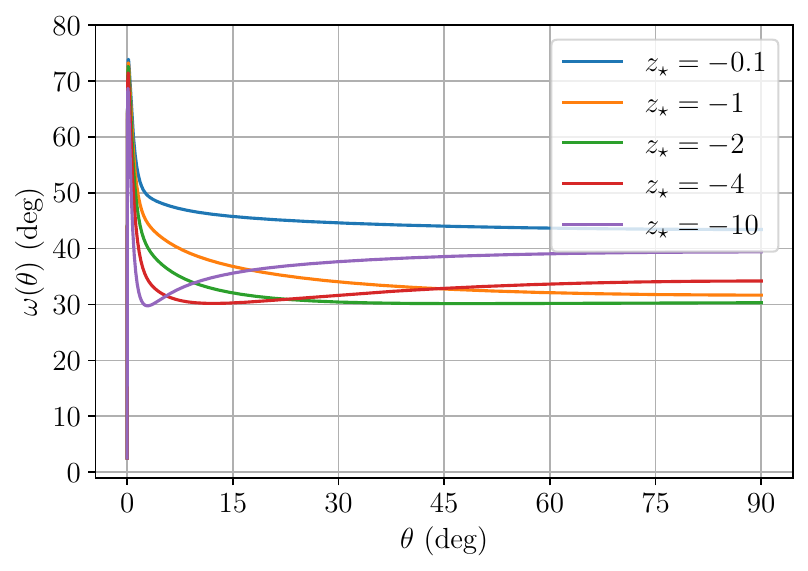}
        \caption{ $M=10$, $z_0=-20$.}
    \end{subfigure}
       \begin{subfigure}{0.49\textwidth}
        \centering
        \includegraphics[width=0.75\linewidth]{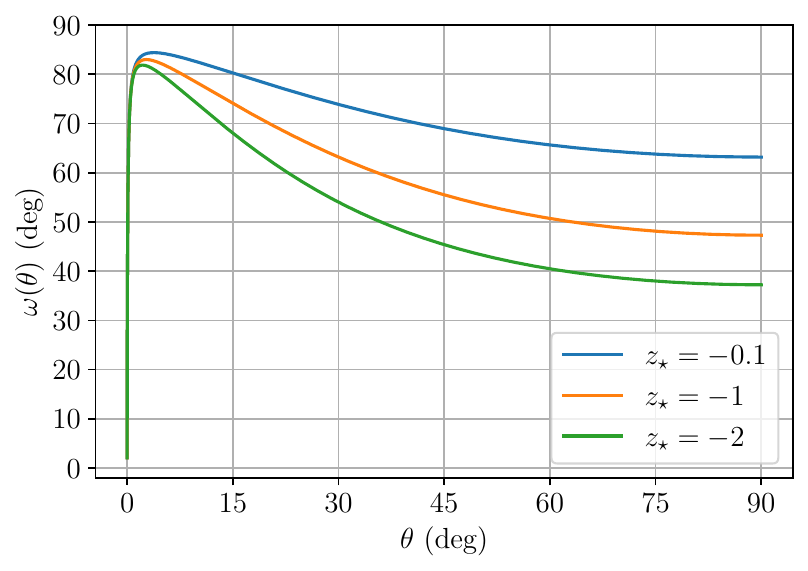}
        \caption{$M=50$, $z_0=-5$.}
    \end{subfigure}
       \begin{subfigure}{0.49\textwidth}
        \centering
        \includegraphics[width=0.75\linewidth]{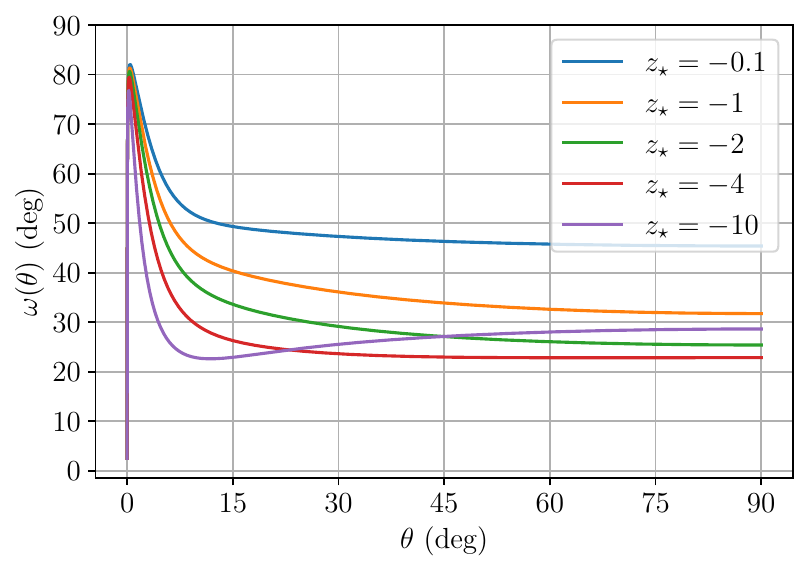}
        \caption{ $M=50$, $z_0=-20$.}
    \end{subfigure}
    \caption{The surface deflection angle in the case of linearly decaying $m_1$ for $|W_{\rm w}|=1$, $\sigma = 0.014$, $H=-80$ and $\mathfrak{m}=0.01$. }
    \label{fig linear deflection}
\end{figure}

The second case we study is of an eddy viscosity that decays linearly in the lower layer to a small value, denoted by $\mathfrak{m}$:
\begin{equation} \label{m_1 linear}
    m_1(z) = \frac{{z_{\star}}\mathfrak{m}-z_0 M}{{z_{\star}}-z_0} + \frac{M-\mathfrak{m}}{{z_{\star}}-z_0}\,z, \qquad z\in [z_0,{z_{\star}}].
\end{equation}
Since the concept of eddy viscosity is inherently associated with the presence of turbulence, which in the present case is generated by wind forcing, it is reasonable to assume that below the Ekman layer, where wind-induced effects are negligible, turbulence should be absent and the viscosity for $z<z_0$ should reduce the molecular viscosity, which is typically several orders of magnitude smaller than the maximum value of the eddy viscosity; therefore, we set $\mathfrak{m}=0.01$. See figures~\ref{fig linear spiral} and~\ref{fig linear deflection} for plots of the solution and the deflection angle for different values of the parameters (as well as figure~\ref{fig linear spiral app} in Appendix~\ref{spirals appendix}).

One remark is in order. The velocity profiles shown in figure~\ref{fig linear spiral}(a) and (c), although differentiable everywhere, exhibit noticeably more irregular behaviour than those in figure~\ref{fig linear spiral}(b). This can be attributed to the eddy-viscosity profile and, in particular, to its degree of smoothness. For the piecewise linear eddy viscosity considered here, the discontinuity in the derivative of \(m_1\) at \(z_0\), namely $\frac{M-\mathfrak{m}}{z_\star-z_0}$, is substantially larger in the cases shown in figure~\ref{fig linear spiral}(a) and (c) than in figure~\ref{fig linear spiral}(b). The latter therefore corresponds to the smoothest velocity profiles among those displayed in figure~\ref{fig linear spiral}. This effect becomes even more apparent when comparing figure~\ref{fig linear spiral} with figure~\ref{fig exp 2d}, which shows the velocity profiles obtained with an exponentially decaying eddy viscosity. In that case, the jump in the derivative of \(m_1\) at \(z_0\) is given by $\mathfrak{m}\frac{\ln\left(\mathfrak{m}/M\right)}{z_0-z_\star}$. For example, with \(M=50\), \(z_0=-5\), \(z_\star=-1\), and \(\mathfrak{m}=0.01\), the jump is approximately \(12.5\) for the piecewise linear eddy viscosity (cf. figure~\ref{fig linear spiral}(c)), whereas it is only about \(0.02\) for the exponentially decaying profile. This substantial reduction in the derivative discontinuity explains the markedly smoother velocity profiles observed in figure~\ref{fig exp 2d}.

\subsubsection{Exponentially decaying eddy viscosity} \label{section exponential}

\begin{figure}
    \begin{subfigure}{0.48\textwidth}
        \centering
        \includegraphics[width=0.49\linewidth]{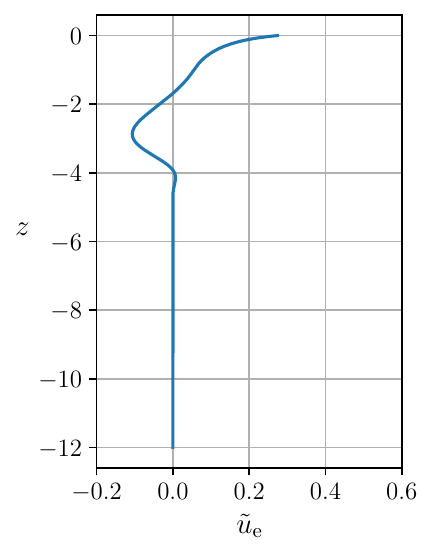}
        \includegraphics[width=0.49\linewidth]{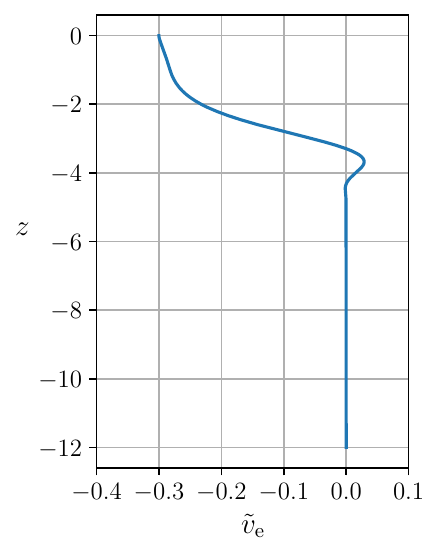}
        \caption{$M=10$, $z_0=-5$, $z_\star=-1$.}
    \end{subfigure}
    \begin{subfigure}{0.48\textwidth}
        \centering
        \includegraphics[width=0.49\linewidth]{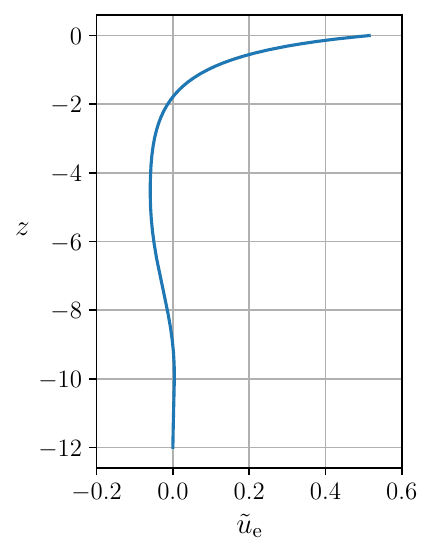}
        \includegraphics[width=0.49\linewidth]{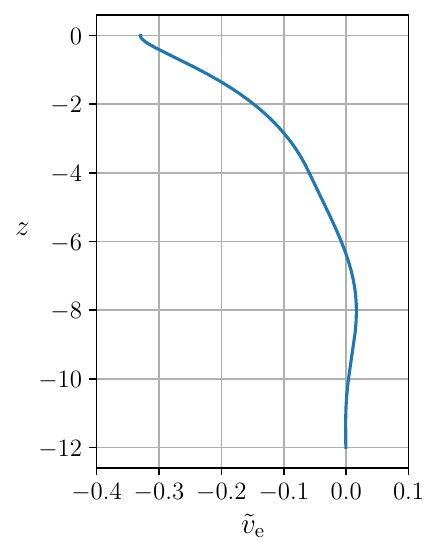}
        \caption{$M=10$, $z_0=-20$ $z_\star=-4$.}
    \end{subfigure}
       \begin{subfigure}{0.48\textwidth}
        \centering
        \includegraphics[width=0.49\linewidth]{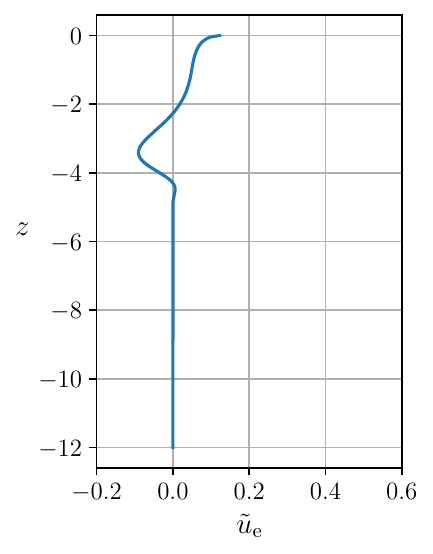}
        \includegraphics[width=0.49\linewidth]{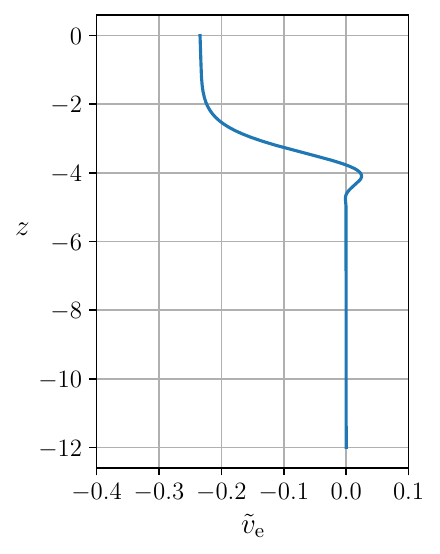}
        \caption{$M=50$, $z_0=-5$, $z_\star=-1$}
    \end{subfigure}
    \hfill
    \begin{subfigure}{0.48\textwidth}
        \centering
        \includegraphics[width=0.49\linewidth]{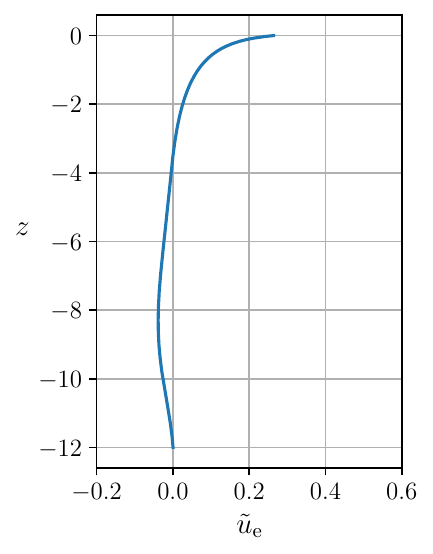}
        \includegraphics[width=0.49\linewidth]{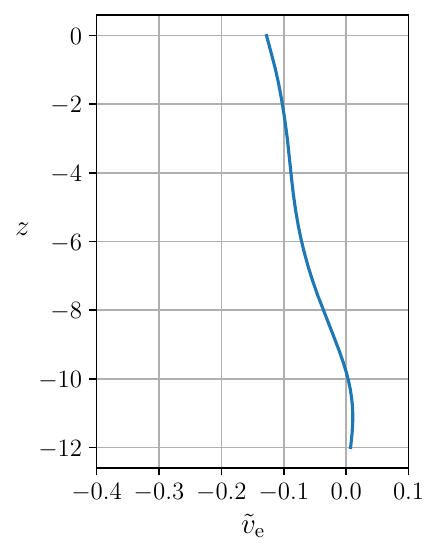}
        \caption{ $M=50$, $z_0=-20$, $z_\star=-4$.}
    \end{subfigure}
    \caption{The velocity field $W = \tilde{u}_\e + \i\tilde{v}_\e$ in the case of exponentially decaying $m_1$ with $|W_{\rm w}|=1$, $\sigma = 0.014$. Moreover, $z_0=-5$, $H=-80$,  and $\mathfrak{m}=0.01$. In all cases, the plots are displayed only up to $z=-12.$}
    \label{fig exp 2d}
\end{figure}

The final case we consider is that, proposed as a model dictated by some turbulence measurements in non-equatorial regions \citep{CroninKessler2009, WenegratMcP2016, Sentchev2023}, of an exponentially decreasing eddy viscosity of the form
\begin{equation}\label{exp eddy}
    m_1(z)=M\E^{q(z-{z_{\star}})}, \qquad z_0 < z < 0,
\end{equation}
where
\begin{equation}\label{q}
    q=\frac{\ln\left(\dfrac{\mathfrak{m}}{M}\right)}{z_0-{z_{\star}}},
\end{equation}
which ensures that $m(z_0)=\mathfrak{m}$. Analogously to the previous case, we set $\mathfrak{m}=0.01$. The solution and the deflection angle are shown in figures~\ref{fig exp 2d} and~\ref{fig exp deflection} for different values of the parameters; see also figure~\ref{fig exp 2d app} in Appendix~\ref{spirals appendix}. Observe that the velocity profiles in figure~\ref{fig exp 2d} are essentially a smoother version of those in figure~\ref{fig linear spiral}.

\begin{figure}    
    \begin{subfigure}{0.49\textwidth}
        \centering
        \includegraphics[width=0.75\linewidth]{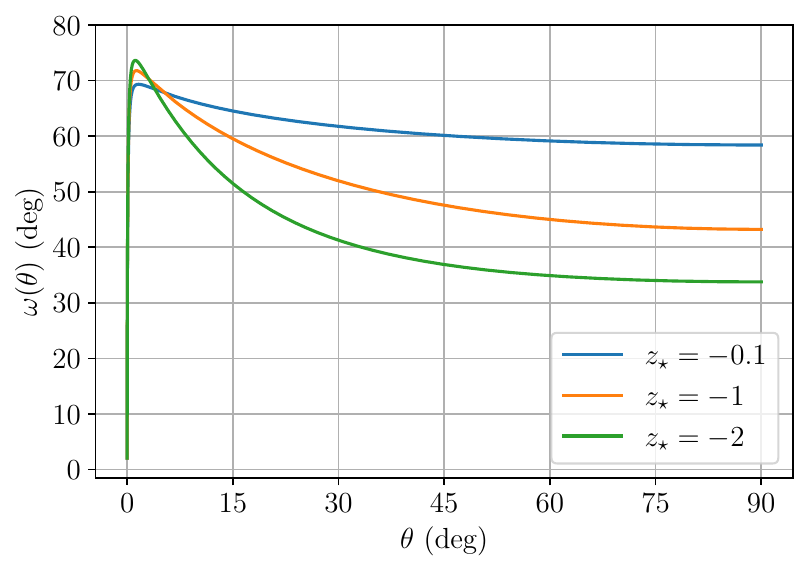}
        \caption{ $M=10$, $z_0=-5$.}
    \end{subfigure}
    \begin{subfigure}{0.49\textwidth}
        \centering
        \includegraphics[width=0.75\linewidth]{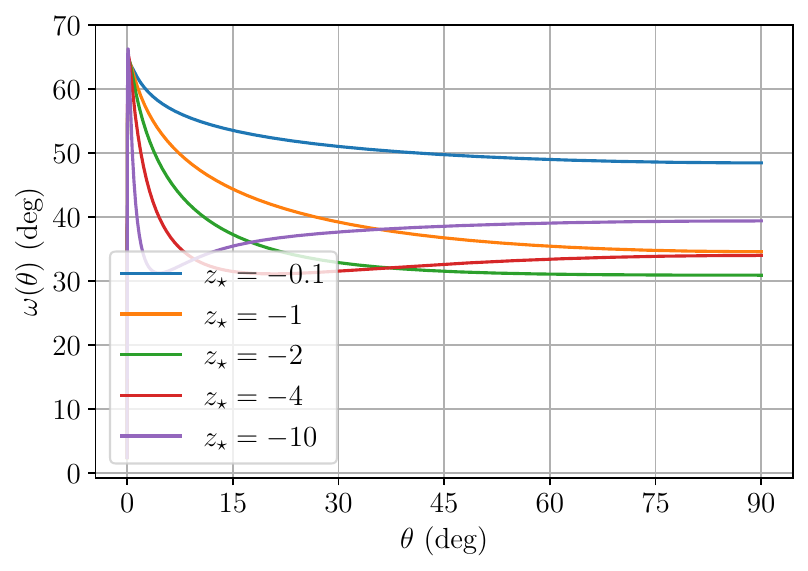}
        \caption{ $M=10$, $z_0=-20$.}
    \end{subfigure}
       \begin{subfigure}{0.49\textwidth}
        \centering
        \includegraphics[width=0.75\linewidth]{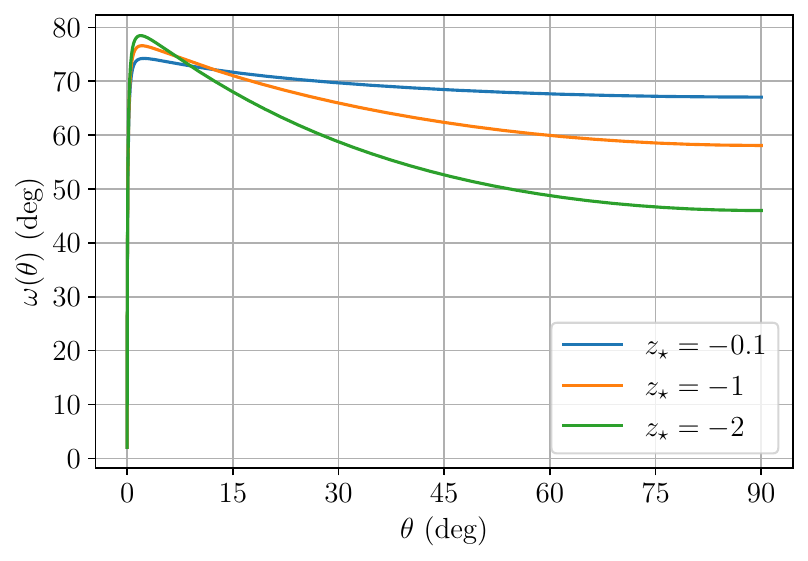}
        \caption{ $M=50$, $z_0=-5$.}
    \end{subfigure}
       \begin{subfigure}{0.49\textwidth}
        \centering
        \includegraphics[width=0.75\linewidth]{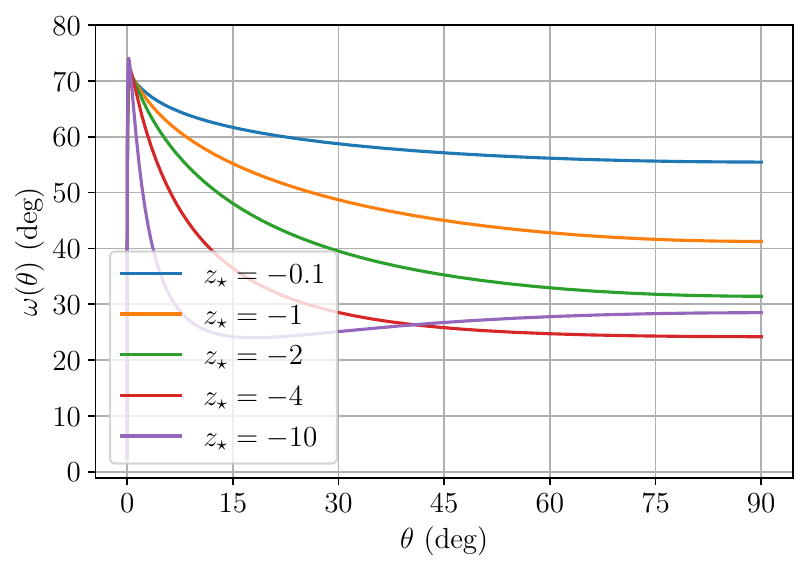}
        \caption{$M=50$, $z_0=-20$.}
    \end{subfigure}
    \caption{The surface deflection angle in the case of exponentially decaying $m_1$ for $|W_{\rm w}|=1$, $\sigma = 0.014$,$H=-80$  and $\mathfrak{m}=0.01$. }
    \label{fig exp deflection}
\end{figure}

\section{Discussion} \label{section discussion}

We conclude with a summary of what this work has achieved and an overview of the possible directions for future research that it opens up.

Starting from the governing equations for a viscous fluid in spherical coordinates, we derived an asymptotic model for the analysis of large-scale wind-drift currents. Unlike classical approximations such as the $f$-plane or $\beta$-plane, our approach retains the full spherical geometry, in the spirit of recent works by \citet{CJ2018,CJ2018b,CJ2019}, and provides a consistent framework for investigating wind-driven flows on large spatial scales, with direct relevance to ocean circulation \citep{WunschFerrari}. The derivation was based on asymptotic expansions with respect to the two small parameters
\[
\varepsilon=\frac{D'}{R'} \qquad \text{and} \qquad \mathscr{R}=\frac{U'}{\Omega'R'},
\]
which characterise the problem and whose smallness reflects the underlying physical regime. Importantly, both parameters emerge naturally from the non-dimensionalisation of the governing equations and do not require any additional assumptions. The scales $R'$ and $\Omega'$ arise directly from the geometry and rotation of the Earth. The characteristic depth scale $D'$ reflects the assumption that the relevant flow occurs near the ocean surface; this leads to the standard scaling in which the vertical velocity is of order $\varepsilon$ smaller than the horizontal velocity components, implying a weak coupling between vertical and horizontal motions (see, for instance, \citet{CJ2018b,CJ2019} and \citet{mioGoverning,c_nlarwa}). The Rossby number $\mathscr{R}$ measures the relative importance of inertial and rotational effects; its smallness identifies a regime in which Earth's rotation dominates over advection.

The asymptotic expansion provided here highlights the fact that, for such flows, the linear terms are the ones determining---at leading order---the dynamics of the wind-driven currents. This theoretical result is related to the fact that, as discussed by \citet{FerrariWunsch} and \citet{W11}, the large-scale motion is deeply connected with the distribution of the potential energy of the ocean, while the predominant contribution to the kinetic energy comes from transport and mixing at much smaller spatial scales; moreover, the time scale for injection/dissipation of kinetic energy (stemming from the nonlinear terms in the equations of motion) is much smaller than the one in \eqref{scaling transformation}, that is, the natural choice $T' = R'/U'$. Moreover, recent studies such as those by \citet{W23,W24}, performing time-averages over many years, argue that there are some simple descriptive, but quantitative, pattern properties of the ocean circulation that are near-globally applicable. Our theoretical analysis can be, in a sense, viewed as a counterpart of these data-driven analyses.

Only two assumptions were made \emph{a priori}: turbulent stresses are parametrised through an eddy viscosity closure, and the eddy viscosity and the density are assumed to depend solely on depth. As we pointed out at the beginning of §~\ref{section scaling}, the eddy viscosity formulation comes with a \emph{caveat}. In the present work, we assume that, on the surface $\{z=0\}$, the eddy viscosity attains a finite value that is subsequently used as the reference scale for the non-dimensionalisation. However, from a physical point of view, the notion of a local eddy viscosity at the ocean surface is not entirely unproblematic. Near the air-sea interface, turbulence is strongly influenced by wave breaking and other air-sea interaction processes, which can induce non-local momentum transport. Consequently, the turbulent stress cannot, in general, be related to the local mean velocity shear through a simple diffusive (eddy viscosity) closure (see, e.g., \citet{Craig}). The surface value of the eddy viscosity should therefore be regarded primarily as a convenient scaling parameter rather than as a quantity with a direct physical interpretation.

Later on, we made a further simplifying assumption, namely that the free surface should be flat. Although, as we pointed out earlier, this assumption can be justified in view of the weak coupling between vertical and horizontal velocities, it is nonetheless an obvious shortcoming, since it neglects physical processes that are known to play a central role in the dynamics of the upper ocean. In particular, the drift in the upper layer is not driven solely by mean ocean currents, but is also influenced by surface waves (see, for example, \citet{rascle2009} and \citet{rohrs2012}), and, in the upper few metres of the ocean, the wave-induced Stokes drift contributes significantly to the overall transport and structure of the flow \citep{stokes1847,monismith2008,rohrs2014wave}. This influence is further enhanced by the dissipation of wave momentum \citep{carniel2009,christensen2009}, which couples with Earth's rotation through the Coriolis--Stokes force. The resulting wave-current interactions can substantially modify both the magnitude and direction of near-surface currents \citep{ursell1950theoretical,Hasselmann,lewis2004,PoltonETAL}. However, in principle, our model can also account for wave motion: the kinematic boundary \eqref{KBC asymptotic} provides an equation for the free surface, which also enters the equations directly through the dynamic boundary conditions \eqref{DBC nondim}. In particular, denoting by $z = \tilde{h}$ the free surface at leading order, \eqref{pressure formula} would read
\begin{equation}
    \tilde{p} = p_{\rm s} + g\int_z^{\tilde{h}}\rho(s)\ds,
\end{equation}
and thus the geostrophic flow would be given by
\begin{equation} \label{geostrophic free surface}
\left.\begin{aligned}
    2\tilde{v}_\g\sin\theta &= \frac{1}{\rho\cos\theta}\left(\frac{\p p_{\rm s}}{\p\varphi} + g\rho(\tilde{h})\frac{\p\tilde{h}}{\p\varphi}\right) \\
    2\tilde{u}_\g\sin\theta &= -\frac{1}{\rho}\frac{\p p_{\rm s}}{\p\theta} - g\frac{\rho(\tilde{h})}{\rho}\frac{\p\tilde{h}}{\p\theta}
\end{aligned}\;\right\} \quad \text{in $\{H < z < \tilde{h}\}$}.
\end{equation}
Clearly, this significantly complicates the analysis, even in the case of constant atmospheric pressure; this is a problem that we hope to tackle in future work.

Even with the simplification of a flat surface, our model has interesting features. One of the key aspects is that it does not rely on tangent-plane approximations; instead, it fully retains the effects of the sphericity of the Earth. To exemplify the potential of this formulation, in §~\ref{section Sverdrup} we discussed the simplest model of ocean circulation, that is, Sverdrup's model \citep{sverdrup47}. Already in this simplified context, the formulation in global spherical coordinates gives rise to additional terms in the formulae due to the variations of~$\theta$. We believe that this approach could be applied successfully also to less simple models and provide a more accurate description of large-scale flows than is possible within the framework of tangent-plane theory; this is something that we plan to pursue in later work.

\begin{table}
    \centering
    \begin{tabular}{c|c|c|c|c|c|c|c|c|c|c}
   {\bf Source}& \rotatebox{90}{\cite{chereskin95}} & \rotatebox{90}{\cite{YM}} \rotatebox{90}{} \rotatebox{90}{\cite{YMetal}}
& \rotatebox{90}{\cite{KimETAL}}
& \rotatebox{90}{\cite{RohrsC2015}} & \rotatebox{90}{\cite{NiilerPaduan}} &  \rotatebox{90}{\cite{Poulain}} & \rotatebox{90}{\cite{SentchevETAL}}& \rotatebox{90}{\cite{BertaETAL}}\\
\hline \shortstack{{\bf Region}\\{}\vspace{4pt} } & \shortstack{California\\(EBC)} & \shortstack{Tsushima\\Strait} & \shortstack{San\\Diego} & \shortstack{North\\Atlantic} & \shortstack{Northeast\\Pacific} &\shortstack{Eastern \\ Medit.} & \multicolumn{2}{c}{\shortstack{Northwestern\\ Mediterranean}}    \\ \hline
       \shortstack{{\bf Deflection} \\ {\bf angle [deg]} } & 70 & 20--60 &  $>$50 & \shortstack{64 \\ 84} & 60 &\shortstack{17--20\\ 28--30} & 12--25 & 25--30
    \end{tabular}
    \caption{Overview of reported measurements of the surface deflection angle of wind-drift currents.}
    \label{tab:deflections}
\end{table}

In §~\ref{section Ekman}, we focused on Ekman flows. There, we split the velocity field into a geostrophic and an Ekman component (the latter being essentially a boundary layer correction that takes into account the effects of viscosity close to the flat surface). It should be noted that this splitting requires the latitude $\theta$ to be non-zero (that is, that we are away from the equator), because the definition \eqref{geostrophic} of the geostrophic flow requires dividing by $\sin\theta$. It is also to be expected that for small values of $\theta$ (say, for $|\theta|$ below $5^\circ$) this approach will not be an accurate description of the real flow, because close to the equator the Coriolis force becomes small and the linear Coriolis terms do not dominate the nonlinear advective terms in the Navier--Stokes equations (contrarily to what happens at higher latitudes), so that a different scaling as in Stommel's theory of wind-driven drift currents \citep{Stommel1960, Webb2018} would be required to capture the emergence of nonlinear effects at leading order; see also the remarks at the beginning of §~\ref{section Ekman}.

Keeping these considerations in mind and assuming for simplicity that the atmospheric pressure is constant (which, together with the assumption of a flat surface, implies that the underlying geostrophic flow vanishes---cf. \eqref{geostrophic free surface}), we proved analytically that the leading-order problem possesses a unique solution that behaves qualitatively like an Ekman spiral, and we were able to characterise the surface deflection angle and the intensity of the surface current, showing in particular that the deflection angle goes to zero as the latitude $\theta$ goes to zero. Then, to obtain some explicit values to compare with earlier solutions and with experimental data available in the literature, we assumed the density to be constant (to allow us to look for explicit solutions) and adopted three different explicit parametrisations for the eddy viscosity. In all cases, the eddy viscosity is taken to decrease with depth from the surface until a certain depth $z_{\star}$, below which we have three profiles: constant, and linearly and exponentially decreasing to a depth $z_0$, below which the eddy viscosity becomes the constant (and very small) molecular viscosity. Choosing reasonable values for the maximum of the eddy viscosity and for the depths $z_{\star}$ and $z_0$, we plotted various profiles of the Ekman spiral and, more importantly, calculated the deflection angles. These can be compared with measurements: in practice, the deflection angle can be observed to be both greater \citep{chereskin95, NiilerPaduan, YM, YMetal, KimETAL, RohrsC2015} and lower \citep{Poulain, YM, YMetal,SentchevETAL, BertaETAL} than 45$^\circ$; see the values in table~\ref{tab:deflections}. We recover both regimes, obtaining highly realistic values, as can be seen comparing figures~\ref{fig constant spiral}, \ref{fig linear spiral} and \ref{fig exp 2d} with table~\ref{tab:deflections}. 

Away from equatorial regions (where our model ceases to be valid), our results from §~\ref{section explicit} seem to point to the following conclusions. For fixed parameters of the eddy viscosity, in most cases the deflection angle decreases with latitude; moreover, as the maximum $M$ of the eddy viscosity increases, the deflection angle increases, and as the depth $z_{\star}$ of the maximum $M$ decreases, the deflection angle increases as well; and finally, the faster the eddy viscosity decays in the interval $[z_0,z_{\star}]$, the higher the deflection angle. 

As we mentioned in §~\ref{section explicit}, from a physical point of view, $M$ represents the maximum efficiency of turbulent momentum transport within the water column. A larger $M$ indicates more vigorous mixing, meaning that the momentum imparted by the wind is redistributed vertically more effectively. The depth $z_{\star}$ at which this maximum occurs characterises how deeply the wind-generated turbulence penetrates before reaching its greatest intensity. For a fixed $M$, a larger $z_{\star}$ implies that turbulent motions remain energetic to greater depths, allowing the influence of the surface forcing to extend further into the ocean interior. On the other hand, a smaller $z_{\star}$ is often associated with a stronger stratification, which suppresses vertical motions and confines the turbulence closer to the surface. In this sense, $z_{\star}$ can also be viewed as an indicator of the balance between wind-driven turbulent production and stratification-induced damping. In particular, we infer that larger deflection angles, such as those observed by \citet{chereskin95} and \citet{RohrsC2015}, are attained for high values of $M$, which corresponds to a more efficient mixing due to the wind, coupled with a strong stratification (i.e., a small $z_{\star}$). In contrast, small deflection angles, such as those measured by \citet{Poulain}, \citet{SentchevETAL} and \citet{BertaETAL}, arise for smaller $M$  and weaker stratification (i.e., larger $z_{\star}$). Additionally, the velocity profiles provided by \citet{SentchevETAL} can be seen to be qualitatively similar to those of figures~\ref{fig linear spiral}(b) and~\ref{fig exp 2d}(b), which indeed correspond to small $M$ and large $z_\star$ ($M=10$, $z_\star=4$).
\begin{figure}
    \begin{subfigure}{0.49\textwidth}
        \centering
        \includegraphics[width=0.75\linewidth]{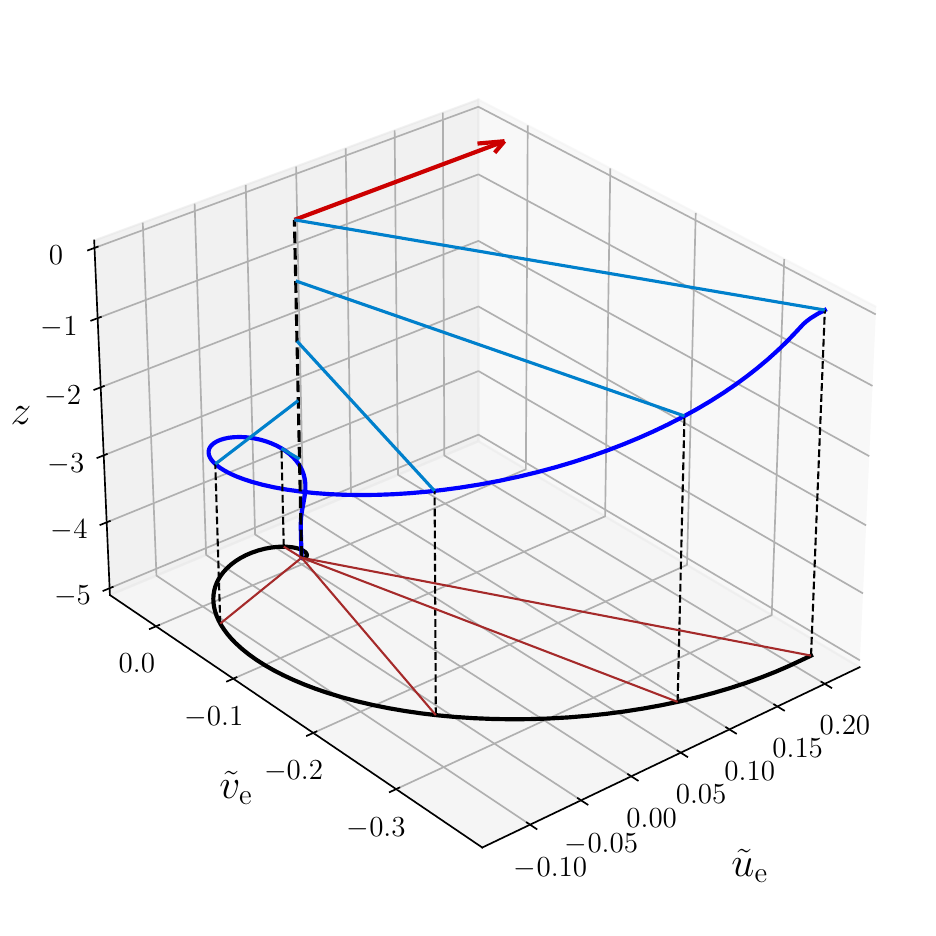}
        \caption{$M=10$, $z_\star=-0.1$.}
    \end{subfigure}
    \begin{subfigure}{0.49\textwidth}
        \centering
        \includegraphics[width=0.75\linewidth]{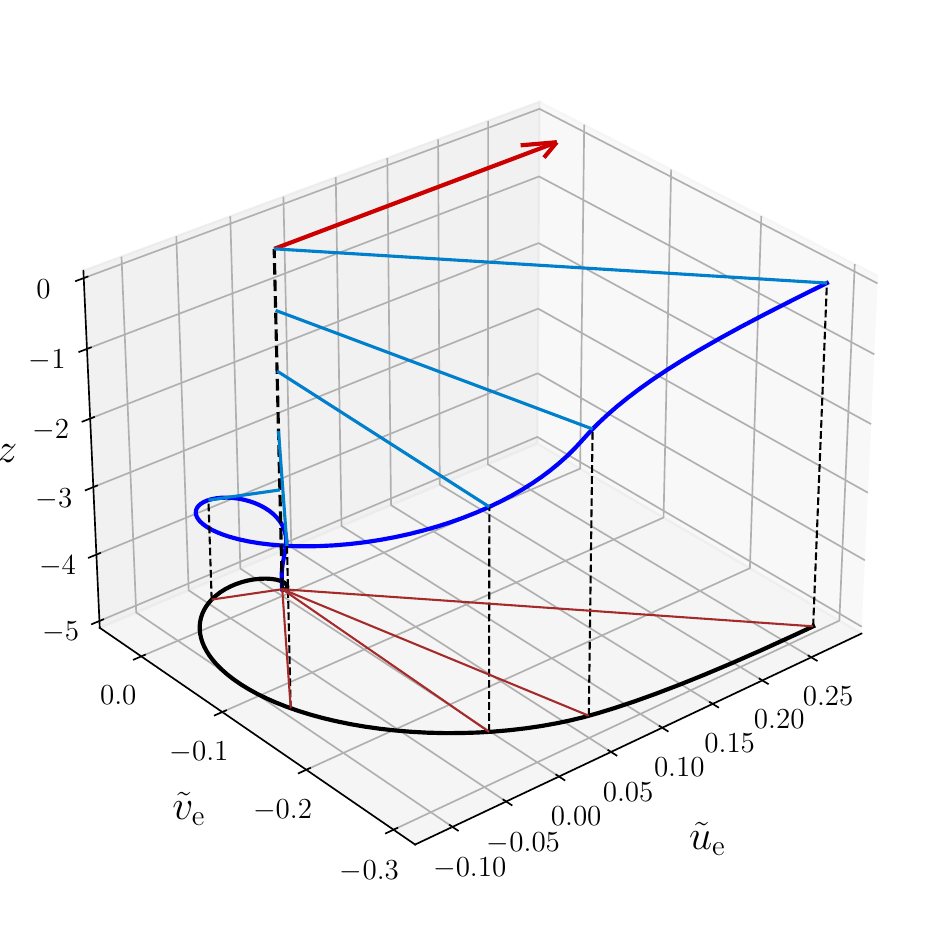}
        \caption{$M=10$, $z_\star=-1$.}
    \end{subfigure}
       \begin{subfigure}{0.49\textwidth}
        \centering
        \includegraphics[width=0.75\linewidth]{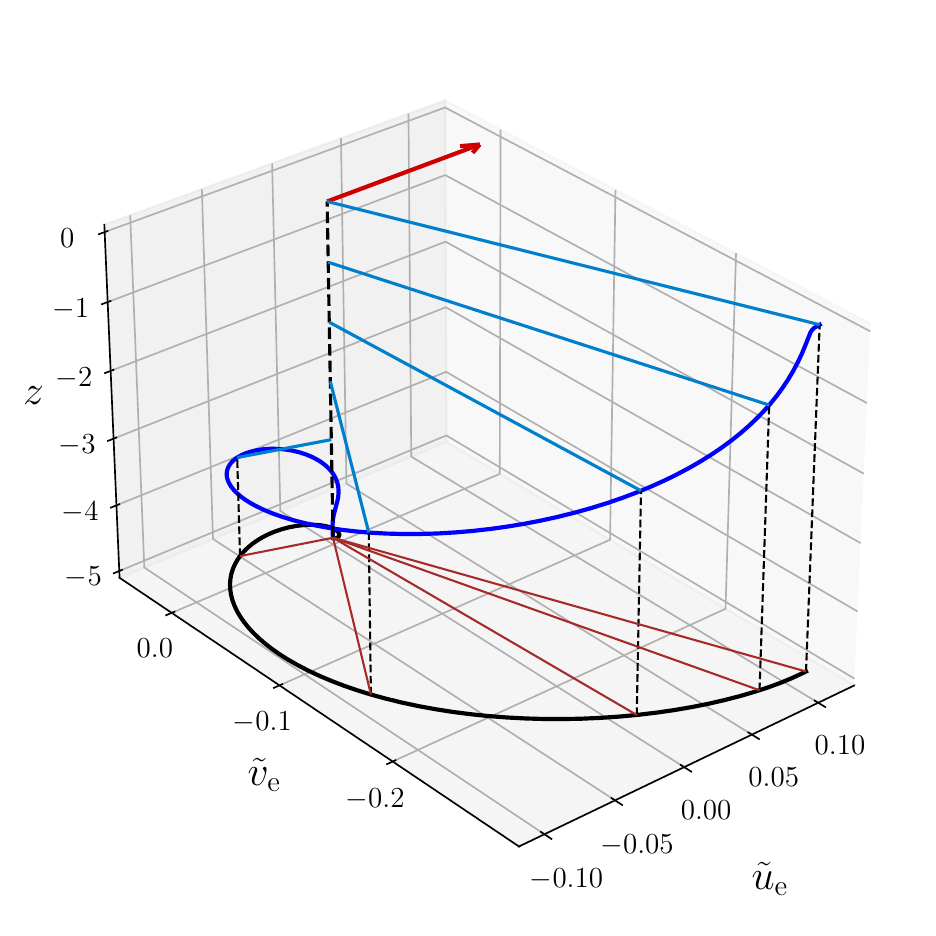}
        \caption{$M=50$, $z_\star=-0.1$}
    \end{subfigure}
       \begin{subfigure}{0.49\textwidth}
        \centering
        \includegraphics[width=0.75\linewidth]{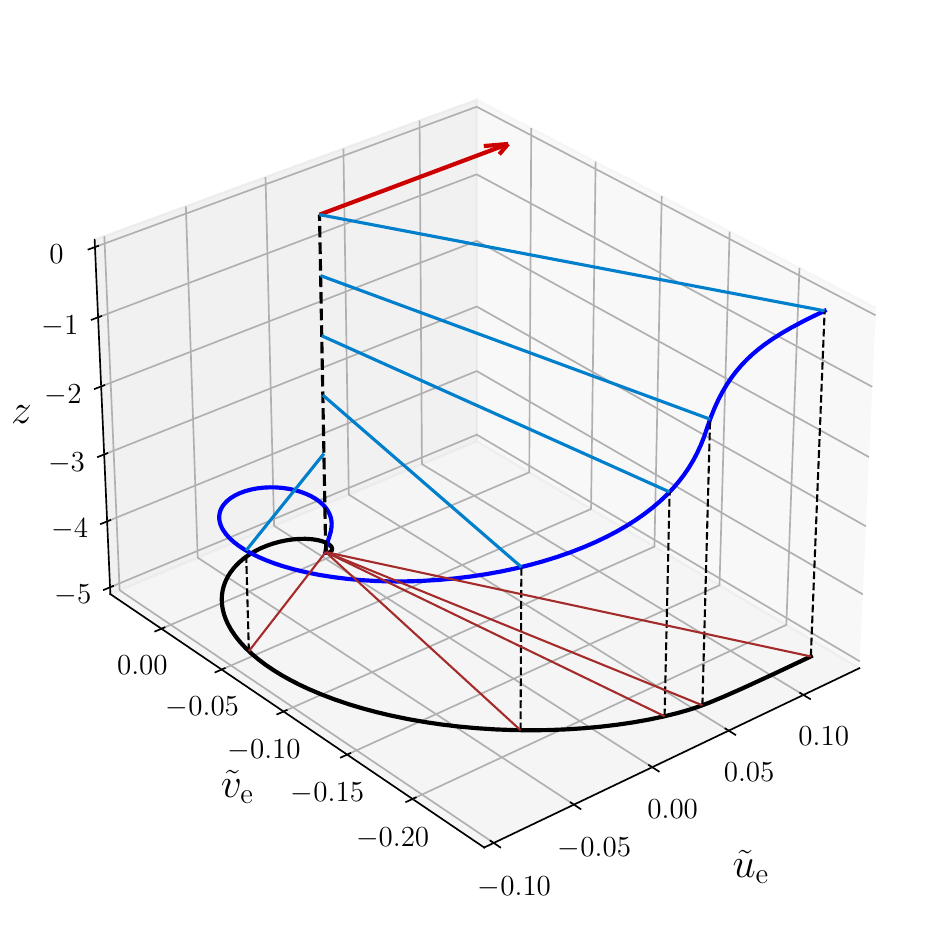}
        \caption{ $M=50$, $z_\star=-1$.}
    \end{subfigure}
    \caption{The case of exponentially decaying $m_1$ with $|W_{\rm w}|=1$, $\sigma = 0.014$. Moreover, $z_0=-5$, $H=-80$,  and $\mathfrak{m}=0.01$. In all cases, the plots are displayed only up to $z=-5.$}
    \label{fig exp flattening}
\end{figure}

Additionally, measurements show that the speed of the Ekman current decays with depth more rapidly than its directional rotation, in contrast to the theoretical predictions from Ekman's classical theory \citep{Wijffels1994}. This effect is known as Ekman spiral flattening or compression \citep{Schudlich1998}. Although it can be easily observed in our figures~\ref{fig constant spiral},~\ref{fig linear spiral} and~\ref{fig exp 2d}, it is interesting to investigate how the Ekman flattening is affected by the parameters $M$ and $z_\star$. Figure~\ref{fig exp flattening} shows the velocity profiles for different two different values of $M$ and two different values $z_\star$, for an exponentially decaying $m_1.$ The plots indicate that the flattening is enhanced as both $M$ and $z_\star$ decrease, corresponding to a reduced mixing and a higher stratification (or, in general, a turbulent motion confined close to the surface).

We conclude with one last remark. In the present paper, we have dealt with the currents generated by the wind over a free surface, namely considering air-sea interactions via the boundary condition \eqref{stress3}. However, a similar analysis could be carried out for the bottom boundary layer or the atmospheric boundary layer. In this case, the ODE \eqref{ode inhom} remains unchanged, whereas the boundary conditions must be modified (see, e.g., \citet{veronis} or \citet{Ste24}). Specifically, for atmospheric flows over a rigid surface at $\{z=0\}$, one obtains
\begin{equation}\label{ode inhom atmospheric}
    \left\{\begin{aligned}
    & \frac{\p}{\p z}\left(m\frac{\p W}{\p z}\right) = 2\i\rho W\sin\theta - \frac{\p}{\p z}\left(m\frac{\p W_\g}{\p z}\right) \qquad \text{in $\{z>0\}$},  \\[0.2em]
    & W = -W_\g \qquad \text{on $\{z=0\}$}, \\[0.2em]
    & \lim_{z\rightarrow +\infty} W(z)=0.
    \end{aligned}\right.
\end{equation}
In the simplest setting, where the geostrophic current $W_\g$ is constant and both the eddy viscosity and the density are uniform, i.e. $m(z)=\rho(z)=1$, the solution to \eqref{ode inhom atmospheric} is
\begin{equation}
W(z)=-W_\g\E^{-\sqrt{|\sin\theta|}\,z}\E^{-\i\sign(\theta)\sqrt{|\sin\theta|}\,z}.
\end{equation}
However, a comprehensive discussion of boundary layer flows near a rigid surface lies beyond the scope of this work.

\begin{bmhead}[Acknowledgements.]
The authors are grateful to the anonymous reviewers for their valuable and insightful comments and suggestions that greatly helped to improve the quality of the paper. C.P. would like to thank Kerry Emanuel and Richard Rotunno for the interesting discussion on this topic during the scientific trimester workshop ``Vortices and vorticity in geophysical flows'' within the scientific trimester ``Mathematical Developments in Geophysical Fluid Dynamics'', held at IHP in Paris.
\end{bmhead}

\begin{bmhead}[Funding.]
C.P. was partially supported by the Austrian Science Fund (FWF) [grant number Z 387-N, grant doi: \href{https://doi.org/10.55776/Z387}{10.55776/Z387}].
\end{bmhead}

\begin{bmhead}[Declaration of interests.]
The authors report no conflict of interest.
\end{bmhead}

\begin{bmhead}[Data availability statement.]
All data used in this work are properly cited throughout the manuscript.
\end{bmhead}

\begin{bmhead}[Author contributions.]
C.P. proposed the topic, developed the governing equations in §~\ref{section governing equations} and §~\ref{section asymptotics}, carried out the theory of §~\ref{section Sverdrup} as well as the theory and numerics of §~\ref{section explicit}, and contributed to writing, alongside minor contributions to the theory of §~\ref{section analytical results}. L.R. developed the governing equations in §~\ref{section governing equations} and §~\ref{section asymptotics} and the theory of §~\ref{section analytical results} and contributed to the theory and numerics of §~\ref{section explicit} and to writing. E.S. contributed to Appendix \ref{appendix proofs}. The revision was carried out by C.P. and L.R.
\end{bmhead}

\begin{bmhead}[Author ORCIDs.]\\
Christian Puntini \href{https://orcid.org/0009-0008-5454-0922}{https://orcid.org/0009-0008-5454-0922};\\
Luigi Roberti \href{https://orcid.org/0000-0001-7678-7389}{https://orcid.org/0000-0001-7678-7389};\\
Eduard Stefanescu \href{https://orcid.org/0009-0001-6507-9691}{https://orcid.org/0009-0001-6507-9691}.
\end{bmhead}

\begin{appendix}

\section{Proofs of the results from §~\ref{section analytical results}} \label{appendix proofs}

We provide here the proofs of the analytical results we formulated in §~\ref{section analytical results}.

\subsection{Proof of Theorem \ref{theorem Ekman spiral}}
For simplicity, we denote a derivative with respect to $z$ by a prime. Let $Y\in C^2((H,0))\cap C([H,0])$ be the solution to the linear initial value problem
\begin{equation} \label{IVP}
    \left\{\begin{aligned}
    & (m(z)Y'(z))' - 2\i\sin\theta\,\rho(z)Y(z) = 0, \qquad H < z < 0, \\[0.2em]
    & Y(H) = 0, \\[0.2em]
    & Y'(H) = 1;
    \end{aligned}\right.
\end{equation}
the existence of a unique global solution to this problem follows from standard theory for second-order linear systems of ordinary differential equations (see, e.g., \citet{Teschl}). Multiplying the first equation of \eqref{IVP} with the complex conjugate $\overline{Y}$ of $Y$, integrating from $H$ to $z$, performing an integration by parts, and taking the third equation in \eqref{IVP} into account, we obtain
\begin{equation} \label{energy}
m(z)Y'(z)\overline{Y(z)} = \int_{H}^z m(s)|Y'(s)|^2\ds + 2\i\sin\theta\int_{H}^z\rho(s)|Y(s)|^2\ds.
\end{equation}
Writing
\[Y(z) = r(z)\E^{\i\phi(z)},\]
plugging into \eqref{energy}, and taking the real and imaginary part, we see that, on the one hand,
\begin{equation} \label{gamma 1}
    m(z)r'(z)r(z) = \int_{H}^z m(s)|Y'(s)|^2\ds > 0,
\end{equation}
hence, since $m(z) > 0$ and $r(z) \ge 0$, we have
\begin{equation} \label{Ekman spiral 1}
r(z) > 0 \qquad \text{and} \qquad r'(z) > 0 \qquad \text{for all $z> H$},
\end{equation}
and, on the other hand,
\begin{equation} \label{gamma 2}
    m(z)\phi'(z)r(z)^2 = 2\sin\theta\int_{H}^z\rho(s)|Y(s)|^2\ds 
    \begin{cases}
        > 0 & \text{if $\theta \in \left(0,\dfrac{\pi}{2}\right)$,} \\[0.7em]
        < 0 & \text{if $\theta \in \left(-\dfrac{\pi}{2},0\right)$,}
    \end{cases}
\end{equation}
thus
\begin{equation} \label{Ekman spiral 2}
\sign(\theta)\phi'(z) > 0 \qquad \text{for all $z > H$}.
\end{equation}
In particular, we have
\[Y(z) \neq 0 \qquad \text{for all $z > H$}.\]
Now, note that a solution $W$ to \eqref{ode z1} must necessarily be of the form
\[W = cY\]
for some $c\in\C$; therefore, we only have to show that we can find a unique $c$ such that the nonlinear upper boundary condition is met. To this end, let us simplify the notation by setting
\[a \coloneqq W(0) \qquad \text{and} \qquad d \coloneqq W_{\rm w} -\sigma a,\]
hence $c = \frac{a}{Y(0)}$ (as we saw, $Y(0) \neq 0$) and
\[W'(0) = \lambda a,\]
where
\begin{equation} \label{lambda 1}
\lambda \coloneqq \frac{Y'(0)}{Y(0)} = \frac{r'(0)}{r(0)} + \i\phi'(0).
\end{equation}
Note that ${\rm Re}(\lambda) > 0$ and $\sign(\theta){\rm Im}(\lambda) > 0$, in view of \eqref{Ekman spiral 1} and \eqref{Ekman spiral 2}.

In order for the second equation in \eqref{ode z1} to be satisfied, we must have
\[\frac{\lambda}{\mu}(W_{\rm w}-d) = |d|d,\]
which, writing
\[d = \chi\E^{\i\delta}\]
for $\chi = |d|$, becomes
\begin{equation} \label{varrho}
    \lambda W_{\rm w} = \chi\E^{\i\delta}(\lambda + \sigma\chi).
\end{equation}
Multiplying both sides of this equation by the respective complex conjugate and rearranging terms, we see that $\chi$ is determined through the positive solutions of the equation
\begin{equation} \label{root}
    p(\chi) = 0,
\end{equation}
where $p$ is the real fourth-order polynomial
\begin{equation}
    p(x) = \mu^2x^4 + 2\mu{\rm Re}(\lambda)x^3 + |\lambda|^2x^2 - |\lambda W_{\rm w}|^2.
\end{equation}
Note that $p(0) < 0$, $p(x) \to \infty$ as $x\to \infty$, and $p$ is strictly increasing on $(0,\infty)$, because
\begin{equation}
    p'(x) = 2x\bigl(2\mu^2 x^2 + 3\mu{\rm Re}(\lambda)x + |\lambda|^2\bigr) > 0 \qquad \text{for $x>0$};
\end{equation}
therefore, there is a unique positive solution $\chi$ of \eqref{root}, from which, in view of \eqref{varrho}, we recover the constant $c$ through
\begin{equation} \label{W(0)}
c = \frac{a}{Y(0)} = \frac{W_{\rm w}-d}{Y(0)} = \frac{\chi}{\sigma\chi + \lambda}\cdot\frac{W_{\rm w}}{Y(0)}
\end{equation}
(note that $\sigma\chi + \lambda \neq 0$, since $\chi > 0$ and $\lambda \in \C\setminus\R$). Finally, since $W$ is a multiple of $Y$, the claimed properties \eqref{Ekman spiral} of the Ekman spiral follow immediately from \eqref{Ekman spiral 1} and \eqref{Ekman spiral 2}.

\subsection{Proof of Theorem \ref{theorem deflection}}

By \eqref{W(0)}, we have
\begin{equation}
    |W(0)|\E^{\i\Theta} = W(0) = \frac{\chi}{\sigma\chi + \lambda}W_{\rm w} = \frac{1}{\mu}\left(1+\frac{\lambda}{\sigma\chi}\right)^{-1}|W_{\rm w}|\E^{\i\Gamma},
\end{equation}
thus
\begin{equation}
    \left|\frac{W_{\rm w}}{W(0)}\right|\E^{\i(\Gamma - \Theta)} = \sigma + \frac{\lambda}{{\chi}},
\end{equation}
whence \eqref{deflection formula} follows immediately. To prove the bounds \eqref{deflection bounds}, note that, with $p$ as in \eqref{polynomial}, since ${\rm Re}(\lambda) \leq |\lambda|$ we have
\begin{equation}
    q(x) \leq p(x) \leq r(x), \qquad x>0,
\end{equation}
where
\begin{equation}
    r(x) = |\lambda|^2 x^2\left(\frac{\mu}{|\lambda|}x + 1\right)^2 - |\lambda W_{\rm w}|^2 \qquad \text{and} \qquad q(x) = {\rm Re}(\lambda)^2 x^2\left(\frac{\mu}{{\rm Re}(\lambda)}x + 1\right)^2 - |\lambda W_{\rm w}|^2.
\end{equation}
Therefore, the unique positive root $\chi$ of $p$ must be smaller than the unique positive root of $q$ and larger than the unique positive root of $r$, which are given by $\alpha/{\sigma}$ and $\beta/{\sigma}$, respectively, and the bounds \eqref{deflection bounds} follow.

\subsection{Proof of Theorem \ref{deflection equator}}

Let $Y_\theta$ be the solution to \eqref{IVP} (where we explicitly denote the $\theta$-dependence via the index $\theta$) and $Y_0$ the solution to
\begin{equation}
    \left\{\begin{aligned}
    & (m(z)Y_0'(z))' = 0, \quad H < z < 0, \\[0.2em]
    & Y_0(H) = 0, \\[0.2em]
    & Y_0'(H) = 1,
    \end{aligned}\right.
\end{equation}
which is easily found to be
\begin{equation}
    Y_0(z) = \int_H^z\frac{m(H)}{m(s)}\ds.
\end{equation}
Then, the function $\hat{Y}_\theta = Y_\theta-Y_0$ solves the problem
\begin{equation}
    \left\{\begin{aligned}
    & (m(z)\hat{Y}_\theta'(z))' = 2\i\sin\theta\,\rho(z)\hat{Y}_\theta(z) + 2\i\sin\theta\,\rho(z)Y_0(z), \qquad H < z < 0, \\[0.2em]
    & \hat{Y}_\theta(H) = 0, \\[0.2em]
    & \hat{Y}_\theta'(H) = 0.
    \end{aligned}\right.
\end{equation}
Integrating the first equation yields
\begin{equation} \label{integral derivative}
    \hat{Y}_\theta'(z) = \frac{2\i\sin\theta}{m(z)}\int_H^z\rho(s)\hat{Y}_\theta(s)\ds + \frac{2\i\sin\theta}{m(z)}\int_H^z\rho(s)Y_0(s)\ds;
\end{equation}
integrating once more, changing variables and taking absolute values, we obtain
\begin{equation}
    |\hat{Y}_\theta(z)| \leq 2|\sin\theta|\int_H^z\left(\int_t^0\frac{\d s}{m(s)}\right)\rho(t)|\hat{Y}_\theta(t)|\dt + 2|\sin\theta|\int_H^z\left(\int_t^z\frac{\d s}{m(s)}\right)\rho(t)|Y_0(t)|\dt.
\end{equation}
Gr\"{o}nwall's inequality then implies that there is a constant $C>0$ such that
\begin{equation}
    |\hat{Y}_\theta(z)| \leq C|\sin\theta| \qquad \text{for all $z\in[H,0]$},
\end{equation}
hence $\lim_{\theta\to 0}\hat{Y}_\theta(z) = 0$ uniformly in $z$, and plugging into \eqref{integral derivative} we see that $\lim_{\theta\to 0}\hat{Y}_\theta'(z) = 0$ uniformly in $z$ as well. Therefore,
\begin{equation}
    \lim_{\theta\to 0}\lambda_\theta = \lim_{\theta\to 0}\frac{Y_\theta'(0)}{\hat{Y}_\theta(0)} = \frac{Y_0'(0)}{Y_0(0)} = \left(\int_H^0\frac{m(0)}{m(s)}\ds\right)^{-1} > 0,
\end{equation}
whence the claim follows by arguing by continuity in \eqref{deflection formula}.

\section{Finding explicit solutions} \label{Appendix explicit}

In order to solve \eqref{ode z} with the eddy viscosity \eqref{eddy piecewise}, we first determine the general solution in the upper layer $[{z_{\star}},0]$. To this end, we use the monotonicity of $m$ on $[{z_{\star}},0]$ to introduce the change of variable
\begin{equation} \label{change of variable}
    X(z) = \sqrt{m(z)} = \sqrt{1+\frac{M-1}{{z_{\star}}}\,z}, \qquad z \in [{z_{\star}},0]
\end{equation}
\citep{CJ2019}. Then, denoting
\begin{equation}
    W(z) = G(X),
\end{equation}
the differential equation \eqref{ode z} is transformed into the Bessel equation 
\begin{equation}\label{41}
    X^2\frac{\d^2 G}{\d X^2}(X) + X\frac{\d G}{\d X}(X) - \frac{8\i  \,{z_{\star}}^2\sin\theta}{(M-1)^2} X^2 G(X) = 0, \qquad X({z_{\star}})=\sqrt{M}<X<1.
\end{equation}
The general solution of \eqref{41} is the so-called cylindrical function given by
\begin{equation}
    G(X)=c_1J_0\left(\frac{{z_{\star}}}{M-1}\sqrt{8|\sin\theta|}\,X\E^{-\i\sign(\theta)\frac{\pi}{4}}\right)+c_2Y_0\left(\frac{{z_{\star}}}{M-1}\sqrt{8|\sin\theta|}\,X\E^{-\i\sign(\theta)\frac{\pi}{4}}\right)
\end{equation}
\citep{PZ}, where $c_1,c_2\in\C$ and $J_0$ and $Y_0$ are the complex-valued Bessel functions of the first and second kind, respectively; $J_0$ and $Y_0$ are also known as Neumann functions \citep{Bessel}. Therefore, given the eddy viscosity \eqref{eddy piecewise}, the general solution of \eqref{ode z} (already taking into account the boundary condition \eqref{BC lower}) is
\begin{equation} \label{W increasing eddy}
    W(z) = \begin{cases}
    c_1 J_0\left(\dfrac{{z_{\star}}}{M-1}\sqrt{8|\sin\theta|\left(1+\dfrac{M-1}{{z_{\star}}}\,z\right)}\E^{-\i\sign(\theta)\frac{\pi}{4}}\right) & \\[0.2em]
    +\; c_2 Y_0\left(\dfrac{{z_{\star}}}{M-1}\sqrt{8|\sin\theta|\left(1+\dfrac{M-1}{{z_{\star}}}\,z\right)}\E^{-\i\sign(\theta)\frac{\pi}{4}}\right), &\quad z \in [{z_{\star}},0], \\[2em]
    c_3 F(z) + c_4 G(z), &\quad z\in[z_0,{z_{\star}}], \\[1em]
    c_5\sinh(\alpha(z-H)), &\quad z\in[H,z_0],
    \end{cases}
\end{equation}
where $\{F,G\}$ is a basis of the solution space of the equation in the layer $[z_0,z_\star]$ and the coefficients ${c_1,\ldots,c_5\in\C}$ must be determined from the boundary conditions \eqref{BC stress} and the requirement that $W$ and $\frac{\p W}{\p z}$ be continuous at the interfaces $z={z_{\star}}$ and $z=z_0$; moreover, we denote
\begin{equation}
    \alpha = (1+\i\sign(\theta))\sqrt{\frac{|\sin\theta|}{\mathfrak{m}}}.
\end{equation}
Enforcing the continuity of $W$ and $\frac{\p W}{\p z}$ at $z=z_0$ yields
\begin{equation}
    c_3 = -\frac{\zeta G(z_0) - \xi \frac{\p G}{\p z}(z_0)}{\mathscr{W}[F,G](z_0)}c_5 \qquad \text{and} \qquad c_4 = \frac{\zeta F(z_0) - \xi \frac{\p F}{\p z}(z_0)}{\mathscr{W}[F,G](z_0)}c_5,
\end{equation}
where we abbreviated
\begin{equation}
    \xi = \sinh(\alpha(z_0-H)) \qquad \text{and} \qquad \zeta = \alpha\cosh(\alpha(z_0-H)),
\end{equation}
and where $\mathscr{W}[F,G]$ denotes the Wronskian of $F$ and $G$,
\begin{equation}
    \mathscr{W}[F,G] = F\frac{\p G}{\p z} - \frac{\p F}{\p z}G,
\end{equation}
which is always non-zero, since $F$ and $G$ are linearly independent. Without loss of generality, we may assume that $\frac{\p G}{\p z}(z_0) \neq 0$. Next, we require the continuity of $W$ and $\frac{\p W}{\p z}$ at $z=z_{\star}$. Recalling the identities
\begin{equation}\label{Bessel derivative}
    \frac{\d J_0(s)}{\d s}= -J_{1}(s) \quad\text{and} \quad \frac{\d Y_0(s)}{\d s}= -Y_{1}(s)
\end{equation}
for the derivatives of the Bessel functions \citep{PZ,Bessel}, a simple computation yields
\begin{equation}
    c_1 = \frac{\beta\hat{\delta}-\hat{\beta}\delta}{\gamma\hat{\delta} - \hat{\gamma}\delta}c_3 \qquad \text{and} \qquad c_2 = -\frac{\beta\hat{\gamma}-\hat{\beta}\gamma}{\gamma\hat{\delta} - \hat{\gamma}\delta}c_3,
\end{equation}
with the abbreviations
\begin{equation}
\begin{aligned}
    &\beta = F({z_{\star}}) + \kappa_{\rm b}G({z_{\star}}), \\[0.2em]
    &\gamma = J_0\left(\frac{{z_{\star}}}{M-1}\sqrt{8M|\sin\theta|}\E^{-\i\sign(\theta)\frac{\pi}{4}}\right), \\[0.2em]
    &\delta = Y_0\left(\frac{{z_{\star}}}{M-1}\sqrt{8M|\sin\theta|}\E^{-\i\sign(\theta)\frac{\pi}{4}}\right),
\end{aligned}
\end{equation}
and
\begin{equation}
\begin{aligned}
    &\hat{\beta} = \frac{\p F}{\p z}({z_{\star}}) + \kappa_{\rm b}\frac{\p G}{\p z}({z_{\star}}), \\[0.2em]
    &\hat{\gamma} = -\sqrt{\frac{2|\sin\theta|}{M}}\E^{-\i\sign(\theta)\frac{\pi}{4}}J_1\left(\frac{{z_{\star}}}{M-1}\sqrt{8M|\sin\theta|}\E^{-\i\sign(\theta)\frac{\pi}{4}}\right), \\[0.2em]
    &\hat{\delta} = -\sqrt{\frac{2|\sin\theta|}{M}}\E^{-\i\sign(\theta)\frac{\pi}{4}}Y_1\left(\frac{{z_{\star}}}{M-1}\sqrt{8M|\sin\theta|}\E^{-\i\sign(\theta)\frac{\pi}{4}}\right),
\end{aligned}
\end{equation}
where
\begin{equation}
    \kappa_{\rm b} = \frac{c_4}{c_3} = -\frac{\zeta F(z_0) - \xi\frac{\p F}{\p z}(z_0)}{\zeta G(z_0) - \xi\frac{\p G}{\p z}(z_0)}.
\end{equation}
Setting
\begin{equation}
    \kappa = \frac{c_2}{c_1} = -\frac{\beta\hat{\gamma}-\hat{\beta}\gamma}{\beta\hat{\delta}-\hat{\beta}\delta} \qquad \text{and} \qquad \vartheta = \frac{{z_{\star}}}{M-1}\sqrt{8|\sin\theta|}\E^{-\i\sign(\theta)\frac{\pi}{4}},
\end{equation}
with the notation of Theorem \ref{theorem deflection} we find that
\begin{equation} \label{59}
    \lambda = -\sqrt{2|\sin\theta|}\E^{-\i\sign(\theta)\frac{\pi}{4}}\frac{J_1(\vartheta) + \kappa Y_1(\vartheta)}{J_0(\vartheta) + \kappa Y_0(\vartheta)}.
\end{equation}
This allows us to determine the coefficient $c_1$: in fact, evaluating \eqref{W increasing eddy} at $z=0$ and substituting into 
\begin{equation} \label{W(0) formula}
    W(0) = \frac{\chi}{\sigma\chi + \lambda}\,W_{\rm w},
\end{equation}
where $\chi$ is the unique positive solution of \eqref{root}, we obtain
\begin{equation} \label{c_1}
    c_1 = \frac{\chi}{\sigma\chi + \lambda}\cdot\frac{W_{\rm w}}{J_0(\vartheta) + \kappa Y_0(\vartheta)}.
\end{equation}
Once $c_1$ is determined, it only remains to substitute it in the earlier identities to find all the remaining coefficients:
\begin{equation}
    c_2 = \kappa c_1, \qquad c_3 = \frac{\gamma\hat{\delta} - \hat{\gamma}\delta}{\beta\hat{\delta} - \hat{\beta}\delta}c_1, \qquad c_4 = \kappa_{\rm b}c_3, \qquad c_5 = -\frac{\mathscr{W}[F,G](z_0)}{\zeta G(z_0) - \xi\frac{\p G}{\p z}(z_0)}c_3.
\end{equation}

\subsection{Constant \texorpdfstring{$m_1$}{TEXT}} \label{Appendix constant}
\noindent
In the case $m_1(z)=M$, we take $z_0 = H$. It is then easy to see that in the layer $[H,z_\star]$ we have
\begin{equation}
    W(z) = c_3\sinh(\tilde{\alpha}(z-H)), \qquad z\in [H,z_\star],
\end{equation}
where
\begin{equation}
    \tilde{\alpha} = (1+\i\sign(\theta))\sqrt{\frac{|\sin\theta|}{M}} \qquad \text{and} \qquad c_3 = \frac{\gamma\hat{\delta} - \hat{\gamma}\delta}{\xi\hat{\delta} - \hat{\xi}\delta}c_1,
\end{equation}
with $\gamma,\hat{\gamma},\delta,\hat{\delta}$ as before and
\begin{equation}
    \xi = \sinh(\tilde{\alpha}(z-H)) \qquad \text{and} \qquad \hat{\xi} = \tilde{\alpha}\cosh(\tilde{\alpha}(z-H)).
\end{equation}
The coefficients $c_1$ and $c_2$ are determined as discussed in the previous section, with the \emph{caveat} that now
\begin{equation}
    \kappa = \frac{c_2}{c_1} = -\frac{\xi\hat{\gamma}-\hat{\xi}\gamma}{\xi\hat{\delta}-\hat{\xi}\delta}.
\end{equation}

\subsection{Linearly decaying \texorpdfstring{$m_1$}{TEXT}}
\noindent
In the case of $m_1$ being given by \eqref{m_1 linear}, we have
\begin{equation}
    F(z) = J_0\left(-\dfrac{{z_{\star}}-z_0}{M-\mathfrak{m}}\sqrt{8|\sin\theta|\left(\dfrac{{z_{\star}}\mathfrak{m}-z_0 M}{{z_{\star}}-z_0} + \dfrac{M-\mathfrak{m}}{{z_{\star}}-z_0}\,z\right)}\E^{-\i\sign(\theta)\frac{\pi}{4}}\right)
\end{equation}
and
\begin{equation}
    G(z) = Y_0\left(-\dfrac{{z_{\star}}-z_0}{M-\mathfrak{m}}\sqrt{8|\sin\theta|\left(\dfrac{{z_{\star}}\mathfrak{m}-z_0 M}{{z_{\star}}-z_0} + \dfrac{M-\mathfrak{m}}{{z_{\star}}-z_0}\,z\right)}\E^{-\i\sign(\theta)\frac{\pi}{4}}\right).
\end{equation}

\subsection{Exponentially decaying \texorpdfstring{$m_1$}{TEXT}}
\noindent
With this choice \eqref{exp eddy} for $m_1$, in the layer $[z_0,z_\star]$, \eqref{ode z} becomes
\begin{equation}\label{ode z2}
    \frac{\p^2 W}{\p z^2}(z)+q\frac{\p W}{\p z}(z)-\frac{2\i\sin\theta}{M}\E^{-q(z-{z_{\star}})}W(z)=0, \qquad z_0 < z < {z_{\star}}.
\end{equation}
Let us define the function $H$ through
\begin{equation}
    W(z) = H(z)\E^{-\frac{q}{2}(z-{z_{\star}})}.
\end{equation}
Then
\begin{equation}
    \frac{\p W}{\p z}(z) = \E^{-\frac{q}{2}(z-{z_{\star}})}\left(H'(z)-\frac{q}{2}H(z)\right)
\end{equation}
and
\begin{equation}
    \frac{\p^2 W}{\p z^2}(z) = \E^{-\frac{q}{2}(z-{z_{\star}})}\left(\frac{\p^2 H}{\p z^2}(z) - q\frac{\p H}{\p z}(z) +\frac{q^2}{4}H(z)\right),
\end{equation}
and \eqref{ode z2} transforms into
\begin{equation} \label{ode H}
    \frac{\p^2 H}{\p z^2}(z) - \left(\frac{2\i\sin\theta}{M}\E^{-q(z-{z_{\star}})} + \frac{q^2}{4}\right)H(z) = 0, \qquad z_0 < z < 0.
\end{equation}
Introducing the change of variables
\begin{equation}
    X = \frac{2}{q}\sqrt{\frac{2\i \sin\theta}{M}} \E^{-\frac{q}{2}(z-{z_{\star}})},
\end{equation}
for which we have
\begin{equation}
    \frac{\d}{\d z} = -\frac{q}{2}X\frac{\d}{\d X} \qquad \text{and} \qquad \frac{\d^2}{\d z^2} = \frac{q^2}{4}\left(X^2\frac{\d^2}{\d X^2} + X\frac{\d}{\d X}\right),
\end{equation}
and denoting
\begin{equation}
    H(z) = G(X),
\end{equation}
\eqref{ode H} transforms into the equation
\begin{equation}
    X^2\frac{\d^2 G}{\d X^2}(X) + X\frac{\d G}{\d X}(X) - (X^2+1)G(X) = 0, \qquad \frac{2}{q}\sqrt{\frac{2\i \sin\theta}{M}} < X < \frac{2}{q}\sqrt{\frac{2\i \sin\theta}{\mathfrak{m}}},
\end{equation}
whose general solution is
\begin{equation}
    G(X) = c_3 I_1(X) + c_4 K_1(X), \qquad \frac{2}{q}\sqrt{\frac{2\i \sin\theta}{M}} < X < \frac{2}{q}\sqrt{\frac{2\i \sin\theta}{\mathfrak{m}}},
\end{equation}
where $c_3,c_4\in\C$ and $I_1$ and $K_1$ are the modified Bessel functions of the first and second kind, respectively, both of order 1. Hence,
\begin{equation}
    F(z) = \E^{-\frac{q}{2}(z-{z_{\star}})}I_1\left(\frac{2}{q}\sqrt{\frac{2\i \sin\theta}{M}} \E^{-\frac{q}{2}(z-{z_{\star}})}\right) \qquad \text{and} \qquad G(z) = \E^{-\frac{q}{2}(z-{z_{\star}})}K_1\left(\frac{2}{q}\sqrt{\frac{2\i \sin\theta}{M}} \E^{-\frac{q}{2}(z-{z_{\star}})}\right).
\end{equation}
Recalling that 
\begin{equation}
   \frac{\d I_1(\mathrm{t})}{\d \mathrm{t}} = \frac{1}{2}(  I_0(\mathrm{t})+ I_2(\mathrm{t})) \qquad \text{and} \qquad \frac{\d K_1(\mathrm{t})}{\d \mathrm{t}} = -\frac{1}{2}(  K_0(\mathrm{t})+ K_2(\mathrm{t})),
\end{equation}
a straightforward computation as described above yields the coefficients $c_1,\ldots,c_5$.

\section{The problem at first order in the expansion with respect to \texorpdfstring{$\rr$}{TEXT}} \label{appendix first order}

In this appendix, we derive the equations for the first-order correction of the leading-order dynamics. Once more setting the coefficients to zero in the asymptotic expansions \eqref{horiziontal expanded}, \eqref{vertical expanded}, \eqref{continuity scaling}, \eqref{DBC expanded 2}, and \eqref{stress expanded}, we obtain the problem
\begin{equation}\label{governing 01}
\left.\begin{aligned}
&\frac{1}{\rho}\frac{\partial}{\partial z}\left(m\frac{\partial\hat{u}}{\partial z}\right) + 2\hat{v}\sin\theta \\
&\quad\quad = \frac{1}{\rho\cos\theta}\frac{\p\hat{p}}{\p\varphi} -\left(\frac{\tilde{u}}{\cos\theta}\frac{\partial\tilde{u}}{\partial\varphi} + \tilde{v}\frac{\partial\tilde{u}}{\partial\theta} + \tilde{w}\frac{\p  \tilde{u}}{\p z}\right) + \tilde{u}\tilde{v}\tan\theta \\
&\frac{1}{\rho}\frac{\partial}{\partial z}\left(m\frac{\partial  \hat{v}}{\partial z} \right) -2\hat{u}\sin\theta \\
&\quad\quad = \frac{1}{\rho}\frac{\p\hat{p}}{\p\theta} -  \left(\frac{\tilde{u}}{\cos\theta}\frac{\partial\tilde{v}}{\partial\varphi} + \tilde{v}\frac{\partial\tilde{v}}{\partial\theta} + \tilde{w}\frac{\p   \tilde{v}}{\p z}\right) - \tilde{u}^2\tan\theta \\
&\frac{\partial\hat{p}}{\partial z} = 0 \\
&\frac{\partial \hat{u}}{\partial \varphi} +  \frac{\partial}{\partial \theta} \left(\hat{v} \cos \theta  \right) + \cos\theta\frac{\p\hat{w}}{\p z} = 0
\end{aligned}\;\right\} \quad \text{in $\{H < z <0\}$},
\end{equation}
with dynamic boundary condition
\begin{equation} \label{pressure first order}
\hat{p} = 0 \qquad \text{on $\{z=0\}$}
\end{equation}
and upper boundary condition due to the wind stress given by
\begin{equation} \label{BC_wind_1}
\begin{pmatrix}\dfrac{\p \hat{u}}{\p z}\\[1em]
\dfrac{\p \hat{v}}{\p z}\end{pmatrix} = -\sigma\left(\sqrt{(u_{\rm a}-\sigma\tilde{u})^2+(v_{\rm a}-\sigma\tilde{v})^2}\begin{pmatrix}
    \hat{u} \\[0.2em]
    \hat{v}
\end{pmatrix} + \frac{\hat{u}(u_{\rm a} - \sigma\tilde{u}) + \hat{v}(v_{\rm a} - \sigma\tilde{v})}{\sqrt{(u_{\rm a}-\sigma\tilde{u})^2+(v_{\rm a}-\sigma\tilde{v})^2}}\begin{pmatrix}u_{\rm a}-\sigma\tilde{u} \\[0.2em]
v_{\rm a}-\sigma\tilde{v}\end{pmatrix}\right) \qquad \text{on $\{z = 0\}$.}
\end{equation}
The equation $\frac{\partial\hat{p}}{\partial z} = 0$, together with \eqref{pressure first order}, implies that $\hat{p}=0$ throughout. Therefore, we are left with the problem
\begin{equation} \label{inhomogeneous Ekman}
\left.\begin{aligned}
\frac{\partial}{\partial z}\left(m\frac{\partial\hat{u}}{\partial z}\right) + 2\rho\hat{v}\sin\theta &= \mathscr{F}(\tilde{u},\tilde{v}) \\[0.2em]
\frac{\partial}{\partial z}\left(m\frac{\partial\hat{v}}{\partial z}\right) + 2\rho\hat{u}\sin\theta &=  \mathscr{G}(\tilde{u},\tilde{v})
\end{aligned}\right\} \qquad \text{in $\{H < z <0\}$},
\end{equation}
where
\begin{equation}
    \mathscr{F}(\tilde{u},\tilde{v}) = - \rho\left(\frac{\tilde{u}}{\cos\theta}\frac{\partial\tilde{u}}{\partial\varphi} + \tilde{v}\frac{\partial  \tilde{u}}{\partial\theta} + \tilde{w}\frac{\p  \tilde{u}}{\p z}\right) + \rho\tilde{u}\tilde{v}\tan\theta
\end{equation}
and
\begin{equation}
    \mathscr{G}(\tilde{u},\tilde{v}) =  - \rho\left(\frac{\tilde{u}}{\cos\theta}\frac{\partial\tilde{v}}{\partial\varphi} + \tilde{v}\frac{\partial   \tilde{v}}{\partial\theta} + \tilde{w}\frac{\p   \tilde{v}}{\p z}\right) - \rho\tilde{u}^2\tan\theta.
\end{equation}
The leading-order velocity $(\tilde{u},\tilde{v})$ enters the equations only through the inhomogeneity, given by $\mathscr{F}$ and $\mathscr{G}$, and the upper boundary condition. Once the solution for $(\hat{u},\hat{v})$ has been obtained from these equations, from the continuity equation we can recover $\hat{w}$ via
\begin{equation}
   \frac{\p\hat{w}}{\p z} =  -\frac{1}{\cos\theta}\frac{\partial \hat{u}}{\partial \varphi} -  \frac{\partial  \hat{v}}{\partial \theta} +\tan\theta\, \hat{v}.
\end{equation}
As with the leading-order problem, it is convenient to introduce the complex notation
\[\hat{W} = \hat{u} + \i\hat{v} \qquad \text{and} \qquad \mathscr{I} = \mathscr{F} + \i\mathscr{G};\]
then, equation \eqref{inhomogeneous Ekman} becomes
\begin{equation} \label{inhomogeneous complex}
    \frac{\p}{\p z}\left(m\frac{\p\hat{W}}{\p z}\right) - 2\i\rho\hat{W}\sin\theta = \mathscr{I}.
\end{equation}
Moreover, with this notation, and using the identity ${\rm Re}(\alpha) = \frac{1}{2}(\alpha + \bar{\alpha})$ for $\alpha\in\mathbb{C}$, we can rewrite the boundary condition \eqref{BC_wind_1} as
\begin{equation}
\begin{aligned}
    \frac{\p\hat{W}}{\p z} &= -\sigma\left(|W_{\rm w} - \sigma W|\hat{W} + \frac{W_{\rm w} - \sigma W}{|W_{\rm w} - \sigma W|}{\rm Re}\left\{(W_{\rm w} - \sigma W\overline{\hat{W}}\right\}\right) & & \\
    &= -\frac{\mu}{2}\left(3|W_{\rm w} - \sigma W|\hat{W} + \frac{(W_{\rm w} - \sigma W^2}{|W_{\rm w} - \sigma W|}\overline{\hat{W}}\right) & & \text{on $\{z=0\}$},
\end{aligned}
\end{equation}
where $W_{\rm w}$ and $W$ are as in \eqref{complex notation}, and the bar denotes complex conjugation. Thus, also taking \eqref{BC nondim 2} into account, we are left with the problem
\begin{equation} \label{problem_first_order}
\left\{\begin{aligned}
    & \frac{\p}{\p z}\left(m\frac{\p\hat{W}}{\p z}\right) - 2\i\rho\hat{W}\sin\theta = \mathscr{I} & &\quad \text{in $\{H<z<0\}$}, \\[0.2em]
    & \frac{\p\hat{W}}{\p z} = -\frac{\mu}{2}\left(3|W_{\rm w} - \sigma W|\hat{W} + \frac{(W_{\rm w} - \sigma W^2}{|W_{\rm w} - \sigma W|}\overline{\hat{W}}\right) & &\quad \text{on $\{z=0\}$}, \\[0.2em]
    & \hat{W} = 0 & &\quad \text{on $\{z=H\}$}.
\end{aligned}\right.
\end{equation}

In order to solve \eqref{problem_first_order}, the standard solution method is the variation of constants. First, we look for a basis $\{\hat{W}_1,\hat{W}_2\}$ of the solution space of the homogeneous equation
\[\frac{\p}{\p z}\left(m\frac{\p\hat{W}}{\p z}\right) - 2\i\rho\hat{W}\sin\theta = 0 \qquad \text{in $\{H<z<0\}$}.\]
Then, a particular solution $\hat{W}_{\rm p}$ of the inhomogeneous equation \eqref{inhomogeneous complex} is given by
\[\hat{W}_{\rm p} = w_1\hat{W}_1 + w_2\hat{W}_2;\]
here, the coefficients $w_1$ and $w_2$ are functions that are determined by integrating
\begin{equation}
\frac{\p w_1}{\p z} = \frac{\hat{W}_2\mathscr{I}}{\mathscr{W}[\hat{W}_1,\hat{W}_2]} \qquad \text{and} \qquad \frac{\p w_2}{\p z} = -\frac{\hat{W}_1\mathscr{I}}{\mathscr{W}[\hat{W}_1,\hat{W}_2]},
\end{equation}
where, as before,
\[\mathscr{W}[\hat{W}_1,\hat{W}_2] = \hat{W}_1\frac{\p\hat{W}_2}{\p z} -\frac{\p\hat{W}_1}{\p z}\hat{W}_2\]
denotes the Wronskian of $\hat{W}_1$ and $\hat{W}_2$ (which is always non-zero, because $\hat{W}_1$ and $\hat{W}_2$ are linearly independent). Without loss of generality, we can choose $w_1(H) = w_2(H) = 0$. The general solution of \eqref{inhomogeneous complex} is then given by
\[\hat{W} = C_1\hat{W}_1 + C_2\hat{W}_2 + \hat{W}_{\rm p},\]
where $C_1,C_2\in\C$ are arbitrary constants, to be determined through the boundary conditions. In fact, assuming without loss of generality that $\hat{W}_2(H) \neq 0$, the third equation in \eqref{problem_first_order}, together with $\hat{W}_{\rm p}(H)=0$, yields
\[C_2 = -\frac{\hat{W}_1(H)}{\hat{W}_2(H)}C_1,\]
hence
\[\hat{W}(z) = C_1\left(\hat{W}_1(z) - \frac{\hat{W}_1(H)}{\hat{W}_2(H)}\hat{W}_2(z)\right) + \hat{W}_{\rm p}(z).\]
Then, plugging into the second equation of \eqref{problem_first_order}, we obtain the following equation for $C_1$:
\begin{equation} \label{C_1}
\alpha C_1 + \beta\overline{C_1} = \gamma,
\end{equation}
where, as before, the bar denotes complex conjugation, and
\begin{equation}
\begin{aligned}
    \alpha &= \hat{W}_1'(0) - \frac{\hat{W}_1(H)}{\hat{W}_2(H)}\hat{W}_2'(0) + \frac{3}{2}\sigma|W_{\rm w} - \sigma W(0)|\left(\hat{W}_1(0) - \frac{\hat{W}_1(H)}{\hat{W}_2(H)}\hat{W}_2(0)\right), \\
    \beta &= \frac{1}{2}\sigma\frac{(W_{\rm w} - \sigma W(0))^2}{|W_{\rm w} - \sigma W(0)|}\left(\overline{\hat{W}_1(0)} - \frac{\overline{\hat{W}_1(H)}}{\overline{\hat{W}_2(H)}}\overline{\hat{W}_2(0)}\right), \\
    \gamma &= -\frac{1}{2}\sigma\biggl[3|W_{\rm w} - \sigma W(0)|\bigl(w_1(0)\hat{W}_1(0) + w_2(0)\hat{W}_2(0)\bigr)\biggr. \\
    &\hspace{1.2cm} \biggl.+ \frac{(W_{\rm w} - \sigma W(0))^2}{|W_{\rm w} - \sigma W(0)|}\left(\overline{w_1(0)}\overline{\hat{W}_1(0)} + \overline{w_2(0)}\overline{\hat{W}_2(0)}\right)\biggr] \\
    &\quad - w_1(0)\hat{W}_1'(0) - w_2(0)\hat{W}_2'(0).
\end{aligned}
\end{equation}
The equation \eqref{C_1} has the unique solution
\[C_1 = \frac{\bar{\alpha}\gamma - \beta\bar{\gamma}}{|\alpha| - |\beta|},\]
as long as $|\alpha| \neq |\beta|$. This condition can be reformulated as follows. Denote
\begin{equation}
    Y(z) = \hat{W}_1(z) - \frac{\hat{W}_1(H)}{\hat{W}_2(H)}\hat{W}_2(z);
\end{equation}
then, assuming without loss of generality that $Y'(H)=1$, the function $Y$ is the solution of \eqref{IVP}. Thus, writing
\begin{equation}
    \lambda = \frac{Y'(0)}{Y(0)}
\end{equation}
and recalling that $\chi = |W_{\rm w} - \sigma W(0)|$ is given by the unique positive solution of \eqref{root}, as in the proof of Theorem \ref{theorem Ekman spiral}, we see that $|\alpha| \neq |\beta|$ is equivalent to
\begin{equation} \label{bad equality}
    \left|\lambda + \frac{3}{2}\sigma\chi\right| \neq \frac{1}{2}\sigma\chi.
\end{equation}
Since
\begin{equation} \label{reverse triangle}
    \left|\lambda + \frac{3}{2}\sigma\chi\right| \ge |\lambda| - \frac{3}{2}\sigma\chi
\end{equation}
and, as we argued in the proof of Theorem \ref{theorem deflection},
\begin{equation} \label{lambda bound}
    \sigma\chi \le \frac{-{\rm Re}(\lambda) + \sqrt{{\rm Re}(\lambda)^2 + 4\sigma|\lambda W_{\rm w}|}}{2} \le \sigma|W_{\rm w}|\frac{|\lambda|}{{\rm Re}(\lambda)},
\end{equation}
where we used the inequality $\sqrt{1+x} \le 1+\frac{1}{2}x$ for $x \ge -1$, combining \eqref{reverse triangle} and \eqref{lambda bound} we have
\begin{equation}
    \left|\lambda + \frac{3}{2}\sigma\chi\right| \ge\left(\frac{{\rm Re}(\lambda)}{|W_{\rm w}|} - \frac{3}{2}\sigma\right)\chi.
\end{equation}
Therefore, the validity of \eqref{bad equality} is guaranteed if
\begin{equation} \label{lambda condition}
    {\rm Re}(\lambda) > 2\sigma|W_{\rm w}|.
\end{equation}
Of course, this in fact boils down to a condition on the eddy viscosity and the wind's intensity, since $\lambda$ depends only on the former. However, finding explicit bounds on ${\rm Re}(\lambda)$ in terms of a general $m$ is difficult, so in practice, this condition would have to be checked case by case. In this place, let us only note that the right-hand side of \eqref{lambda condition} is very small, of order of magnitude $10^{-2}$, and in the explicit cases that we considered earlier, the condition is satisfied by a large margin.
\noindent
It is also possible to provide estimates for the first-order correction of the horizontal velocity field given by \eqref{problem_first_order}. To this end, we recall the notion of the \emph{logarithmic matrix norm}.
\begin{definition}
For $n\in\N$, let $\|\cdot\|_*$ denote a matrix norm on $\C^{n\times n}$ induced by a vector norm $\|\cdot\|$ on $\C^n$. For a matrix $A\in\C^{n\times n}$, we define its \emph{logarithmic matrix norm $\mu_*[A]$ associated to $\|\cdot\|_*$} as
\[\mu_*[A] = \lim_{h\searrow 0}\frac{\|\mathbb{I}+hA\|_*-1}{h},\]
 where $\mathbb{I}$ denotes the identity matrix in $\C^{n\times n}$.
\end{definition}
It is easy to show that $\mu_*[A]$ is well-defined for all $A\in\C^{n\times n}$. For more properties of the logarithmic matrix norm, we refer to \citet{Soe24}. These logarithmic matrix norms were first applied to the classical Ekman problem by \citet{Mar22}. In fact, they can be especially useful, among other things, in determining bounds on solutions to ordinary differential equations. The following result, which will be useful to us, can be found in §~6.6 of \citet{Soe24}.
\begin{lemma} \label{log bound}
    For $T>0$, let $A:[0,T)\to\C^{n\times n}$ and $f:[0,T)\to\C^n$ both be continuous. Suppose that $u\in C^1((0,T),\C^n)\cap C([0,T),\C^n)$ solves the inhomogeneous linear initial value problem
    \[u'(t) = A(t)u(t) + f(t), \qquad u(0)=u_0.\]
    Denote
    \[L(t) = \int_0^t\mu_*[A(\tau)]\dtau.\]
    Then
    \[\|u(t)\| \leq \E^{L(t)}\|u_0\| + \E^{L(t)}\int_0^t\E^{-L(s)}\|f(s)\|\ds.\]
\end{lemma}
In order to apply Lemma \ref{log bound}, we rewrite the first equation of \eqref{problem_first_order} as
\begin{equation}
    \frac{\p x}{\p z}(z) = A(z)x(z) + f(z), \qquad z\in (H,0),
\end{equation}
where
\[x = \begin{pmatrix}
    \hat{W} \\[0.2em]
    \frac{\p\hat{W}}{\p z}
\end{pmatrix}, \qquad A = \begin{pmatrix}
    0                       & 1 \\[0.2em]
    \frac{2\i\sin\theta\,\rho}{m}  & -\frac{m'}{m}
\end{pmatrix}, \qquad \text{and} \qquad f = \begin{pmatrix}
    0 \\[0.2em]
    \frac{\mathscr{I}}{m}
\end{pmatrix}.\]
Choosing the norm $\|x\|_1 = |x_1| + |x_2|$ and denoting $A = (a_{jk})_{1\le j,k\le 2}$, the corresponding logarithmic matrix norms can be computed explicitly as
\begin{equation} \label{mu1}
\mu_1[A(z)] = \max_k\biggl\{{\rm Re}(a_{kk}(z)) + \sum_{j\neq k}|a_{jk}(z)|\biggr\} = \max\left\{2|\sin\theta|,\,1 - \frac{m'(z)}{m(z)}\right\}
\end{equation}
\citep{Soe24}. By Lemma~\ref{log bound} and \eqref{mu1}, we have
\begin{equation}
    \bigl|\hat{W}(z)\bigr| + \left|\frac{\p\hat{W}}{\p z}(z)\right| \leq \left|\frac{\p\hat{W}}{\p z}(H)\right|\exp\left(\int_{H}^z\mu_1[A(s)]\ds\right) + \int_{H}^z\exp\left(\int_t^z\mu_1[A(s)]\ds\right)\frac{|\mathscr{I}(t)|}{m(t)}\dt.
\end{equation}
Note that this bound depends only on the eddy viscosity $m$ and, through $\mathscr{I}$ and $\frac{\p\hat{W}}{\p z}(H)$, the leading-order solution $W$ found in §~\ref{section analytical results}.

\section{Ekman spirals}\label{spirals appendix}
In this appendix, we include some 3D plots of the Ekman spirals for the three profiles of the eddy viscosity considered in  §~\ref{section explicit}. More precisely, figure~\ref{fig constant spiral app} refers to the case of constant $m_1$, whereas figure~\ref{fig linear spiral app} shows the case of a linearly decaying $m_1$ and, lastly, figure~\ref{fig exp 2d app} displays the case of an exponentially decaying $m_1$.

\afterpage{%
\begin{figure}
    \centering
    \begin{subfigure}{0.38\textwidth}
        \centering
        \includegraphics[width=\linewidth]{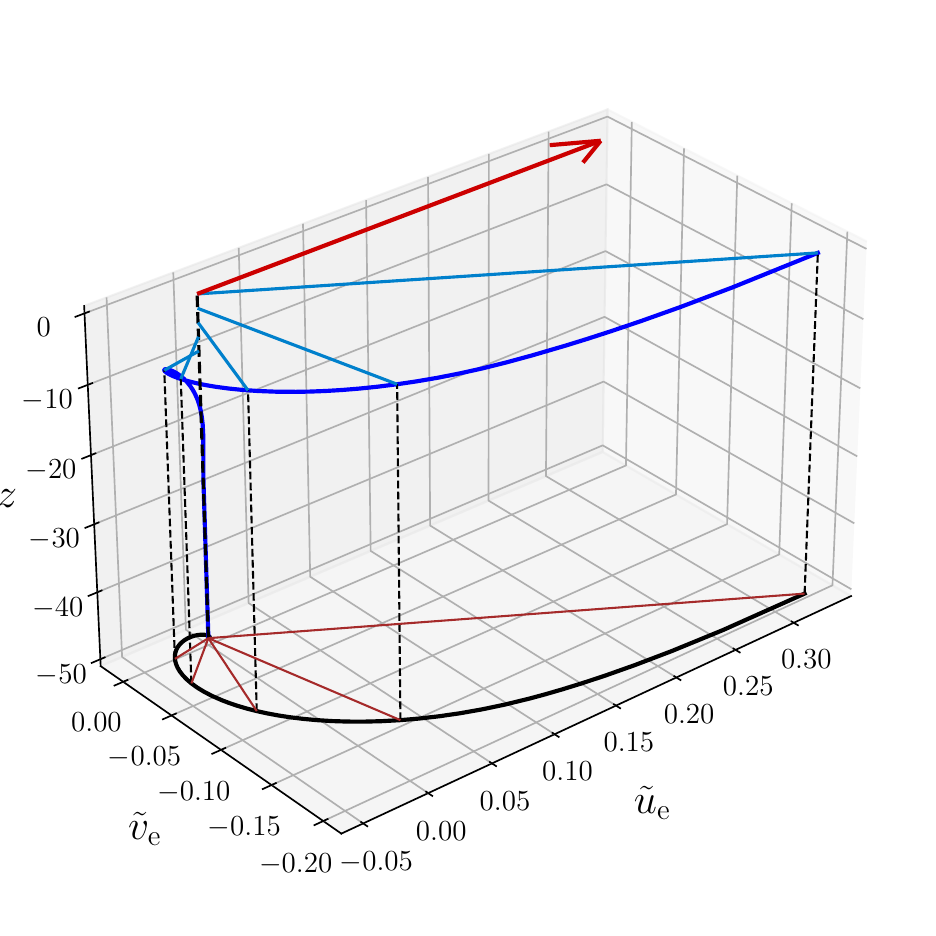}
        \caption{ $M = 10$, ${z_{\star}} = -1$.}
    \end{subfigure}
    \begin{subfigure}{0.38\textwidth}
        \centering
        \includegraphics[width=\linewidth]{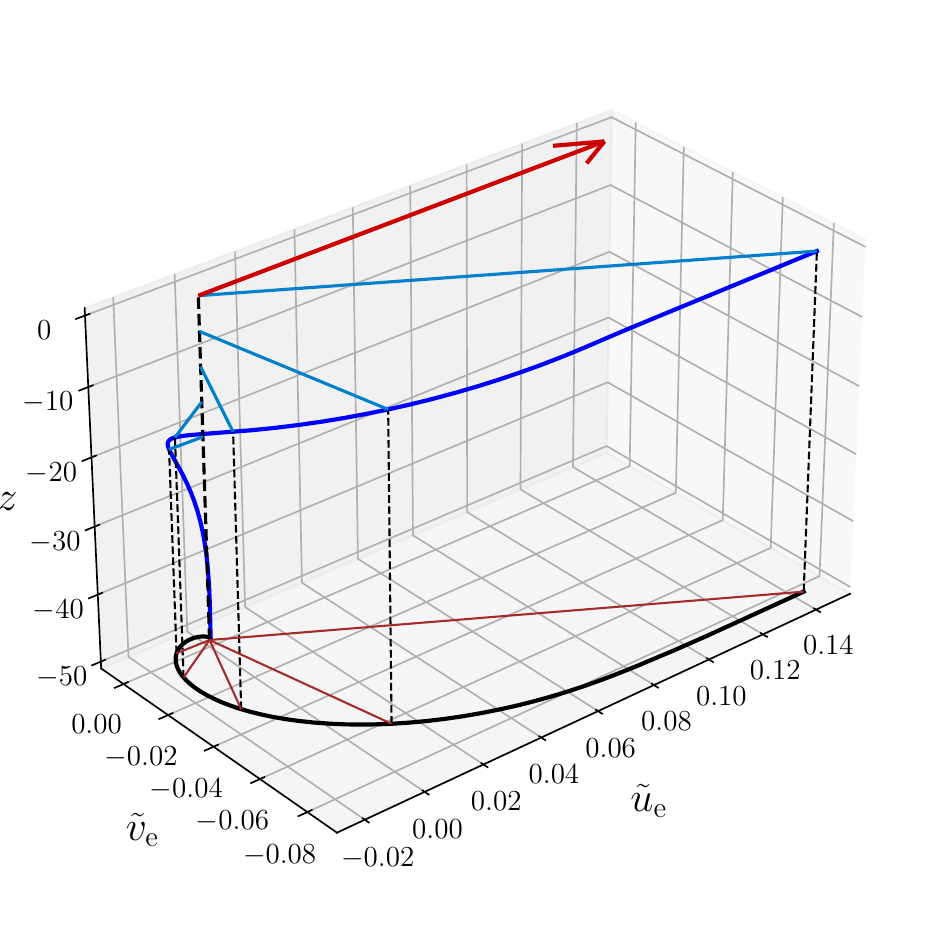}
        \caption{ $M = 50$, ${z_{\star}} = -1$.}
    \end{subfigure}
  \caption{The case of constant $m_1$ for $|W_{\rm w}|=1$, $\sigma = 0.014$ (corresponding to $U_\a' = 10\,{\rm m\,s^{-1}}$) and $H=-80$. }
    \label{fig constant spiral app}
\end{figure}

\begin{figure}
    \centering
    \begin{subfigure}{0.38\textwidth}
        \centering
        \includegraphics[width=\linewidth]{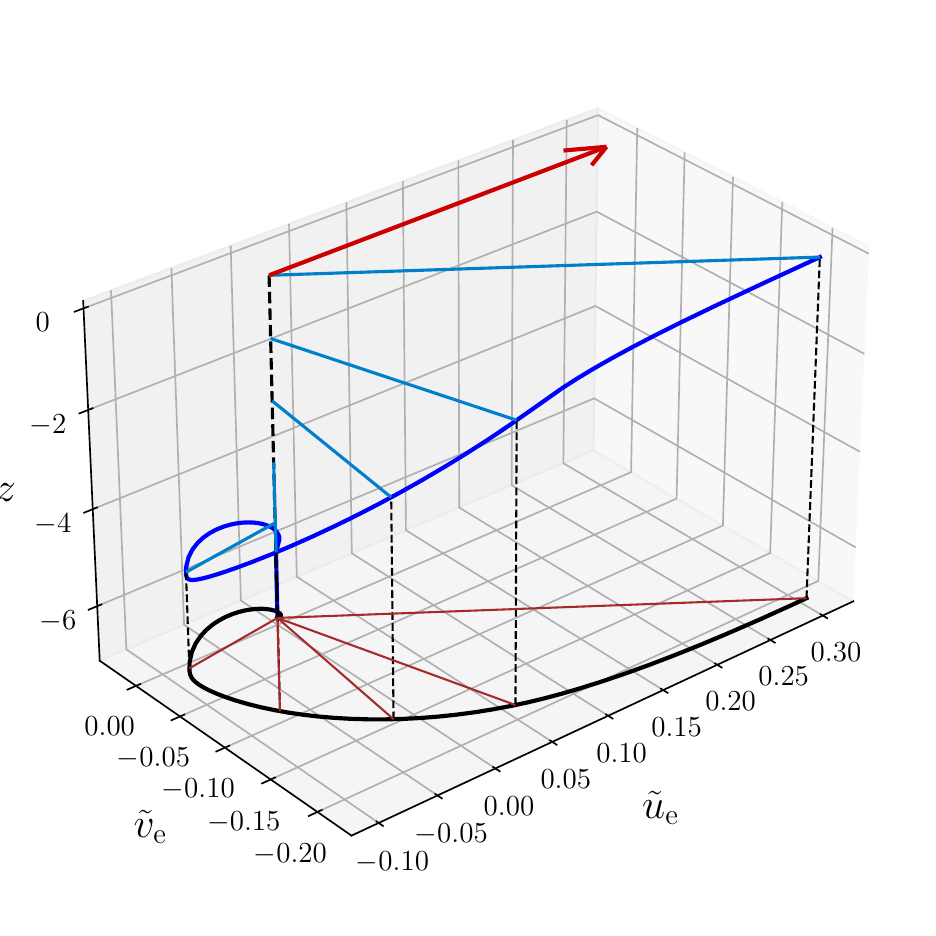}
        \caption{$M=10$, $z_0=-5$, $z_{\star}=-1$.}
    \end{subfigure}
    \begin{subfigure}{0.38\textwidth}
        \centering
        \includegraphics[width=\linewidth]{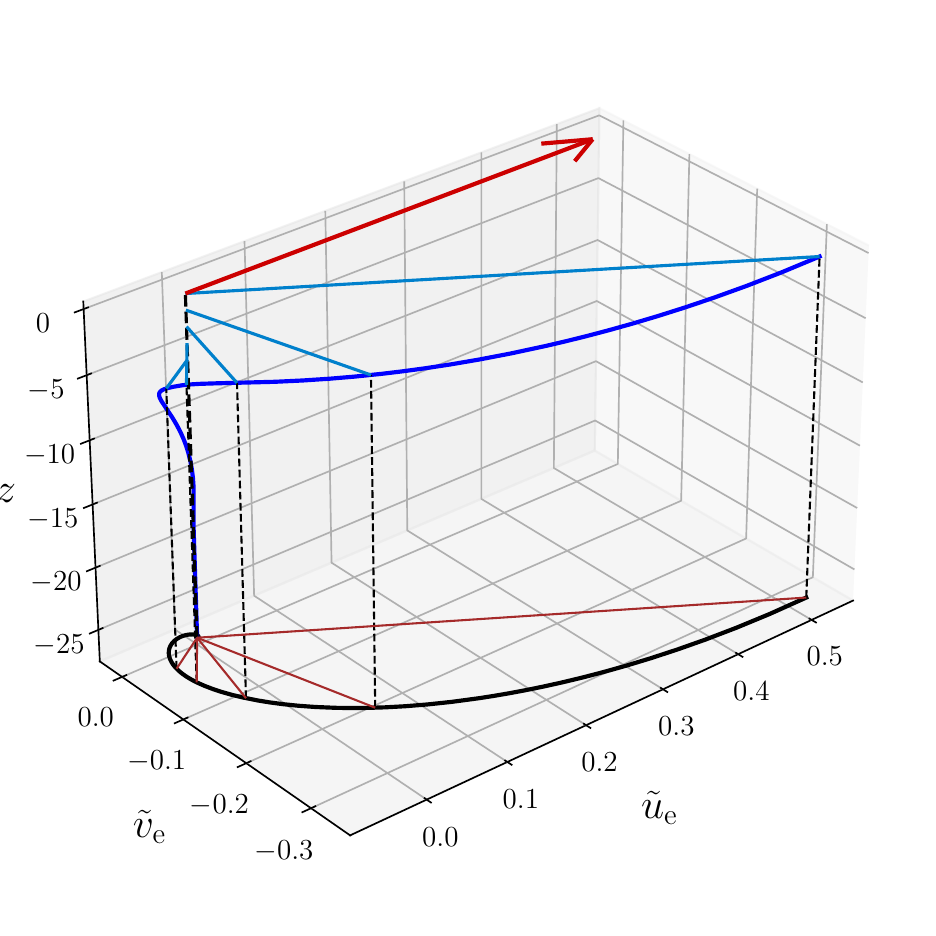}
        \caption{ $M=10$, $z_0=-20$, $z_{\star}=-4$.}
    \end{subfigure}
       \begin{subfigure}{0.38\textwidth}
        \centering
        \includegraphics[width=\linewidth]{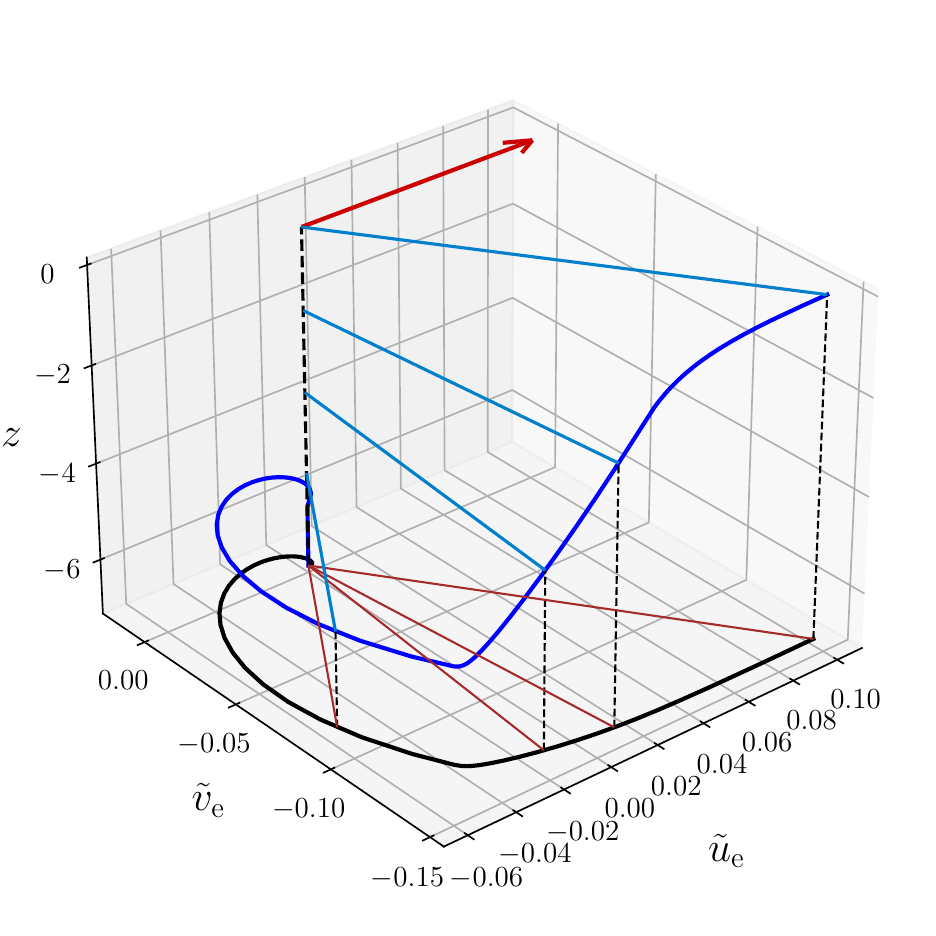}
        \caption{$M=50$, $z_0=-5$, $z_{\star}=-1$.}
    \end{subfigure}
       \begin{subfigure}{0.38\textwidth}
        \centering
        \includegraphics[width=\linewidth]{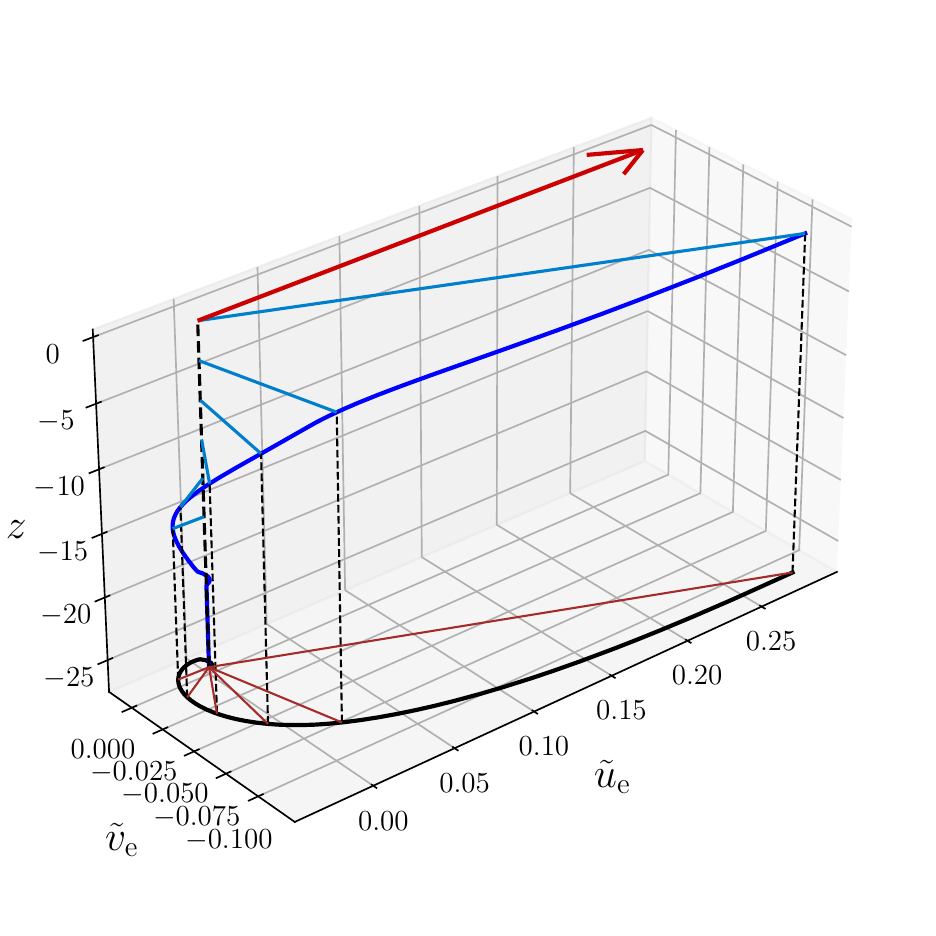}
        \caption{ $M=50$, $z_0=-20$, $z_{\star}=-4$.}
    \end{subfigure}
    \caption{The case of linearly decaying $m_1$ for $|W_{\rm w}|=1$, $\sigma = 0.014$, $H=-80$ and $\mathfrak{m}=0.01$. }
    \label{fig linear spiral app}
\end{figure}

\begin{figure}
    \centering
    \begin{subfigure}{0.38\textwidth}
        \centering
        \includegraphics[width=\linewidth]{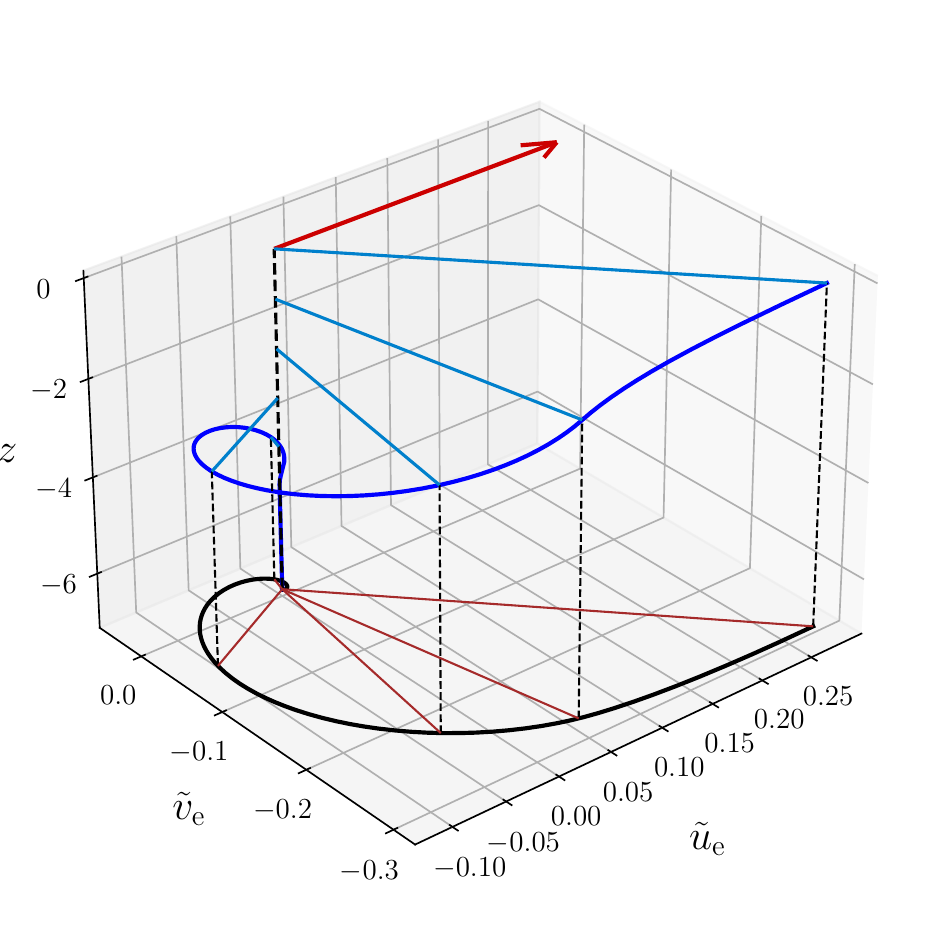}
        \caption{ $M=10$, $z_0=-5$, $z_{\star}=-1$.}
    \end{subfigure}
    \begin{subfigure}{0.38\textwidth}
        \centering
        \includegraphics[width=\linewidth]{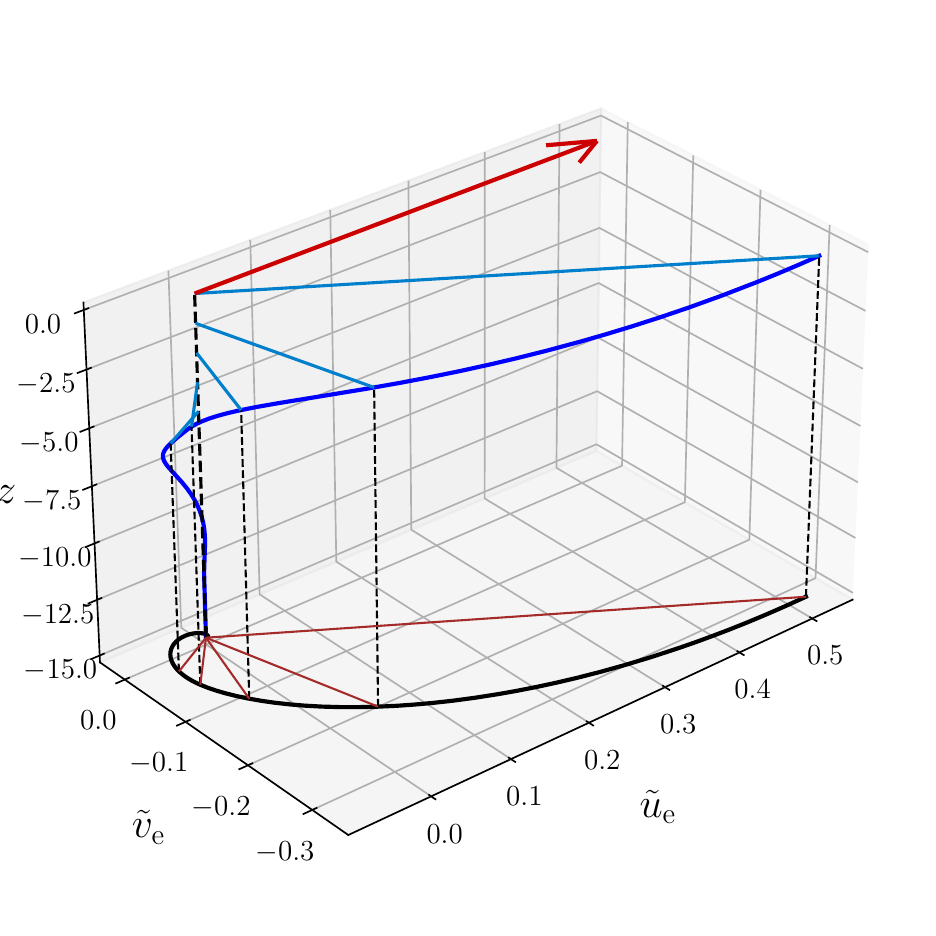}
        \caption{$M=10$, $z_0=-20$, $z_{\star}=-4$.}
    \end{subfigure}
       \begin{subfigure}{0.38\textwidth}
        \centering
        \includegraphics[width=\linewidth]{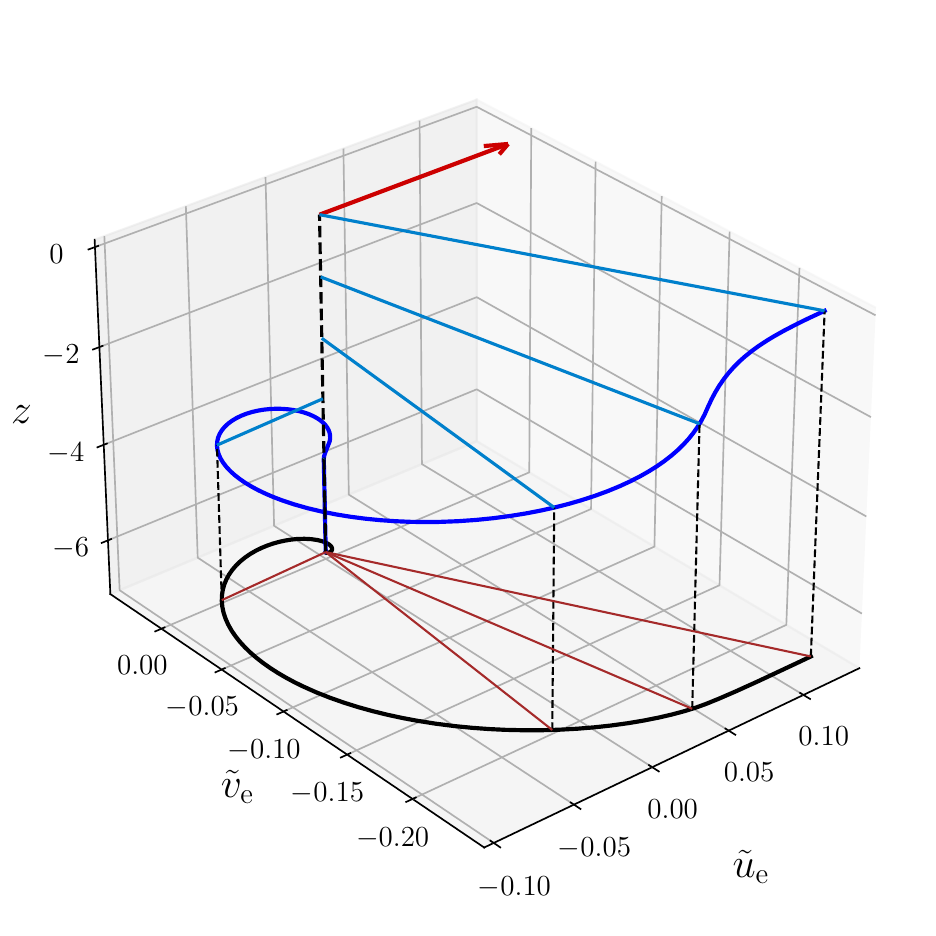}
        \caption{ $M=50$, $z_0=-5$, $z_{\star}=-1$.}
    \end{subfigure}
       \begin{subfigure}{0.38\textwidth}
        \centering
        \includegraphics[width=\linewidth]{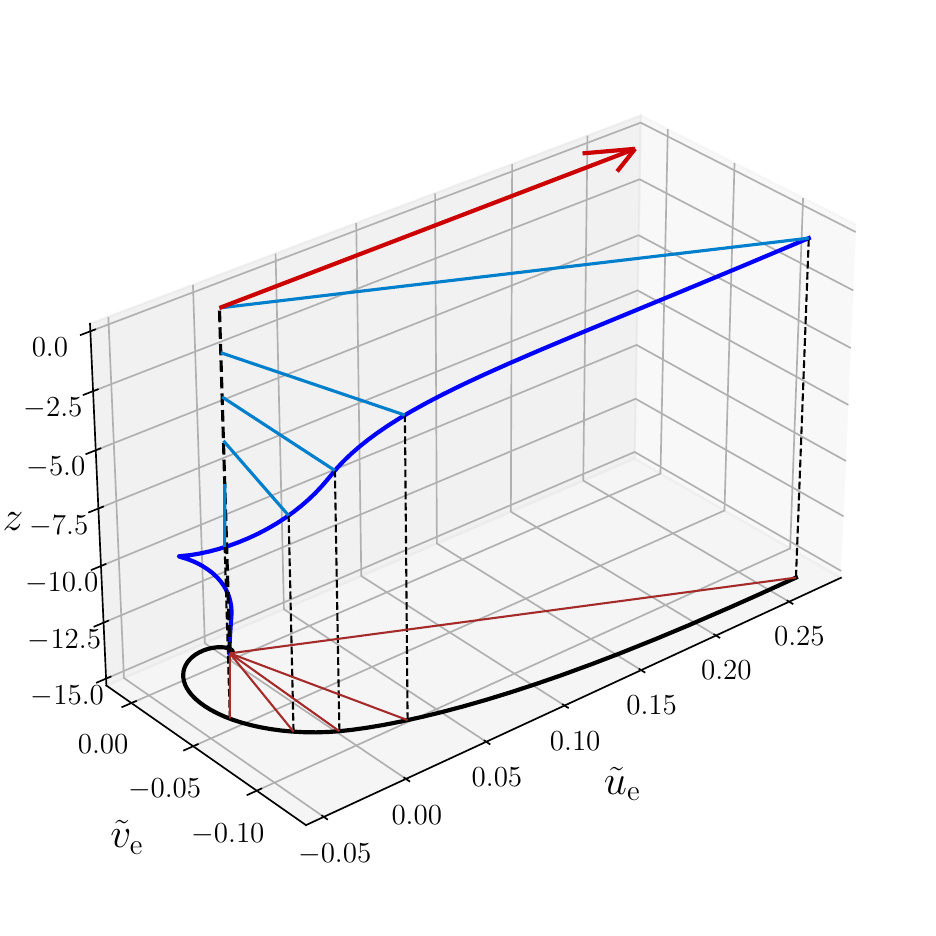}
        \caption{$M=50$, $z_0=-20$, $z_{\star}=-4$.}
    \end{subfigure}
    \caption{The case of exponentially decaying $m_1$ for $|W_{\rm w}|=1$, $\sigma = 0.014$,$H=-80$ and $\mathfrak{m}=0.01$. }
    \label{fig exp 2d app}
\end{figure}
\clearpage
}

\end{appendix}

\newpage

\bibliographystyle{jfm}
\bibliography{bib_REV}

\end{document}